\documentclass{CQG_version/iopjournal}

\usepackage{orcidlink}
\usepackage{wrapfig}
\usepackage{graphicx}
\usepackage{xcolor}
\usepackage{comment}
\usepackage{float}
\usepackage{inputenc}
\usepackage{multicol}
\usepackage{caption}
\usepackage{soul}
\usepackage{amsmath}
\usepackage{mathrsfs}
\usepackage{acronym}
\usepackage{url}
\usepackage{subfig}
\usepackage{mwe}
\usepackage[normalem]{ulem}
\usepackage{ragged2e}
 \usepackage{booktabs}
\usepackage{amsmath,amssymb}

\begin{document}

\articletype{Paper} %e.g. Paper, Letter, Topical Review...

\title{Reducing the Virgo site infrastructure noise in preparation of the O4 observing run}

\author{Irene Fiori$^1$\orcidlink{0000-0002-0210-516X}, 
Federico Paoletti$^2$\orcidlink{0000-0001-8898-1963}, 
Roberto Passaquieti$^{2,3}$\orcidlink{0000-0003-4753-9428}
Maria Concetta Tringali$^{1,*}$\orcidlink{0000-0001-5087-189X}}
Lorenzo Pierini$^4$\orcidlink{0000-0003-0945-2196}
Francesca Bucci$^5$\orcidlink{0000-0003-1726-3838}
Massimo Lenti$^{5,6}$\orcidlink{0000-0002-2765-3955}
Alessandro Longo$^{6,7}$\orcidlink{0000-0003-4254-8579}
\vspace{0.5cm}

\affil{$^1$ European Gravitational Observatory (EGO), I-56021 Cascina, Pisa, Italy}

\affil{$^2$ INFN, Sezione di Pisa, I-56127 Pisa, Italy}

\affil{$3$ Universit\`{a} di Pisa, I-56127 Pisa, Italy}

\affil{$^4$ INFN, Sezione di Roma, I-00185 Roma, Italy}

\affil{$^5$ INFN, Sezione di Firenze, I-50019 Sesto Fiorentino, Firenze, Italy}

\affil{$^6$ Universit\`{a} di Firenze, Sesto Fiorentino I-50019, Italy}

\affil{$^7$ Universit\`{a} degli Studi di Urbino “Carlo Bo”, I-61029 Urbino, Italy}

\affil{$^*$ Author to whom any correspondence should be addressed.}

\email{maria.tringali@ego-gw.it}


\begin{abstract}

The heating, ventilation and air conditioning systems serving the experimental halls of the Virgo gravitational wave interferometer generate low-frequency noise -- namely below 100 Hz -- of seismic, acoustic, and electromagnetic origin. Such disturbances have repeatedly affected the interferometer sensitivity throughout its operational history, with particularly notable impacts during the third observing run. 
In preparation for the fourth run, a comprehensive investigation was carried out to identify the most critical noise sources within this infrastructure and to trace their transmission paths into the experimental areas. This manuscript presents the methodology and results of the noise characterization campaign, together with the design, implementation and assessment of targeted mitigation measures. 
%The outcomes demonstrate a significant reduction of noise at each stage of intervention \textcolor{red}{(non nel target... pero' magari localmente si...)}. 
The technical solutions adopted, along with the operational best practices developed, provide valuable guidance for the design of low-noise environments in future gravitational-wave observatories.

\end{abstract}

\section{Introduction}
\label{sec:Introduction}

%\subsection*{
%\textcolor{red}{Testo vecchio $\rightarrow$} \\
%Impact on detector sensitivity (ad es. BNS aumenta spegnendo AHUs in O3, citations) and Risk mitigation.\\
%Magnetic noise impact on detector sensitivity.}
%REFERENCES: Seismic studies for Fermilab Future Collider Physics -https://inspirehep.net/literature/451995 
%
%The characterization and mitigation of the HVAC noise is a major task in current detectors. Facility noise is a significant project risk, and designing low noise infrastructure is a primary goal for new detectors project.

A low noise environment is mandatory in scientific laboratories that host ultra-high precision measuring instruments,
including facilities dedicated to metrology, nanotechnology and scanning tunneling microscopy, particle accelerators, and in particular, gravitational wave (GW) detectors~\cite{Baklakov_1998, Waysand_2006, Que_2023, Amann_2020, Badaracco_2021}.
Sound, vibration, and electromagnetic (EM) noise produced by laboratory service machinery (e.g. air handling units, air compressors, vacuum devices, electricity system) can significantly impact the operation and performance of the measuring instrument. 
%The noise of interest is typically below approximately 100~Hz, a frequency range that, despite being largely imperceptible to humans, poses significant challenges for precision experiments. 
The noise of interest typically lies below approximately 100~Hz (hereafter referred to as {\it low-frequency}), a range that, although largely imperceptible to humans, poses significant challenges for precision experiments.
Notably, this frequency domain remains relatively underexplored in the scientific literature, both in terms of detailed characterization of noise sources and the development of effective mitigation techniques.

Gravitational wave interferometers detect signals ~\cite{GWTC-1,GWTC-2,GWTC-2-1,GWTC-3} by measuring with extreme precision the differential length of two kilometer-scale arms. Currently, the
%the second-generation detectors 
%\textcolor{red}{(abbiamo dimenticato GEO?)} \textcolor{blue}{GEO non entra in presa dati, dipene cosa vogliamo dire. Comunque l'hanno dismesso}
Virgo, LIGO, KAGRA and GEO~\cite{Virgo, Nardecchia, LIGO,KAGRA, GEO} detectors are in operation, while  next-generation observatories, such as Einstein Telescope~\cite{Punturo2010} and Cosmic Explorer~\cite{CosmicExplorer,CosmicExplorer_website}, are entering the design phase.
%\textcolor{red}{(Cosmic Explorer e' stato approvato?)} \textcolor{blue}{C'è un consorzio}

Experimental areas of current detectors 
span volumes from a few hundred to a few thousand cubic meters and are maintained under controlled conditions of air temperature, cleanliness, and relative humidity. This is achieved through heating, ventilation and air conditioning (HVAC) systems, managing air fluxes of a few thousand ${\rm ~m}^3/{\rm hr}$ and consisting 
%These HVAC plants typically consist 
of moving electro-mechanical machinery (e.g., pumps, fans, compressors) that generate vibrational, acoustic and EM noise. These disturbances, which often exhibit a pronounced low-frequency spectrum, can propagate into experimental areas and affect the detector sensitivity or its operation, as demonstrated by several studies~\cite{HVAC_Fiori2008,Envpaper2020, Nguyen_2021, KAGRA_O3GK}.

%During the AdV Phase I upgrade break, in the years 2020 to 2023, 
From 2020 to 2023, Virgo suspended operations to install and commission the Advanced Virgo plus, Phase I upgrades. During this period, a study and mitigation campaign was carried out to characterize and reduce noise originating from the detector's HVAC facilities.
The dual objective was to minimize the risk of noise impact on the upgraded Virgo detector 
and to draw lessons applicable to the next generation of higher-performance GW detectors.
 
The activity focused on two identical HVAC systems serving the experimental halls located at the end of each interferometer arm, with the aim of extending the most effective solutions to the other HVAC systems on site. The work involved characterizing the noise generated by the various components of the HVAC plants, identifying the associated propagation paths into the experimental halls, and implementing effective mitigation measures.
Because invasive interventions were not feasible—owing to cost constraints and the risk of interference with commissioning activities—only minor structural modifications were pursued. These were designed to be implemented within a few hours or distributed over a small number of working days, thereby minimizing HVAC downtime.

%The main goal of the study was to mitigate the acoustic, seismic, and magnetic noise within the experimental halls, where the most sensitive components of the interferometer are located. In particular, vibration noise affecting the hall floor—where the test mass and optical bench suspensions are placed—was a primary concern.
%A dedicated set of sensors was installed and integrated into the Virgo environmental monitoring network. Specific tests and diagnostic techniques were devised to study the noise propagation paths. These tests enabled the assessment of individual mitigation actions by comparing noise levels before and after implementation. Whenever possible, interventions were applied sequentially to allow for the disentangling of their individual effects.

The manuscript is organized as follows.
Section~\ref{sec:Virgo_and_noise} introduces the Virgo detector and its noise specifications. Section~\ref{sec:HVAC_overview} provides an overview of the Virgo HVAC system. Section~\ref{sec:Objectives_and_methodology} describes the objectives of the study, the tools employed and the methods adopted. Section~\ref{sec:vibro_acoustic_mitigations} details the implementation and results of the vibro-acoustic noise mitigation actions, while Section~\ref{sec:noise_investigations} discusses noise characterization studies of vibro-acoustic and magnetic emissions, which provided insights into further noise reduction strategies. Finally, Section~\ref{sec:Conclusions} summarizes the conclusions and outlines future perspectives.

\section{The Virgo detector and the noise challenges}
\label{sec:Virgo_and_noise}

Located in Cascina, near Pisa (Italy), Virgo is a 3-km dual-recycled Michelson interferometer with Fabry–Perot cavities in the arms~\cite{Virgo,Nardecchia}.
Gravitational waves induce differential changes in the arm lengths on the order of $10^{-18},\mathrm{m}$, producing measurable interference patterns at the interferometer output. To detect such minute signals, Virgo employs state-of-the-art technologies. Among these are the ultra-high-vacuum system enclosing the laser beam path—which also provides shielding from ambient acoustic noise—and the multistage suspension system (the \emph{Super Attenuator}, SA), which isolates the test-mass mirrors (TM) and optical benches from ground vibrations by up to 12 orders of magnitude, starting from a few hertz~\cite{SA}.
%To detect such minute signals Virgo employs ultra-high vacuum systems, multistage seismic isolation, low-noise optics, and active control systems. However, its sensitivity can still be significantly affected by environmental noise, such as seismic, acoustic, and electromagnetic disturbances, including those generated by the site infrastructure.

The residual noise of the Virgo instrument, which defines the shape of its sensitivity curve, consists of multiple sources \cite{Nardecchia}. Some sources are intrinsic to the interferometer design - e.g. thermal fluctuations in the mirrors and suspensions, shot noise, radiation pressure, Newtonian noise \cite{Harms2019, Tringali2020,Koley2023, Fiorucci2018} - while others are related to possible imperfections in its complex technical apparatus -  e.g. control noise - or to the influence of noise fields in the surrounding environment.

Environmental noise can affect GW interferometers in multiple ways. Acoustic and vibrational fields can enhance the motion of surfaces (e.g. vacuum chambers) accidentally intercepted by stray laser beams, leading to back-scattered light that couples to the main beam~\cite{Soni_2021,Envpaper2020}. %Ambient EM fields may instead couple directly to magnetic actuators used for the active control of the seismic isolation system, or induce noisy currents in signal cables and low-noise electronics~\cite{Cirone_2019, Nguyen_2021,LIGO_DetcharO2}.\\
Ambient electromagnetic fields may induce noisy currents in signal cables and low-noise electronics or 
couple directly to magnetic actuators used for the active control of the seismic isolation system, and in the case of Virgo, to the magnets glued onto the mirror test masses~\cite{Cirone_2019, Nguyen_2021,LIGO_DetcharO2,AHU_magnetic_emission,Handbook_Fiori}. These effects motivate stringent noise control within the experimental buildings, particularly near the most sensitive interferometer components, such as the test masses and the clean rooms hosting the input laser system and the readout systems.

\begin{figure}[htp!]
    \centering
    \includegraphics[width=0.8\linewidth]{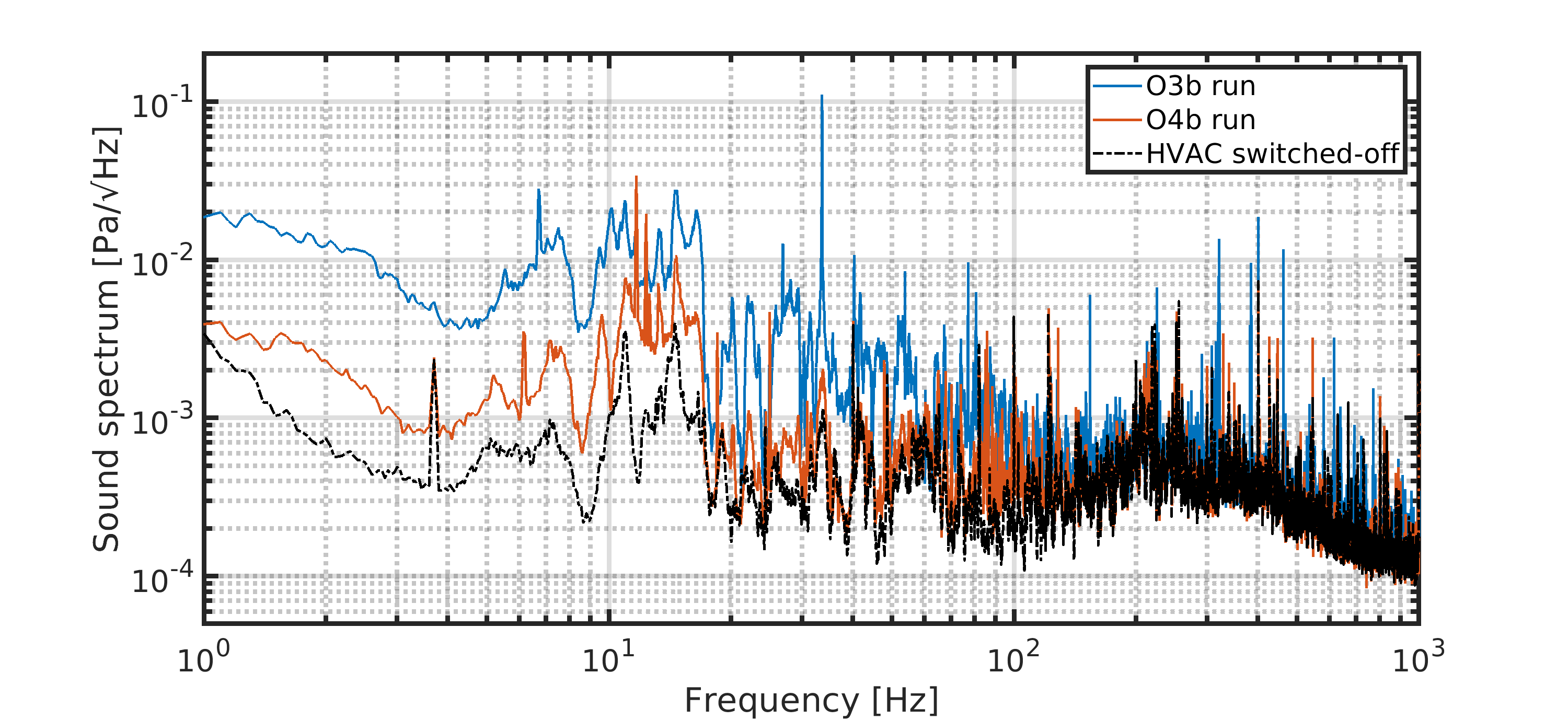}
    \caption{Acoustic spectrum of the NEB experimental hall. The blue and red curves correspond to the acoustic noise during the O3b run and O4b run, respectively. The black curve indicates the acoustic noise level when the HVAC system is switched-off. }
   % \caption{VECCHIA caption Acoustic spectrum in the experimental hall during HVAC switch-off. The red and blue curves correspond to the acoustic spectra with the HVAC system switched off and on, respectively. The inset shows a zoomed-in view highlighting narrow spectral lines at specific frequencies: 6.7 Hz (belts), 24.9 Hz (motor), 26.8 Hz (the second harmonic of the fan), and 34.05 Hz (the fifth harmonic of the belt frequency). The line at 33.55 Hz was investigated XXXXX .\textcolor{blue}{Promomeria: ventilatore vecchio con belt 1790 cm, fan 280 cm, motor 150 cm, rpm 1440., inverter 50 Hz}}
    \label{fig:NEBmic_AHUswitchedOFF}
\end{figure}

\begin{figure}[htp!]
    \centering
    \includegraphics[width=0.8\linewidth]{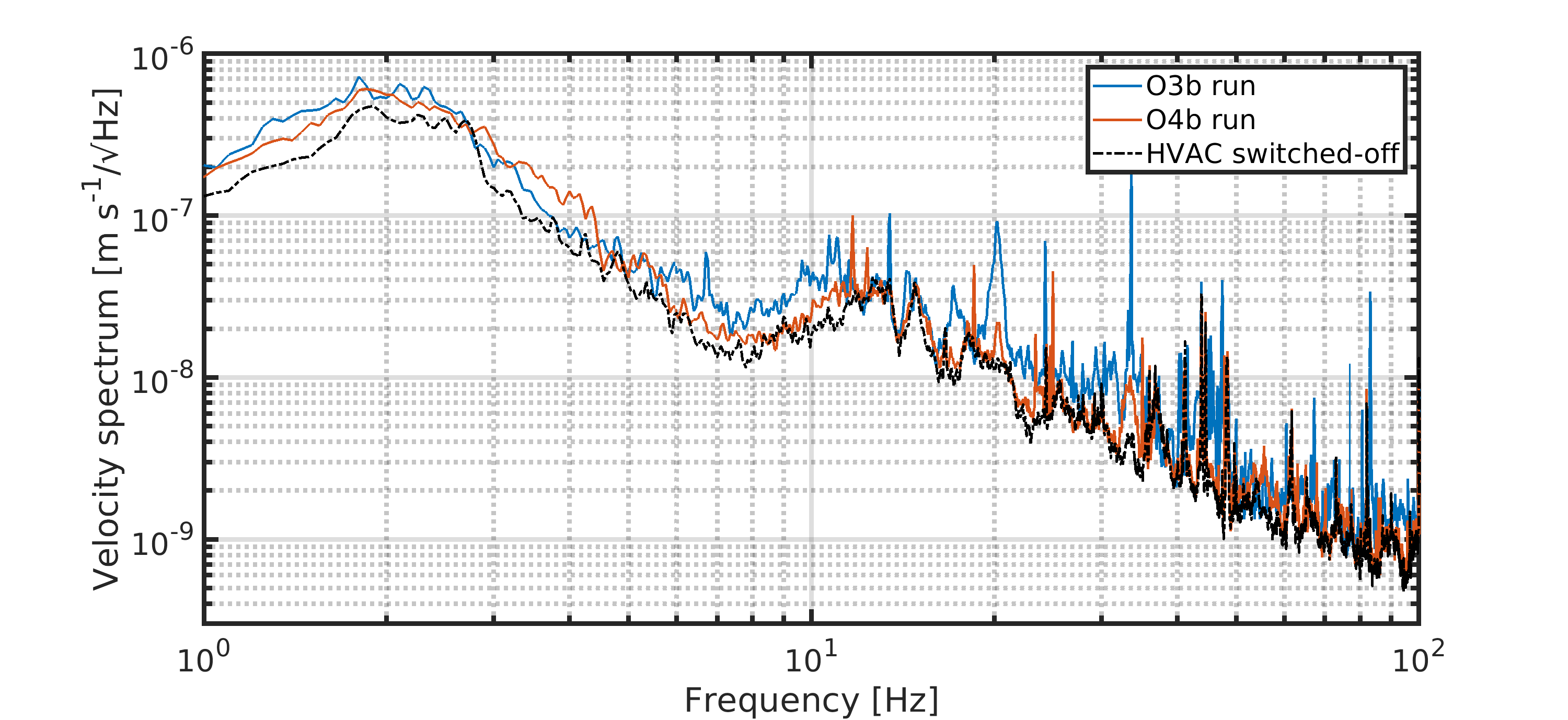}
    \caption{Seismic spectrum (vertical channel) of the NEB experimental hall floor.
    The blue and red curves correspond to the seismic noise during the O3b and O4b runs, respectively. The black curve shows the seismic noise level when the HVAC system is switched off. The vertical axis of the seismometer was used for the plot. 
    %\caption{VECCHIA CAPTION:Seismic noise of the Virgo NEB experimental hall floor, vertical axis, showing the contribution of the HVAC system. The red and blue curves correspond to the seismic spectra with the HVAC system switched off and on, respectively. The inset shows a zoomed-in view highlighting narrow spectral lines at specific frequencies associated to HVAC devices: 13.4 Hz (AHU fan), 24.9 Hz (AHU motor), 26.8 Hz (second harmonic of the AHU fan), and 34.05 Hz (fifth harmonic of the AHU belt frequency). \textcolor{red}{The line at 33.55 Hz was investigated XXXXX.}
    }
    \label{fig:NEBseis_AHUswitchedOFF}
\end{figure}

%This highlights the need to minimise noise levels inside the experimental halls, particularly near the most sensitive parts of the interferometer, such as the test masses and the clean rooms that house the input laser and the vacuum chamber containing the photo-diodes used for the detection of gravitational waves.
%
%\textcolor{red}{Una versione piu' sintetica proposta da Chat:
HVAC plants are a significant source of broadband low-frequency acoustic, seismic, and EM disturbances, often characterized by both broadband contributions and discrete spectral features. Figures~\ref{fig:NEBmic_AHUswitchedOFF} and~\ref{fig:NEBseis_AHUswitchedOFF} show representative acoustic\footnote{The HVAC induced noise dominates the acoustic noise of the hall in the 1-100~Hz frequency range. Below 1~Hz the dominant acoustic contribution is due to influences of wind and presence of air fluxes, which we do not explore here. Above approximately 100~Hz the dominant acoustic noise is due to cooling fans of electronic modules and racks located inside the experimental hall.} and seismic\footnote{The HVAC induced vibration dominates the seismicity of the experimental building floor in the 10-100~Hz frequency range.
Other seismic sources are 
sea-induced microseism~\cite{Cessaro1994} in the 0.1 to 1~Hz range, road traffic in the 1 to 10~Hz range~\cite{Fiori2004, Koley2017}.}
noise spectra measured in the Virgo North End Building. Turbulent air flows within ducts and fan casings generate colored noise, especially at low frequencies, and can excite structural resonances of mechanical components and acoustic modes of rooms, appearing as broad spectral peaks. Rotating machinery introduces narrow spectral lines at rotation frequencies and harmonics, while inverters generate magnetic lines at their operating frequency. Contributions may be intermittent, following the duty cycle of devices  such as chillers and boilers. These phenomena are further illustrated and discussed in Sections~\ref{sec:vibro_acoustic_mitigations} and~\ref{sec:noise_investigations}.

\section{The HVAC systems of Virgo}
\label{sec:HVAC_overview}

%%% provo a fare parte piu' generica prima e poi dettagli su NE/WE HVAC (e citare Figura 1)
The Virgo detector is equipped with a total of nine HVAC systems that serve clean areas of varying volumes and cleanliness requirements \footnote{Not all HVAC systems serve ISO-classified cleanroom areas.}. These include: the experimental area in the central building (CEB, 8370~m$^3$); the CEB clean room (507~m$^3$); the experimental halls of the North and West end buildings (NEB and WEB, 4995~m$^3$); the laser clean room (INJ 507~m$^3$, ISO 7);
the detection clean room (DET, 468~m$^3$, ISO 7); the two Filter Cavity clean areas (57~m$^3$, ISO 7) and the Mode-Cleaner building (MCB, 782~m$^3$).
All clean zones are maintained within a few tenths of a degree from the temperature set point
%\textcolor{red}{(typically $20 \pm 0.1 ^\circ $C)}
%at a temperature of (20 ± 0.1)°C, and (except for the CEB and MCB) 
and a slight overpressure of a few Pa.
The laser and detection laboratories have additional constraints on relative humidity (RH $\sim 50\%  \pm 5\%$). 

\begin{figure}[hb!]
    \centering
    \includegraphics[width=0.55\linewidth]{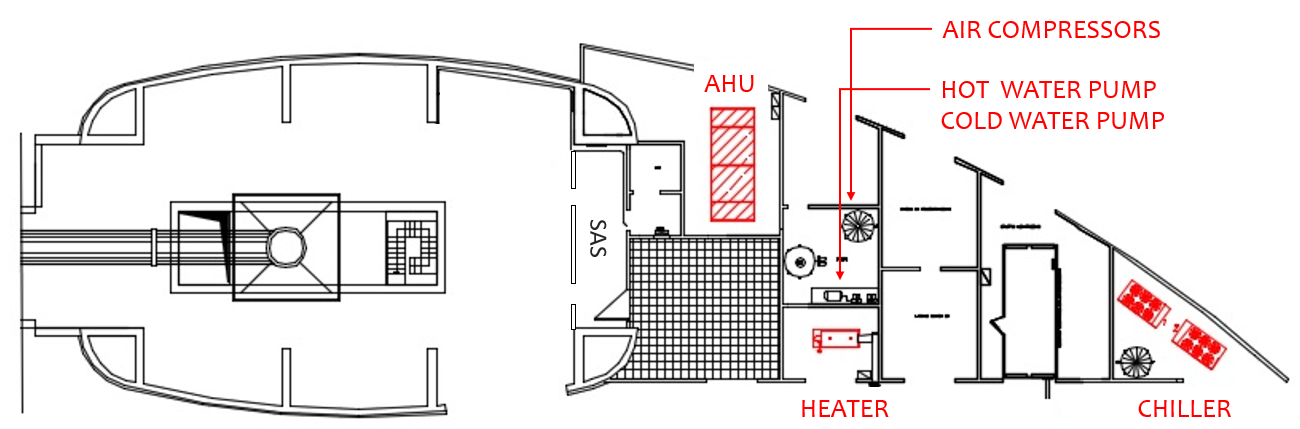}
    \includegraphics [width=.39\textwidth]{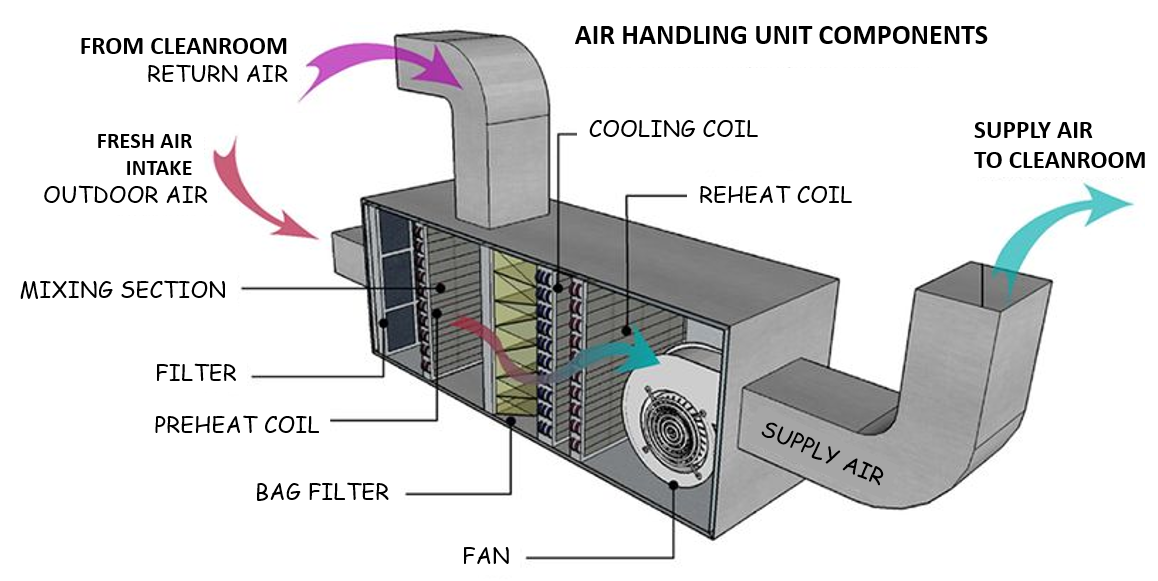}
    \caption{Left. Plan view of North end building and the technical area which hosts HVAC devices. Right. Sketch drawing of an AHU.}
    \label{fig:NEB_AHU}
\end{figure}

Most HVAC systems were installed over 25 years ago and were not designed with low-noise performance as a primary goal.
All systems follow a similar architectural concept: each air handling unit (AHU, see Figure~\ref{fig:NEB_AHU}, right) consists of a multi-compartment stainless-steel enclosure and a centrifugal supply fan, with some units also featuring a return fan. The fan is driven by an electric motor coupled through a belt–pulley transmission\footnote{The DET and INJ AHU systems of the Virgo interferometer employ direct-drive fans. These systems are outside the scope of this manuscript.}. The fan and motor are housed within a dedicated section of the enclosure, hereafter referred to as the \emph{fan compartment}.
The fan speed is regulated by a variable frequency drive (VFD), also known as {\it inverter}, which allows 
the optimal setting of the airflow\footnote{All HVAC are set at a constant speed operation, thus avoiding unpredicted change of associated noise.}.
Air is distributed to and extracted from the clean area through single-layer 0.8 mm thick stainless steel ducts, which branch into smaller ducts that terminate in air vents, typically equipped with high-efficiency particulate air (HEPA) filters.
The air thermal control is achieved via heat exchangers fed by hot and cold water circuits, serviced by boilers and chillers, respectively. Hydraulic pumps provide water circulation.

The oldest AHUs — those serving the CEB experimental hall, the CEB clean room, and the MCB — are located inside or directly adjacent to the experimental areas. For example, the AHU of MCB shares the same floor as the vacuum chamber, while the AHU of CEB hall shares a basement level with the experimental hall walls. This configuration offers poor seismic isolation and limits the possibility of effective noise mitigation due to the restricted installation spaces.

\noindent
The AHUs serving the laser and detection clean rooms and the filter cavity areas have been installed outside the buildings and placed on dedicated platforms consisting of a sand pit topped with a reinforced concrete slab, offering improved vibration isolation.

At the time of this study access to the Virgo central experimental area was restricted due to ongoing commissioning activity. Therefore, the two similar HVAC plants serving the NEB and WEB experimental halls were adopted as a test bench for the investigation and mitigation actions described in this manuscript.

%Figure \ref{fig:NEB_AHU} (left) illustrates a plan view of NEB with the layout of HVAC devices. 
%\noindent
%The NEB and WEB buildings host the end test masses of the Fabry–Perot cavities. They are identical in construction and layout: each is 17 m high, with a footprint 17 m wide and 25 m long along the arm direction. 
%A small room, the Safety Access (SAS) room, is placed at the entrance and is separated from the experimental hall by a light partition wall.

Figure~3~(left) shows a plan view of the NEB, including the layout of the HVAC equipment. The NEB and WEB house the end test masses of the Fabry–Perot cavities and are identical in construction and layout. Each building is 17~m high, with a footprint 17~m wide and 25~m long along the arm direction. A small Safety Access (SAS) room is located at the entrance and is separated from the experimental hall by a light partition wall.

\noindent
Each end building has a double foundation system: a tower platform, supported by deep piles reaching stable gravel layers at a depth of 52~m, and a building platform, supported by shorter piles extending to a depth of 30~m. The tower platform supports the vacuum chamber, the SA, and the clean room, while the building platform carries the walls and the roof. The two platforms are separated by a gap of a few centimeters; this configuration provides seismic decoupling above approximately 10 Hz~\cite{Tringali2020}.

\noindent
A separate platform with a shallow foundation hosts the technical area, located at the rear of the experimental hall, as shown in Figure~\ref{fig:NEB_AHU}~(left). This area includes dedicated rooms housing the electrical and thermo-mechanical infrastructure—such as power transformers, uninterruptible power supply (UPS) units, diesel generators, and water pumps—while the chilled and hot water units for the HVAC system are installed outdoors but remain part of the same area.

\begin{figure}[htp!]
    \centering
    \includegraphics[width=0.35\linewidth]{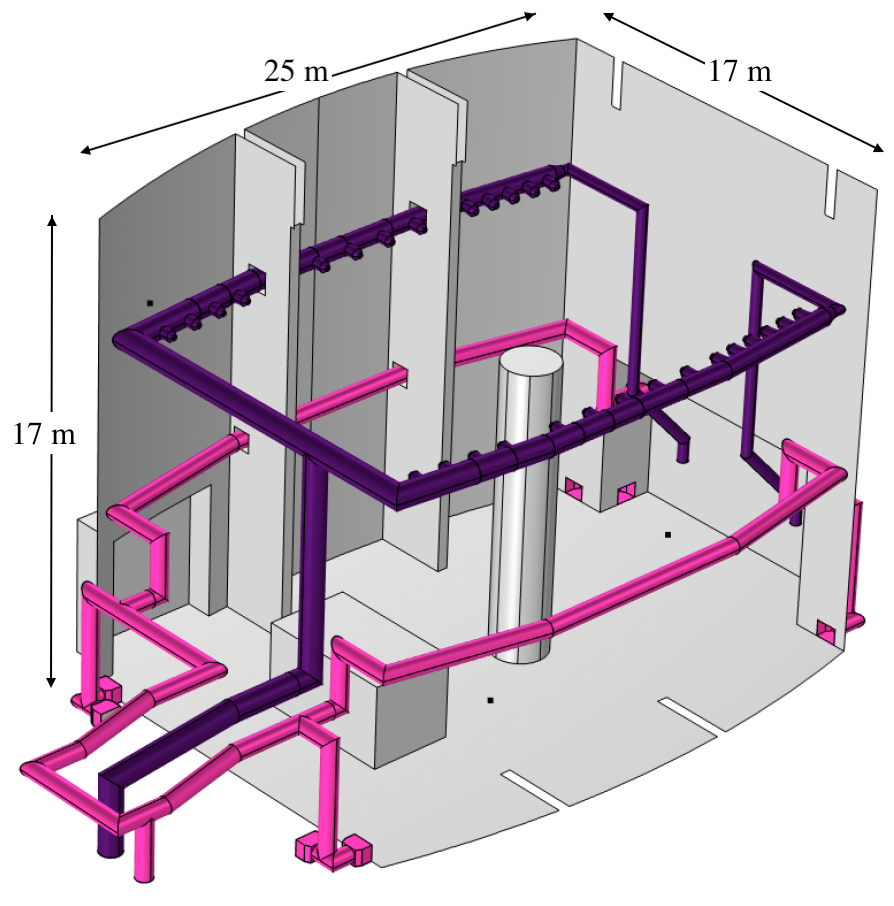}
    \includegraphics[width=0.45\linewidth]{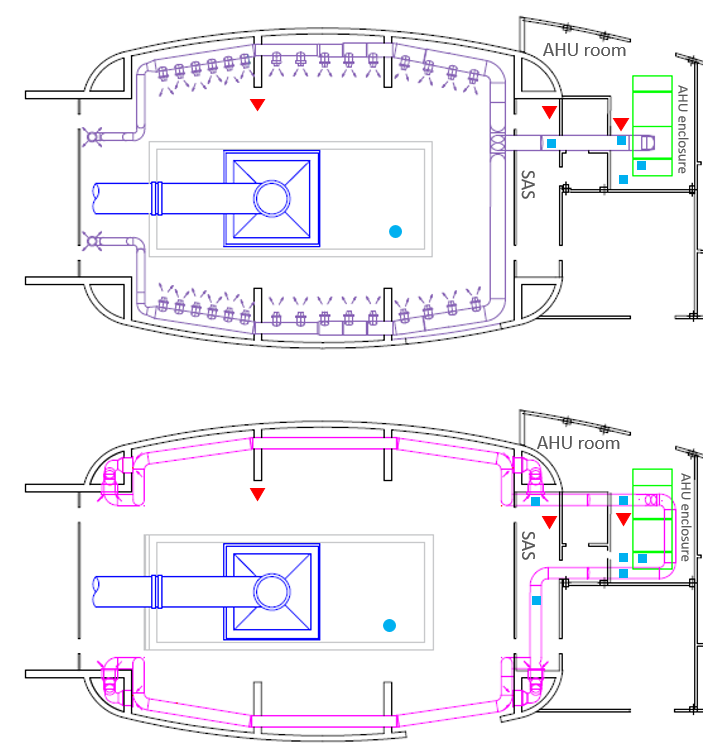}
    \caption{Left. 3D model of the supply (purple) and return (magenta) air ducts within the NEB experimental hall \cite{Lionel2023}. Right. Plan views of NEB experimental hall showing the same air duct network (top - supply, bottom - return) and the technical area housing the AHU. Also indicated is the temporary installation of sensors inside the AHU room and on the air ducts: triangles represent microphones and rectangles represent accelerometers. The location of the permanent sensors in the experimental hall is also shown: one microphone (red triangle) and one triaxial velocimeter (blue circle). The tower platform is shown as a grey rectangle, the vacuum chamber and pipe are indicated in blue.}
    \label{fig:3Dmap}
\end{figure}

\noindent
The airflow generated by the air handling unit (AHU) is distributed into the experimental rooms through two duct networks: the return network, terminating in six outlets (magenta in Figure~\ref{fig:3Dmap}), and the supply network, terminating in 32 outlets (violet in Figure~\ref{fig:3Dmap}). These outlets act as localized acoustic sources that can excite room modes~\cite{Lionel2023}. The HVAC system maintains the experimental halls at a temperature of $22 \pm 0.2^{\circ}\mathrm{C}$, and an overpressure of 7~Pa, thereby improving air cleanliness.

\section{Objectives and methodology}
\label{sec:Objectives_and_methodology}
%One strategy that paid off was to optimize the machine's operating point, with the goal of minimizing the noise input into the halls, while still matching required values of environmental parameters - temperature stability and overpressure - needed to preserve the cleanliness levels. For this we acted on the fan speed and control PIDs.
%% potremmo prendere uno spunto dal lavoro di Roberto per tuning di DET.
%\textcolor{red}{Vogliamo dedicare una sezione al lavoro di Roberto sul tuning del controllo della HVAC di DET?}
% ottimizzazione parametri di lavoro
%Una strategia che ha dato buoni frutti e' stata quella di ottimizzare il punto di lavoro della macchina, con l'obiettivo di minimizzare il rumore immesso nella hall pur rispettando i requisiti ambientali (stabilita' temperatura e sovrappressione - necessaria per preservare il livello di pulizia). Per questo abbiamo agito sulla velocita' del ventilaore e i PID di controllo. (\textcolor{red}{ Vogliamo mettere qualcosa sul lavoro di Roberto in DET?})

%%%%%%%%%%%%%%%%%%%%%%%%%%%%%%%%%%%%%%%%%%%%%%%%%%%%%%%%%%%

The goal of the noise mitigation campaign was to minimize HVAC-induced noise inside the NEB and WEB experimental halls, ideally reaching %approaching to 
the level observed when the systems are switched off. 

The campaign relied on the effective monitoring of all sources and the multiple noise paths, as well as of the hall ambient parameters and HVAC operation.
%
%\noindent
A permanent distributed monitoring sensor network is
deployed inside the Virgo experimental and technical areas~\cite{Envpaper2020}. It includes fast\footnote{Fast and slow sensors are sampled at $\geq 1$kHz and 1~Hz, respectively.} acoustic, seismic, and magnetic probes for measuring the ambient noise, and several slow probes to track HVAC operation (e.g., fan speeds, air-flows, air velocity) and environmental parameters (e.g., hall overpressure and temperature). 
During the study campaign, additional temporary sensors were installed near HVAC devices and along suspected noise transmission paths—for example, accelerometers on ducts, microphones in the AHU and SAS rooms, see Figure \ref{fig:3Dmap}, right.
This setup allows a systematic assessment of mitigation actions by comparing noise levels and environmental conditions before and after each intervention. 
%This setup enables a systematic evaluation of mitigation actions by comparing noise levels and environmental conditions before and after each action.
%Slow sensors enable to monitor the stability and consistency of ambient parameters, such as temperature, overpressure inside the hall, after each action.
%Figure \ref{fig:3Dmap} illustrates the sensors deployment at NEB.

In order to achieve effective mitigation, both noise sources and propagation paths must be addressed. 
The adopted strategy consisted in first correcting the most evident issues and verifying the impact of each action through switch-off tests.
Selective switch-off of devices also allowed to identify those responsible for the most significant noise contributions and to associate specific spectral components to each source. 
%\ref{fig:NEBmic_AHUswitchedOFF},  Figure  \ref{fig:NEBseis_AHUswitchedOFF} and Figure \ref{fig:water_system}.
 
The investigation of noise propagation paths is challenging, as they are often mutually coupled.
This could be the case of air turbulence at the AHU fan stage, which may generate sound waves propagating inside the air ducts while simultaneously inducing mechanical vibrations that travel through the ducts or the floor structures.
%This could be the case of air turbulence at the AHU fan stage that may generate sound waves propagating inside the air ducts, and at the same time induce mechanical vibrations traveling through the ducts or the floor structures. 
An experiment described in Section \ref{subsec:Ducts_disconnection} involved interrupting selected paths and studying the resulting changes in the measured noise.
%An experiment adopted, described in Section \ref{subsec:Ducts_disconnection} consists of interrupting selected paths and studying the resulting changes in the measured noise.

%and a procedure known as noise injection. The latter involves exciting one or more paths by introducing controlled noise levels using loudspeakers or shakers.\\

Not all mitigation actions produced a measurable impact. It should be noted that the evidence of noise reduction observed at witness sensors following a given mitigation also depends on the sequence in which the actions are implemented. In principle, measures providing the largest contribution should be applied first; however, the optimal sequence is not always known or predictable in advance, nor is it always practically feasible. As a consequence, for some actions only the cumulative effect of multiple interventions could be evaluated.

Finally, all mitigation actions must preserve the functionality of the HVAC systems, ensuring that environmental parameters remain within their operating specifications. Long-term testing was therefore required to validate both the achieved noise reduction and the HVAC system performance over seasonal timescales.

\section{Vibro-acoustic mitigations}
\label{sec:vibro_acoustic_mitigations}

%\textcolor{red}{Spostiamo il paragrafo seguente nella sezione 2 ?}
\begin{comment}
The overall noise reduction achieved is illustrated by 
Figure \ref{fig:NEBmic_AHUswitchedOFF} and Figure \ref{fig:NEBseis_AHUswitchedOFF}, respectively for the acoustic and the seismic noise fields of the NEB experimental hall. The WEB is similar.
These Figures show the noise spectra measured during the O3b run (from November 1st, 2019 to March 27th, 2020))
and the quieter noise spectra measured during the O4b run (2024-2025) as a result of the mitigation campaign. Also shown is the noise levels corresponding to the HVAC system being turned off, which quantifies the residual contribution.
This Section describes the individual mitigation steps leading to this result.
\end{comment}

The achieved reduction in acoustic and vibrational noise is illustrated in Figures~\ref{fig:NEBmic_AHUswitchedOFF} and~\ref{fig:NEBseis_AHUswitchedOFF}, which show, respectively, the acoustic and seismic noise fields measured in the NEB experimental hall (the WEB case is similar). Spectra from the O3b observing run (November 1, 2019–March 27, 2020) are compared with the quieter spectra obtained during the O4b run (April 10, 2024–January 28, 2025), following the mitigation campaign. For reference, the noise level measured with the HVAC system switched off is also reported, providing an estimate of the residual contribution.
This section describes the individual mitigation steps that led to these results.

\subsection{Seismic decoupling actions}
\label{sec:decoupling}

%% springs under AHU
A well-established approach to mitigating vibration transmission through structural connections involves placing an elastic, energy-dissipating element (e.g., damped springs) between the active source (e.g., motor) and the receiving structure (e.g., floor) \cite{AHUspring}.
A set of soft springs under the AHU's fan and motor assembly was already in place. 
Additionally, 
%~\textcolor{red}{(MOSTRIAMO FOTO? Diciamo che ne abbiamo comunque verificato l'efficacia? Abbiamo stima della frequenza di taglio?)}. 
a set of damped springs was inserted underneath the AHU enclosure. The springs were sized at a cutoff frequency of approximately 3~Hz. 
%\textcolor{blue}{(Valore verificato con Davide solo per le molle sotto la box e non per quelle interne!)}.
This action did not produce a noticeable effect in the experimental hall, most likely because of the predominance of other seismic transmission paths within the AHU enclosure, which had not yet been addressed at the time of this intervention, as discussed below.

%\noindent
% - ducts disconnection from SAS wall (openings)
Other rigid connections associated with the AHU include air ducts and water pipes; the latter are discussed in Section \ref{sec:Seismic_water}. One supply and one return air duct exit the AHU room. The return duct splits into two branches within the AHU room, while the supply duct divides inside the experimental hall, as illustrated in Figure \ref{fig:3Dmap}.

Visual inspection revealed that the ducts present additional unintended points of contact with the surrounding walls, acting as efficient structure-borne vibration transmission paths (i.e. vibration short-circuits). One such vibration short-circuit was identified at the penetration of the lightweight partition wall separating the SAS room from the hall.
This second intervention, while significantly reducing the vibration of the partition wall, did not produce noticeable changes in the noise level within the experimental hall.
%A further intervention would have involved decoupling the ducts from the walls of the hall, replacing the rigid supports with elastic supports, though an invasive intervention also because of to the need to work at heights.

%% sleeves
%\noindent  
%(\textcolor{red}{Questo sarebbe da verificare, mi pare all'epoca avessimo un unico accelerometro sui condotti dentro AHU room, ricordo che si valuto' a mano che i condotti ora vibrano anche di piu', almeno alle basse frequenze}) 

\begin{figure}[ht!]
    \centering
    \includegraphics[width=0.35\textwidth]{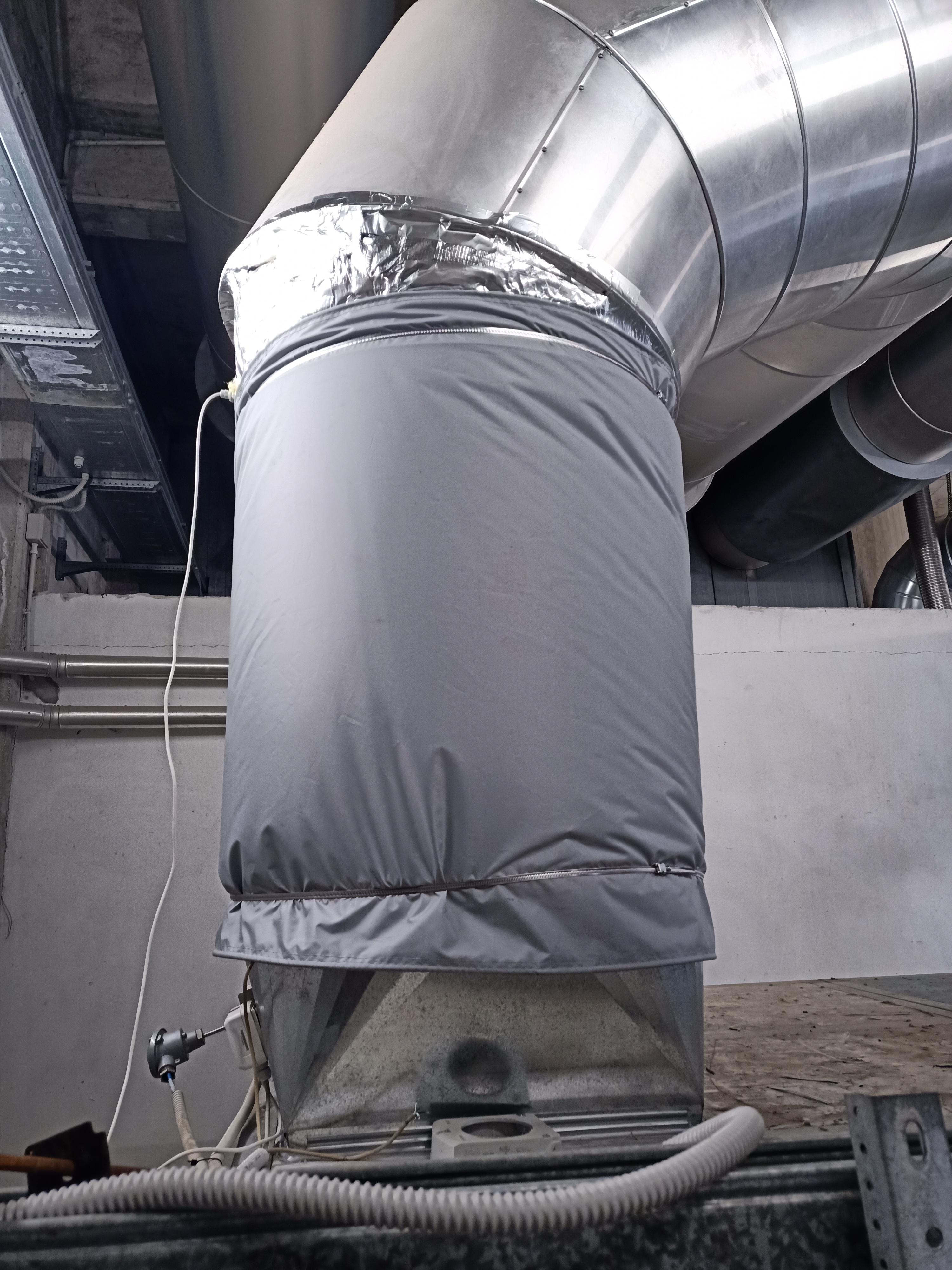}
    \includegraphics[width=0.35\textwidth]{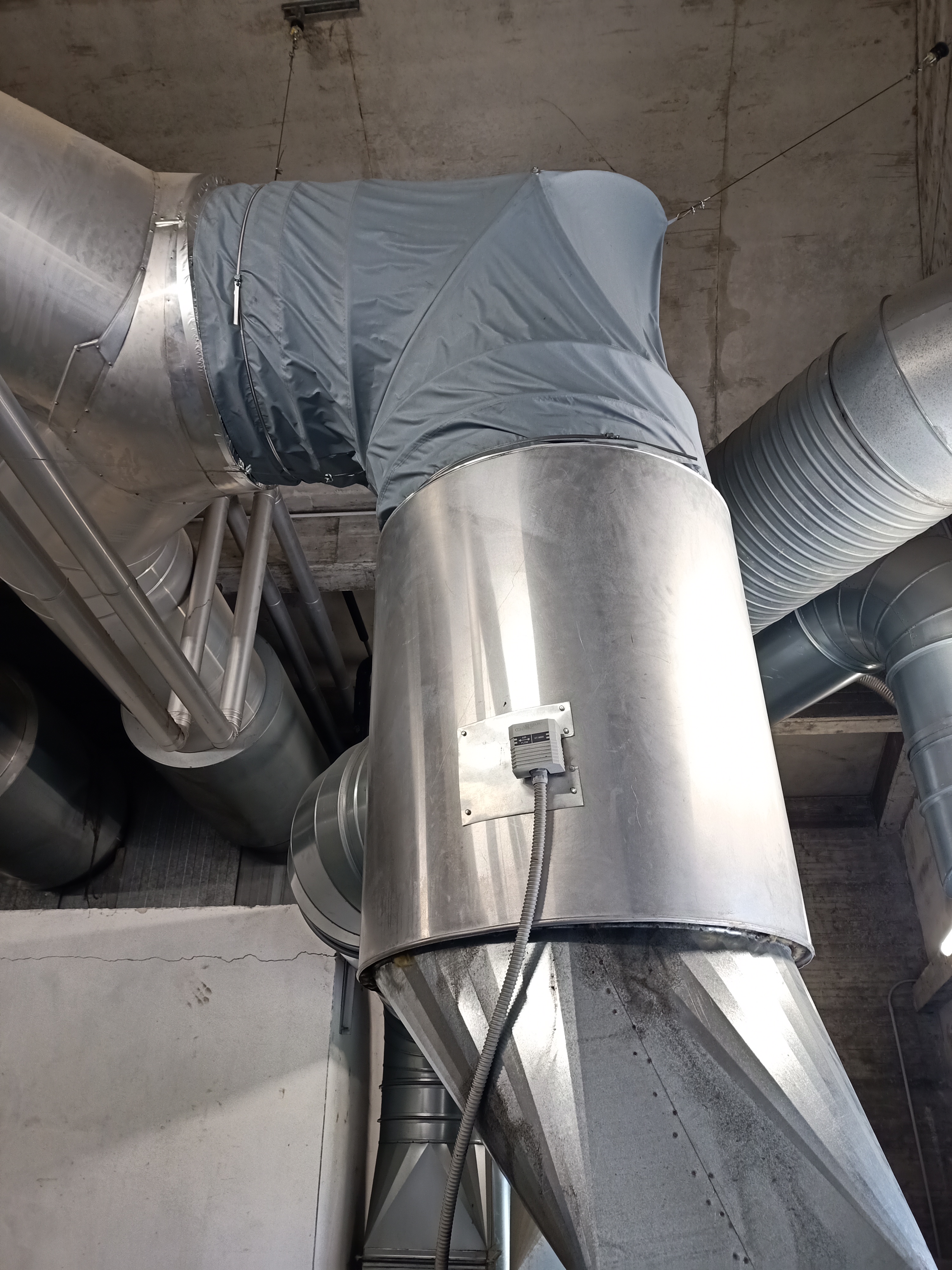}
    \caption{Pictures of the light textile sleeves connecting the supply (left) and return (right) ducts to the AHU enclosure.}
    \label{fig:Sleeves}
\end{figure}

\begin{figure}[ht!]
    \centering
 \includegraphics[width=0.48\textwidth]{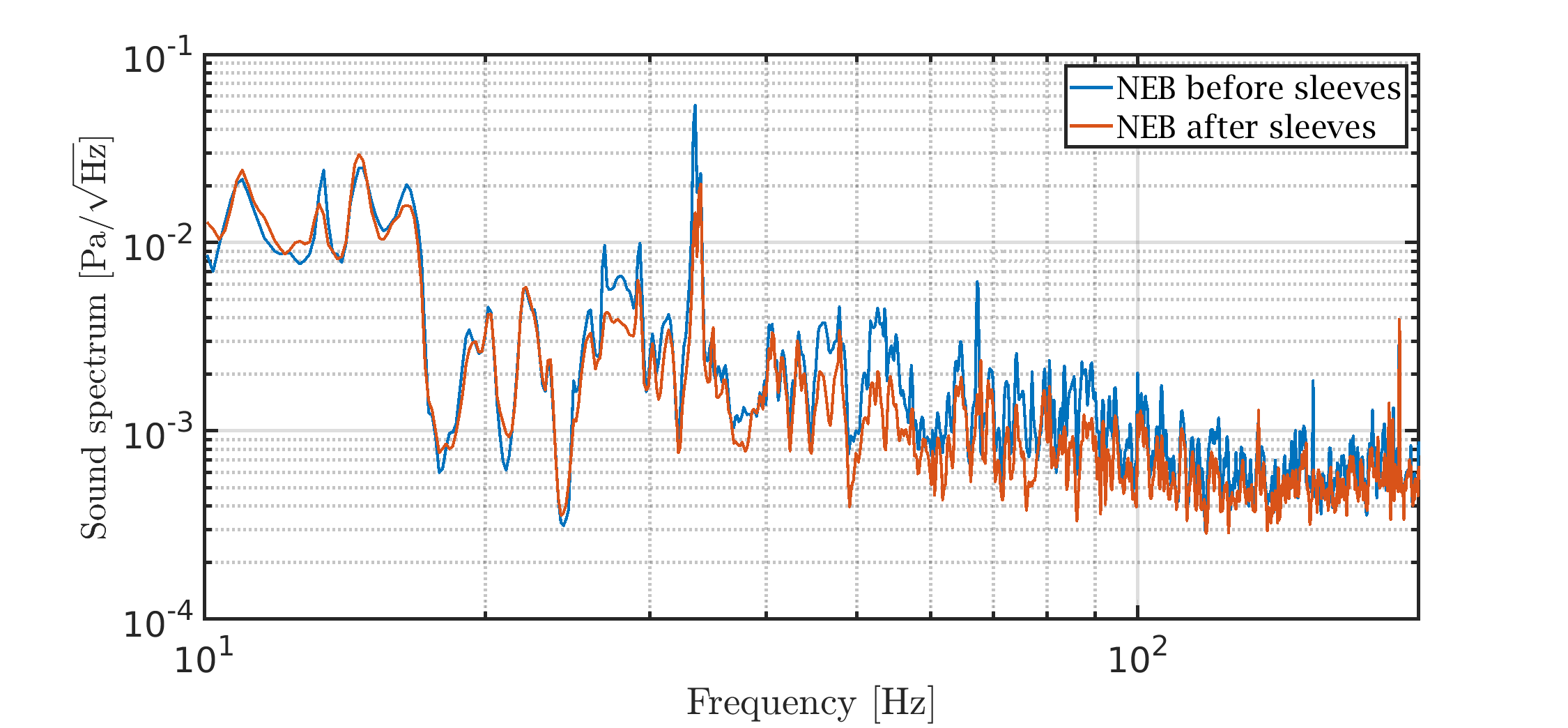}
  \includegraphics[width=0.48\textwidth]{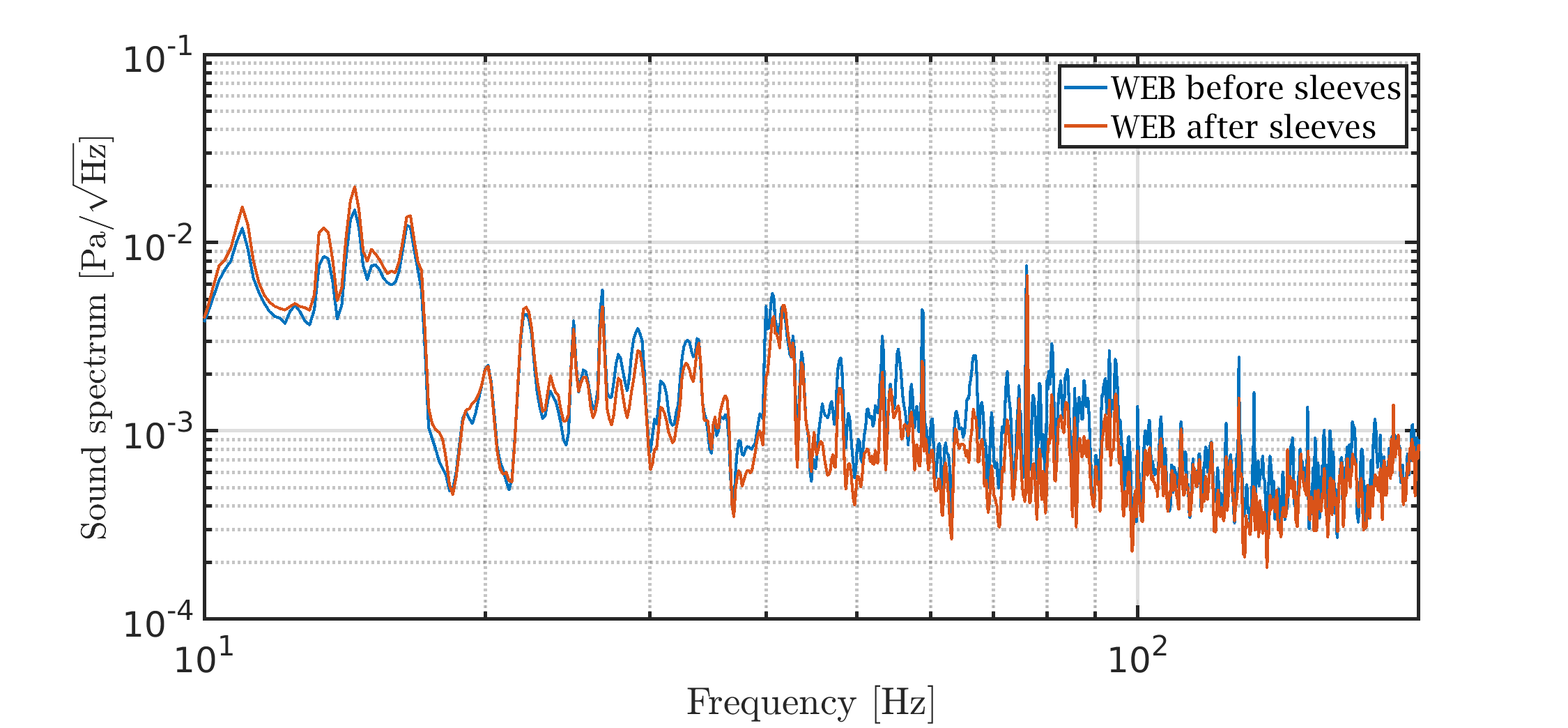}    
 \includegraphics[width=0.48\textwidth]{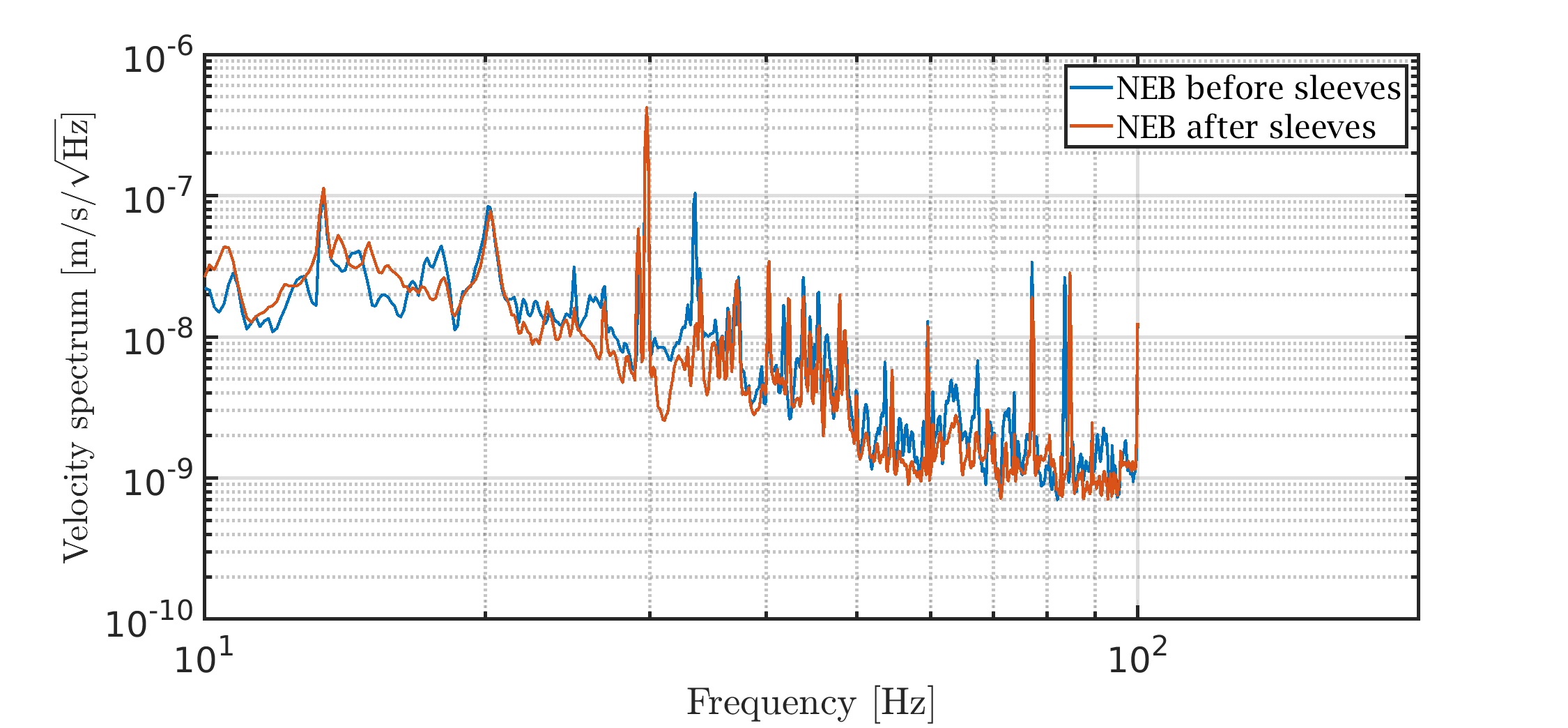} 
 \includegraphics[width=0.48\textwidth]{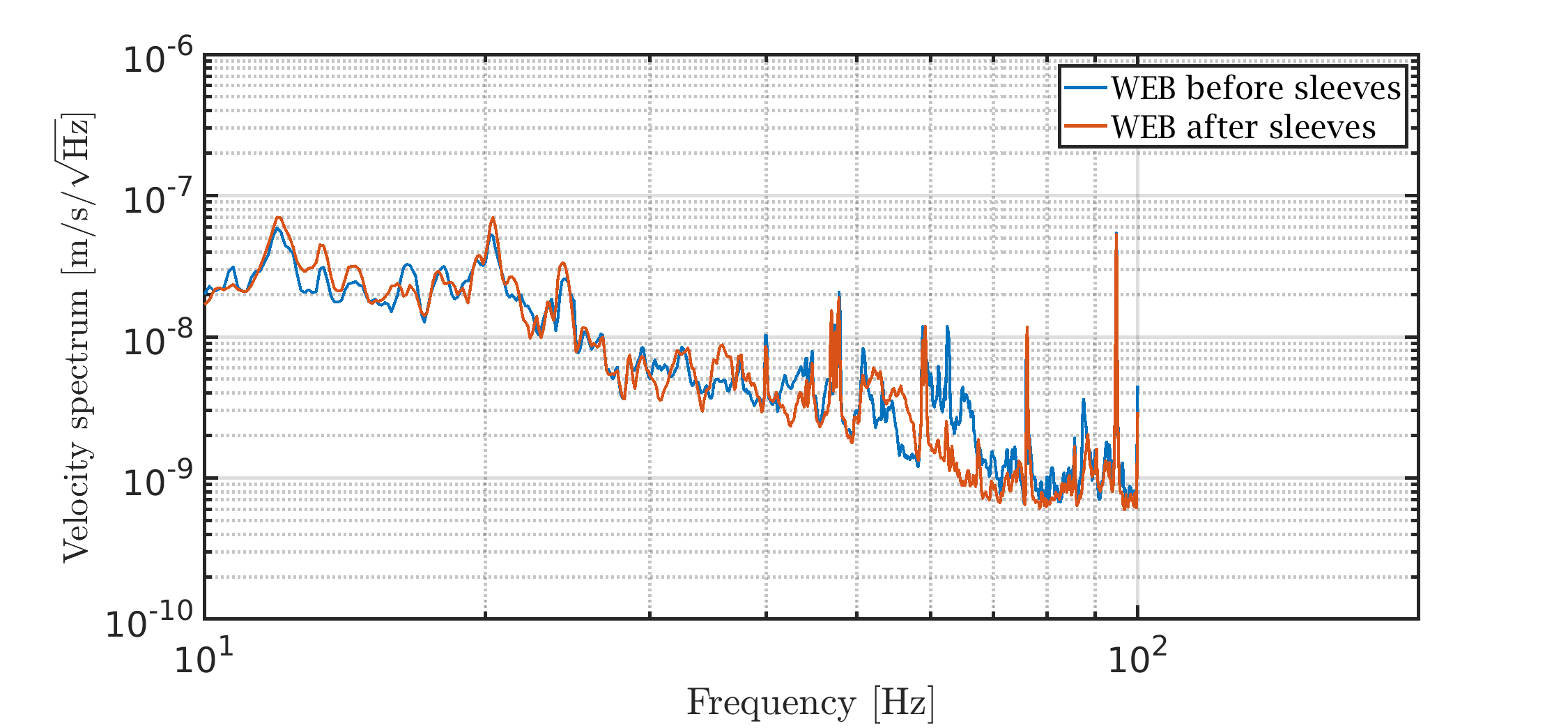}    
\caption{Noise reduction achieved by the sleeves.  Top-left: microphone in NEB hall; Top-right: microphone in WEB hall; Bottom-left: seismometer (vertical channel) at NEB hall floor; Bottom-right: seismometer (vertical channel) at WEB hall floor. Seismometer spectra are limited to 100 Hz, since a flat instrument response is specified only up to 50 Hz.}
    \label{fig:Sleeves_grafico}
\end{figure}

The third action in the sequence consisted of removing the vibration short-circuit between the AHU enclosure and the air ducts by disconnecting the ducts from the AHU enclosure and inserting one-meter-long soft fabric sleeves in the supply and return ducts \cite{Maniche}, as illustrated in Figure~\ref{fig:Sleeves}.
This action did not substantially change the vibration levels of the ducts themselves, which actually increased at low frequency, as expected.
Instead, it produced a sizable decrease in the acoustic noise in the hall between 50 and 100~Hz, as well as some reduction in floor vibration, as shown in Figure~\ref{fig:Sleeves_grafico}.
This suggests that displacing the duct inlet by one meter from the fan reduced the solid angle of acceptance for the air-borne sound waves generated by the fan and propagating within the ducts.
% This could be the case of air turbulence at the AHU fan stage, which may generate sound waves propagating inside the air ducts while simultaneously inducing mechanical vibrations that travel through the ducts or the floor structures.png}

%================================================================================
\subsection{Structural dumping actions}
\label{sec:damping}

%% structural damping of AHU enclosure 
% 1-pannelli interno (14 settembre 2021)
% https://logbook.virgo-gw.eu/virgo/?r=53144
% 2- pannelli esterno 11-18 gen. 2022
% https://logbook.virgo-gw.eu/virgo/?r=54513

%%% diciamo cosa e' structural damping e quale percorso di rumore intende curare 
% covering of fan box e fan case (uniformare nomi)
% covering of air ducts (menzionare risonanze ducts)  SAS e hall
% commentiamo risultati: 
% fan box +enclosure
% ducts in SAS 
% ducts in hall

%%%% dove mettiamo isolamento stanza AHU?  forse nel terzo insieme a nuovoventilatore?
%%%%  .... trattamenti acustici?

%%% damping of AHU enclosure
%This could be the case of air turbulence at the AHU fan stage, which may generate sound waves propagating inside the air ducts while simultaneously inducing mechanical vibrations that travel through the ducts or the floor structures.

Vibrations may be amplified at the mechanical resonant modes of structures, which can then act as acoustic transducers radiating sound into the surrounding environment. To mitigate this effect, mechanically excited components—here the air ducts and the air handling unit enclosure—should be designed to be sufficiently rigid to avoid low-frequency resonances, and any remaining resonances should exhibit a low quality factor. This can be achieved by coupling the vibrating elements to viscoelastic materials providing additional damping. In the following mitigation actions based on these principles are described and their effectiveness discussed.

\begin{figure}[ht!]
    \centering
    \includegraphics[width=0.25\linewidth]{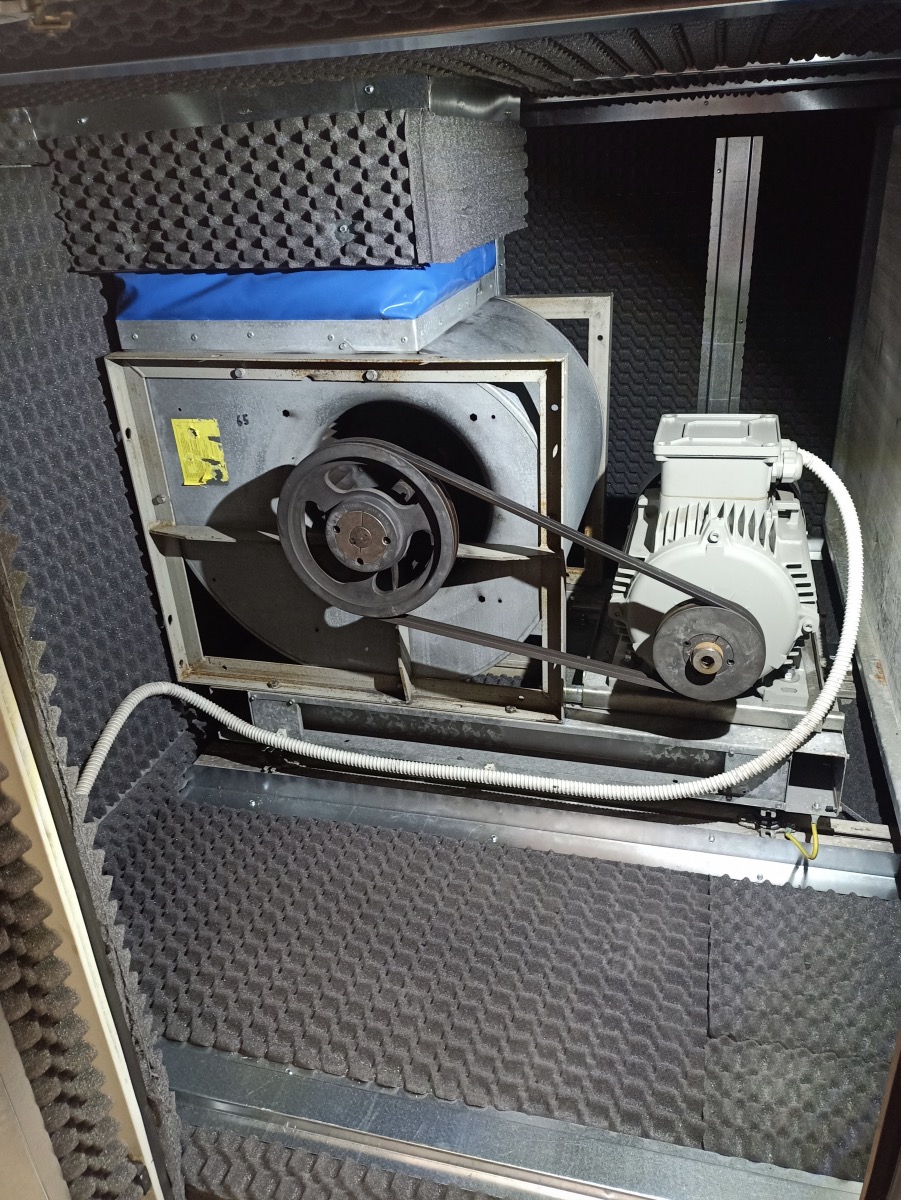}
    \includegraphics[width=0.44\linewidth]{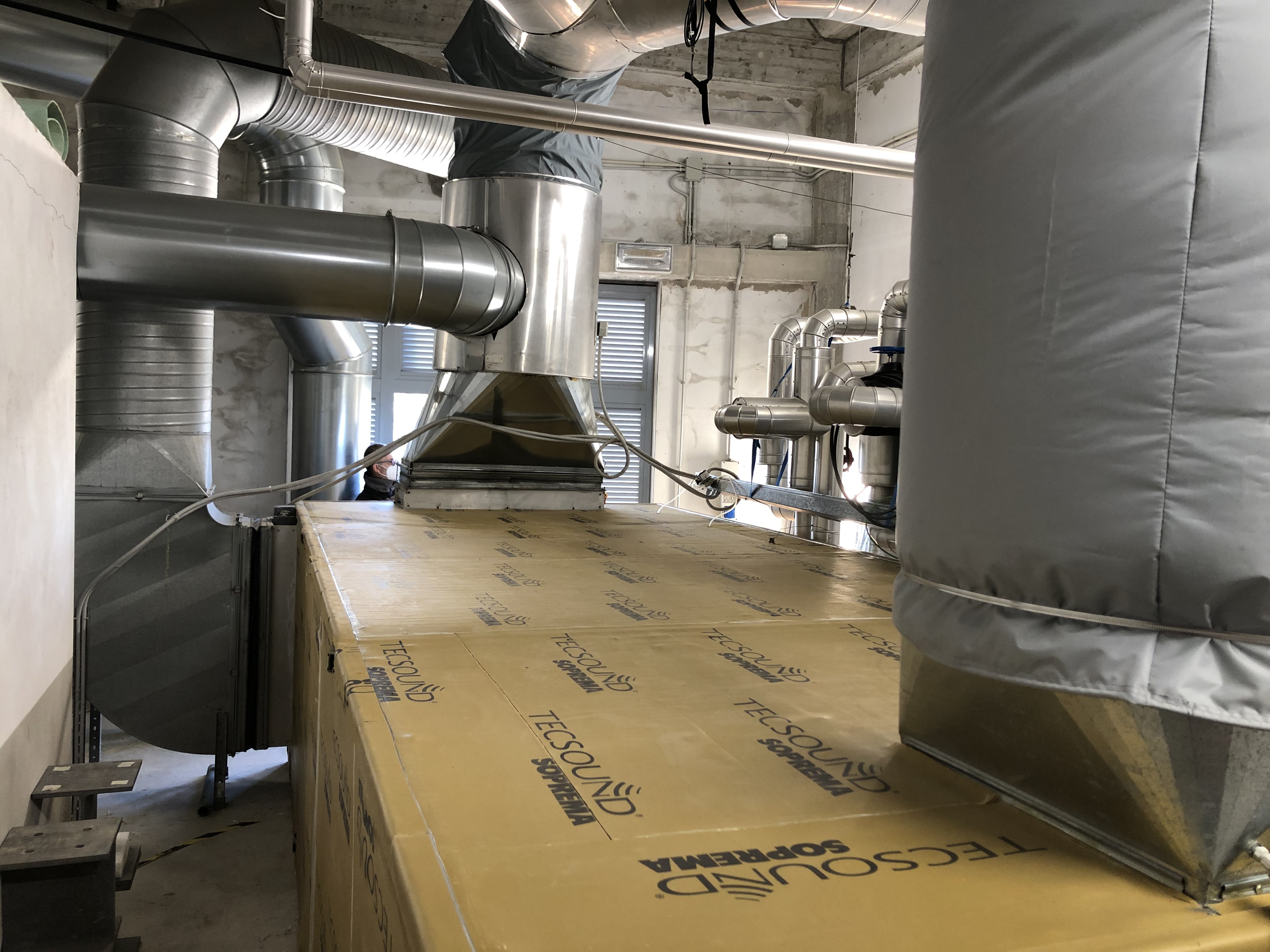}
    \includegraphics[width=0.34\linewidth]{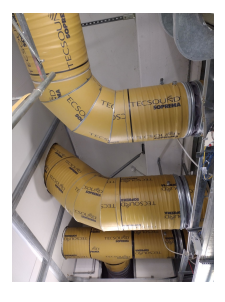}
    \includegraphics[width=0.31\linewidth]{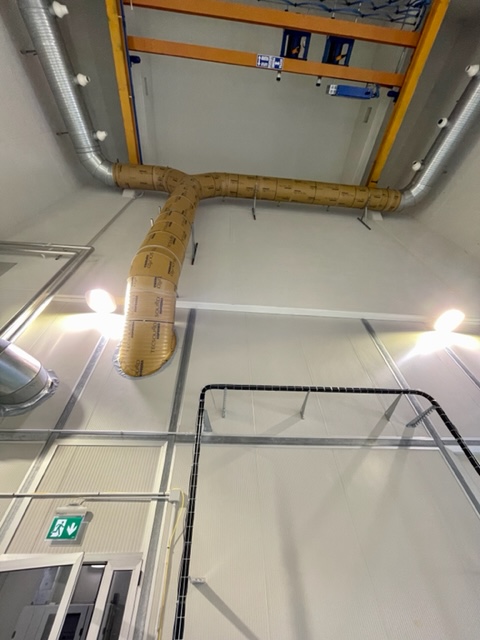}
\caption{Structural damping of the AHU enclosure and air ducts. Top-left: covering of the interior of the fan compartment; Top-right: covering of the exterior of the AHU enclosure. Bottom-left: covering of the air ducts inside the SAS; Bottom-right: covering of the supply duct branch inside the NEB hall.}
    \label{fig:soundproofing}
\end{figure}

\begin{figure}[ht!]
    \centering
    \includegraphics[width=0.45\linewidth]{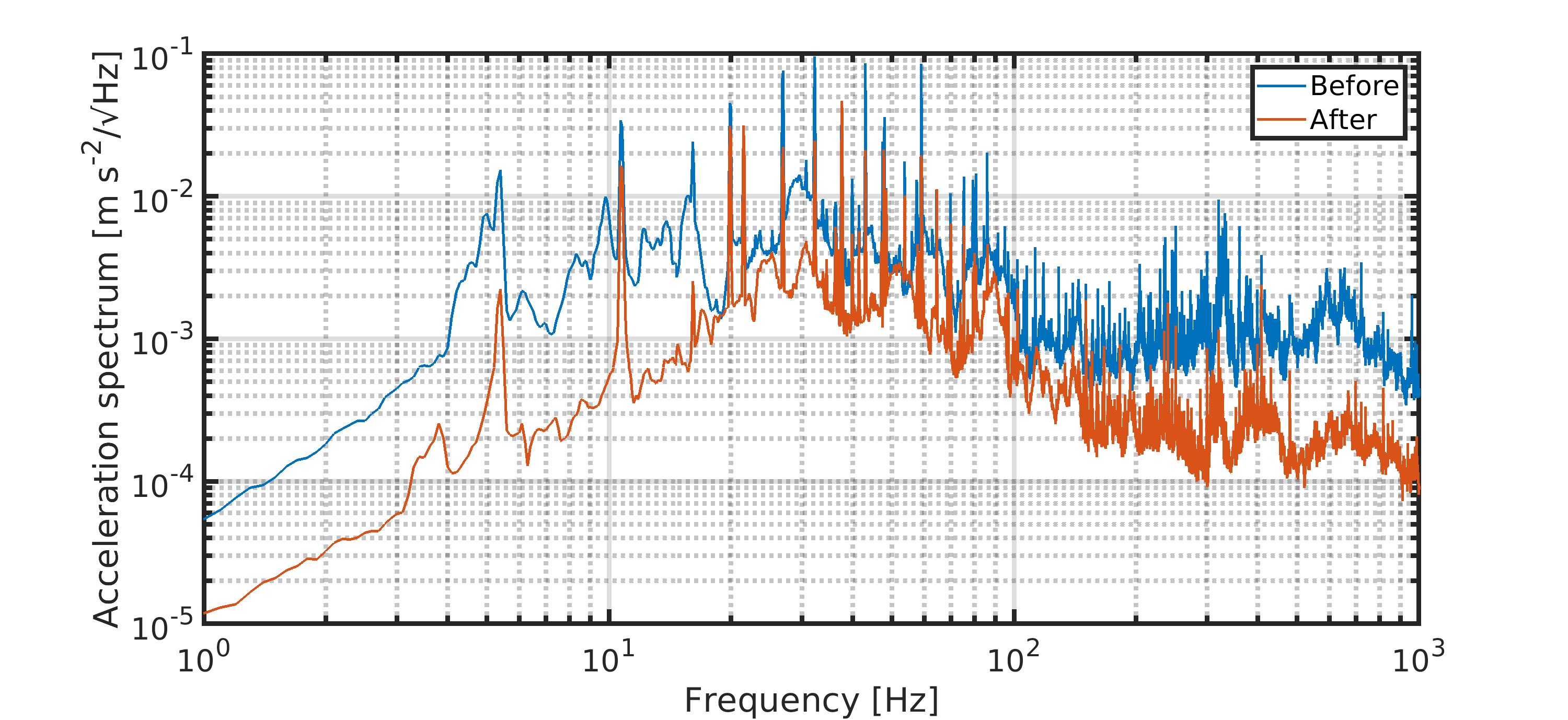}
    \includegraphics[width=0.45\linewidth]{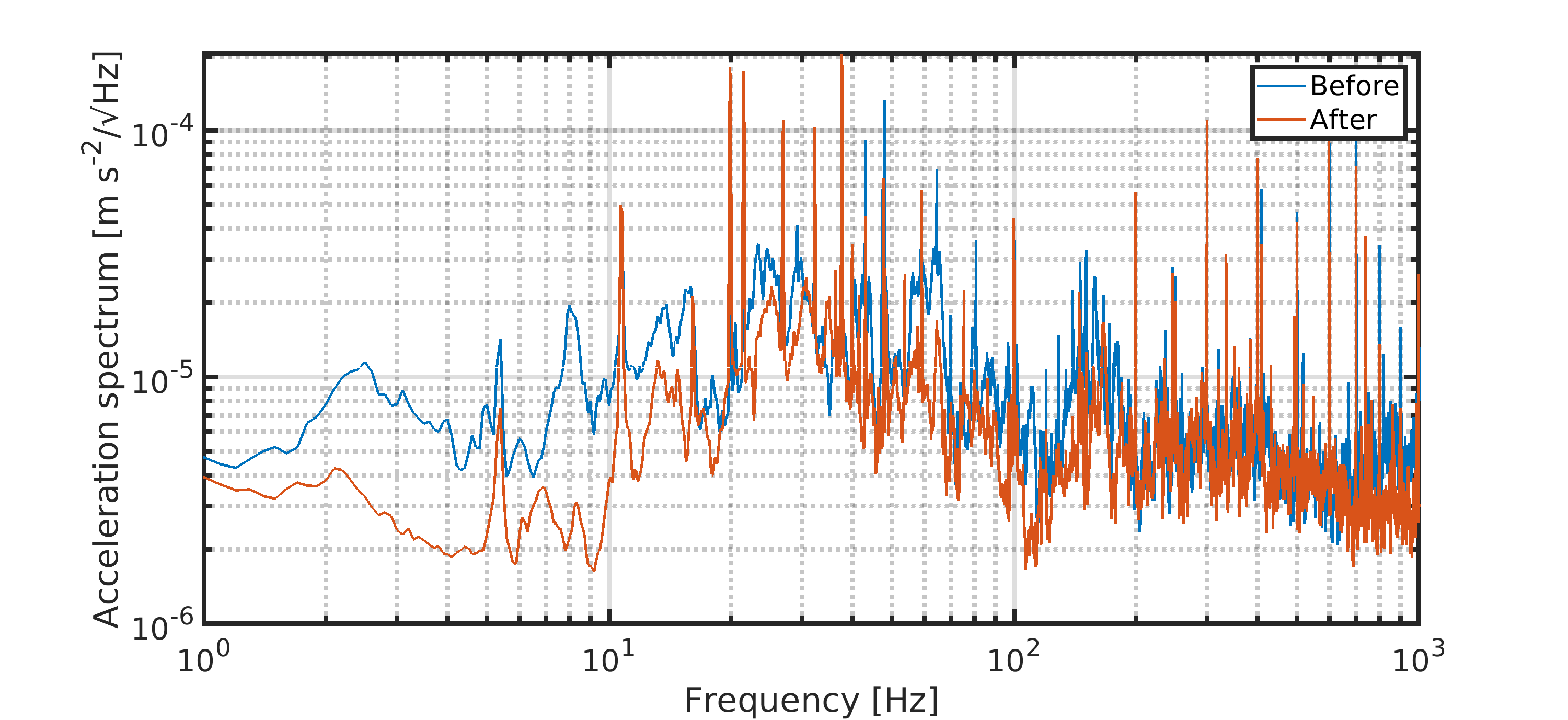}
    \includegraphics[width=0.45\linewidth]{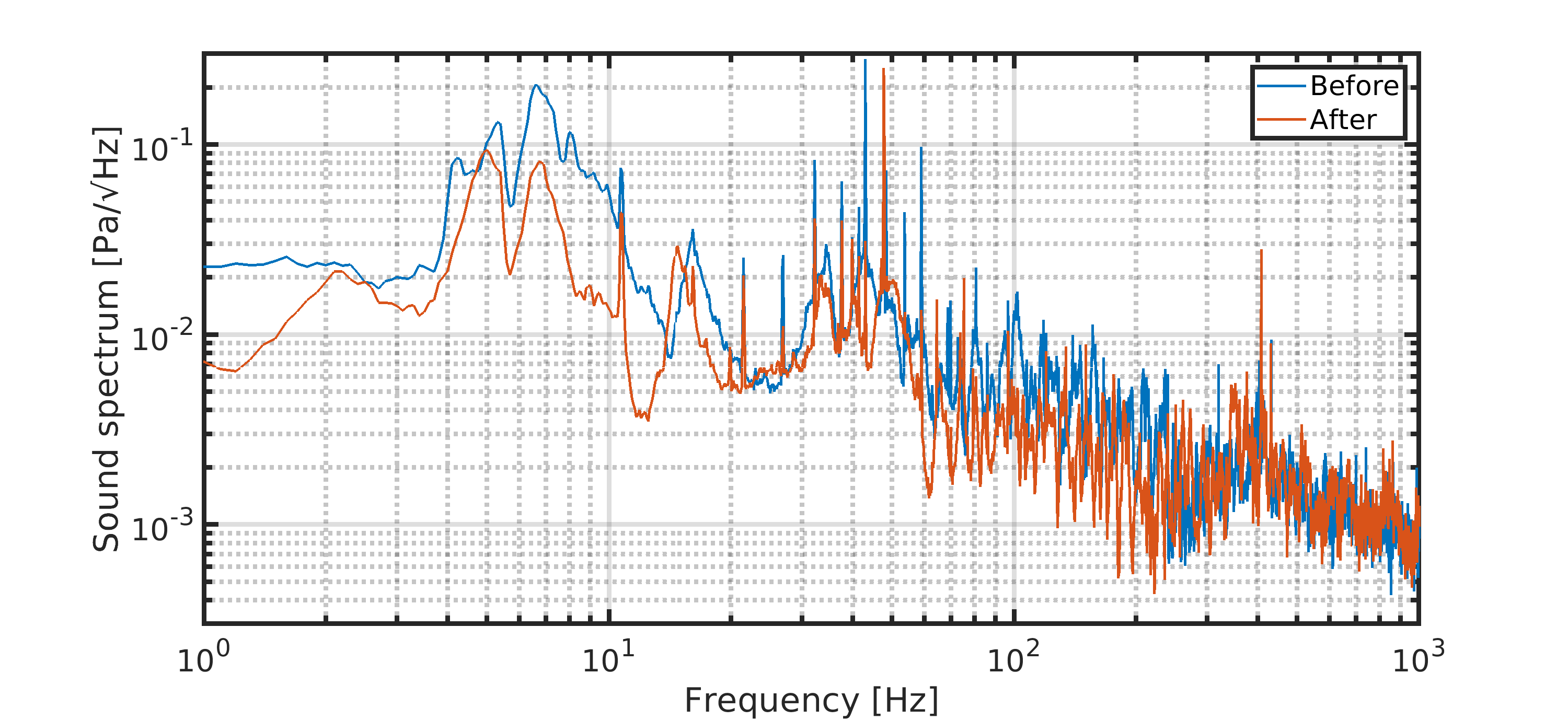}
    \caption{Vibration spectra before (blue curves) and after (red curves) covering of the AHU enclosure. The left plot shows the vibration level of the enclosure, while the right plot displays the vibration transmitted to the floor of the AHU room. A clear reduction is observed in both cases.
    Vibration and acoustic spectra before (blue) and after (red) covering of the AHU enclosure. Top-left: vibration of the enclosure. Top-right: vibration transmitted to the floor of the AHU room. Bottom: acoustic noise measured by the microphone. A clear noise reduction is observed in all sensors.}
    \label{fig:NEB_AHUroom}
\end{figure}

A structural damping of the AHU enclosure was performed to reduce its vibrations, as illustrated in the top row of Figure \ref{fig:soundproofing}. The intervention consisted of two distinct actions \cite{DampingInterno, DampingEnclosure}. First, the interior of the fan compartment was lined with panels made of two layers of self-extinguishing, non-drip, expanded polyurethane resin with an inserted elastomeric high density barrier (surface density 5 {kg/m$^2$}, sound reduction index of R=28.5 dB at 500 Hz). Subsequently, the exterior of the AHU enclosure was lined with a viscoelastic, high-density, polymer-based material, with 10 kg/m$^2$ surface density. % TECSOUND SOPREMA SY100.
The damping of the fan compartment was effective, substantially reducing vibrations of the AHU enclosure above 100 Hz and vibrations of the floor in proximity of the AHU itself. Covering of the enclosure reduced its vibrations even further, starting from a few 10 Hz, up. A reduction of acoustic noise in the 40–100 Hz range was observed inside the AHU room, as shown in Figure  \ref{fig:NEB_AHUroom}, and inside the SAS room, as reported in \cite{DampingEnclosure_results}. However, the achieved mitigation effects were local, with no measurable impact inside the experimental hall.

%%% DUCTS damping

\begin{figure}[ht!]
    \centering
    \includegraphics[width=0.45\linewidth]{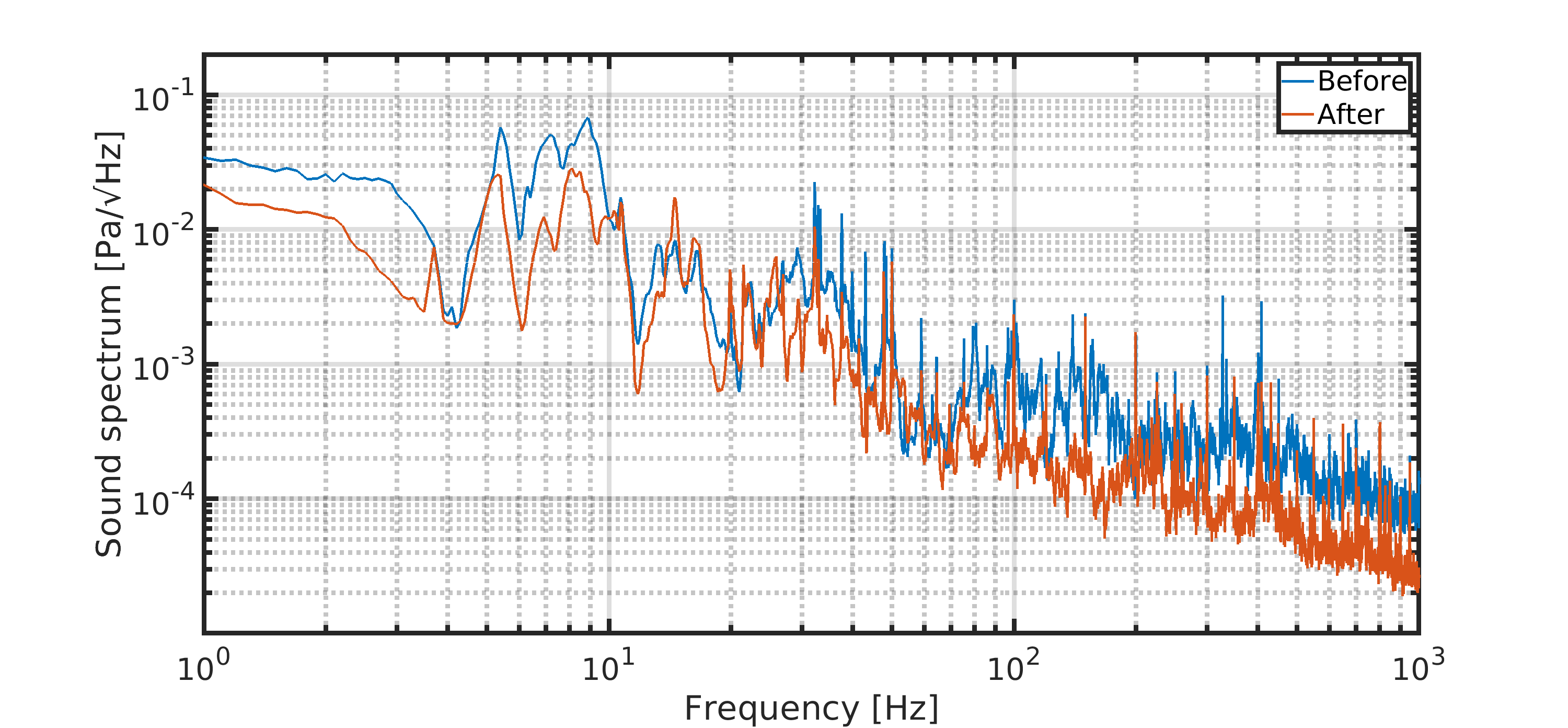}
    \includegraphics[width=0.45\linewidth]{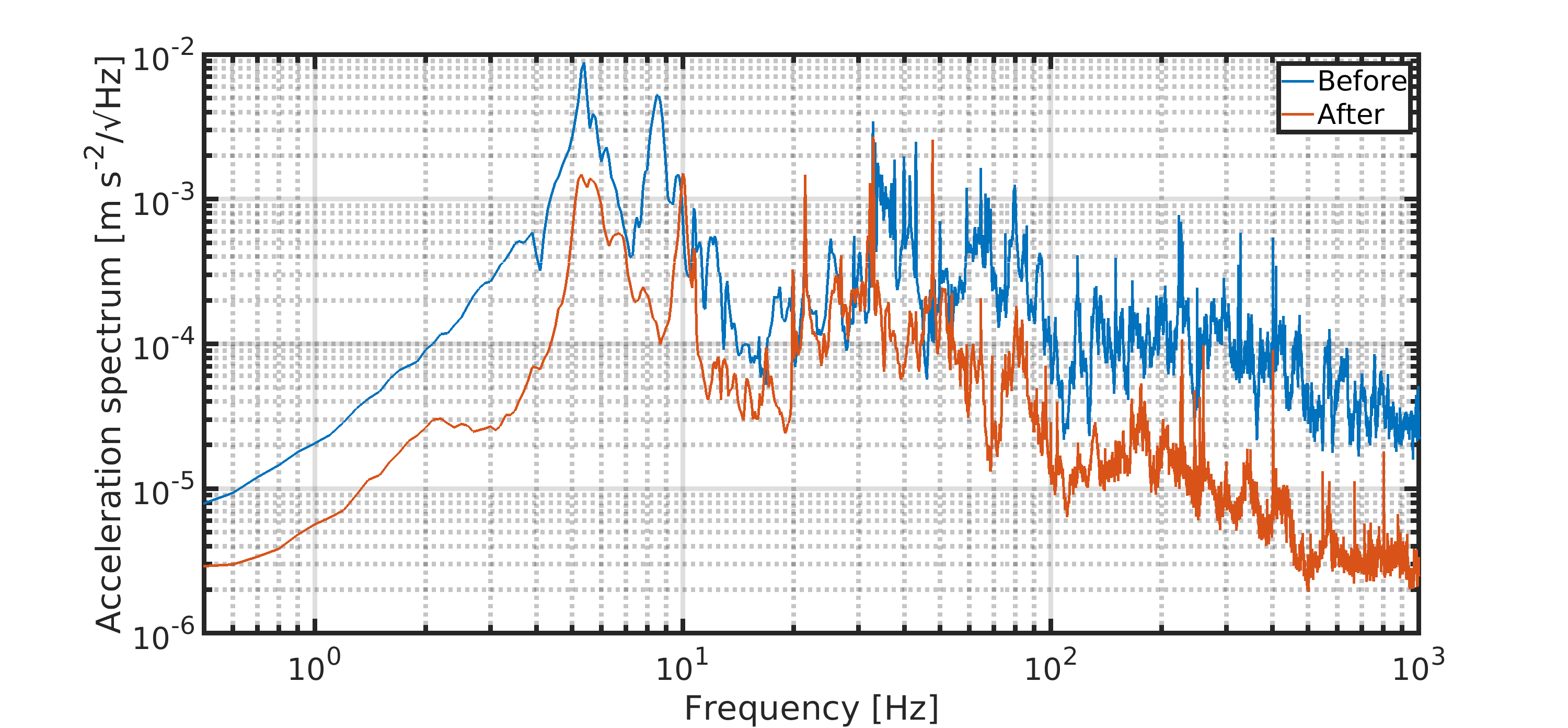}
    \includegraphics[width=0.45\linewidth]{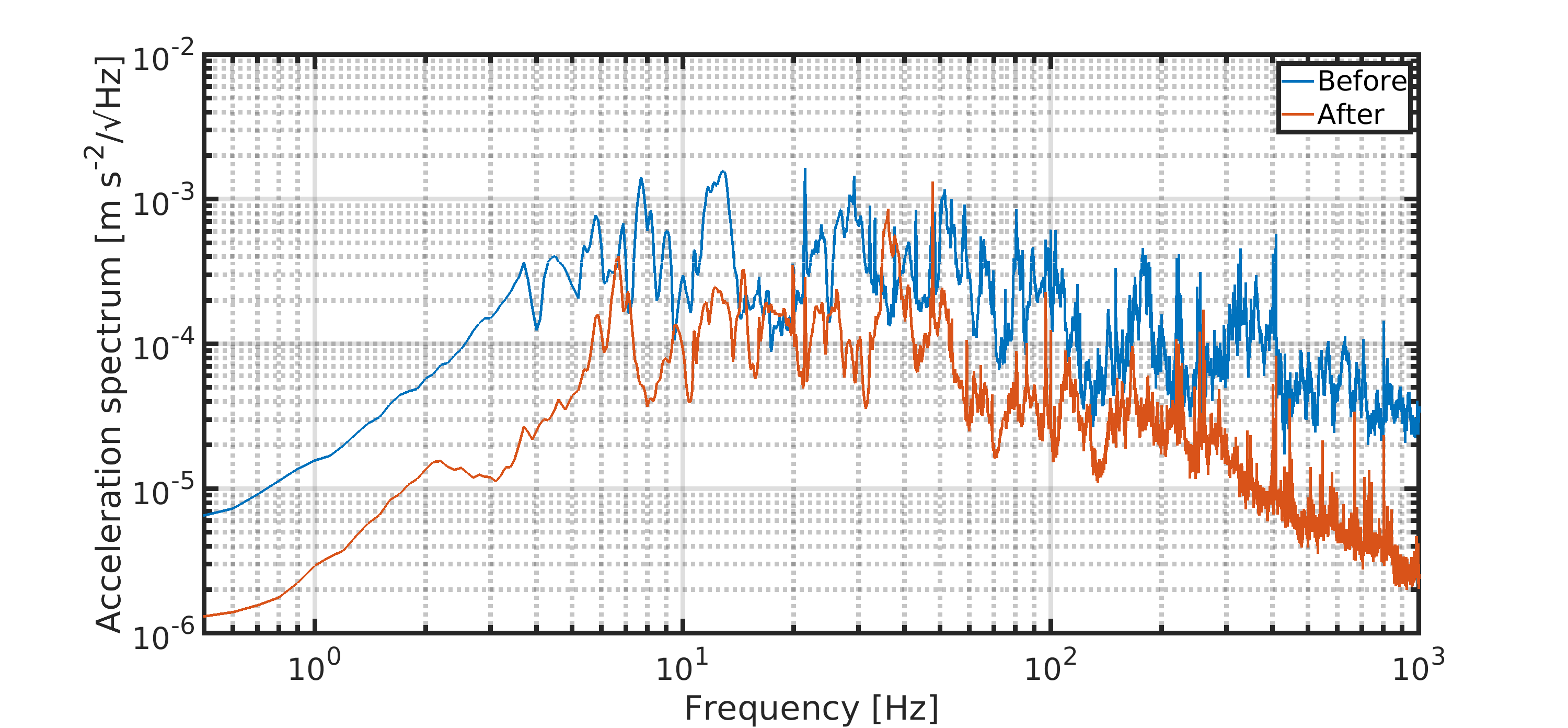}
    \includegraphics[width=0.45\linewidth]{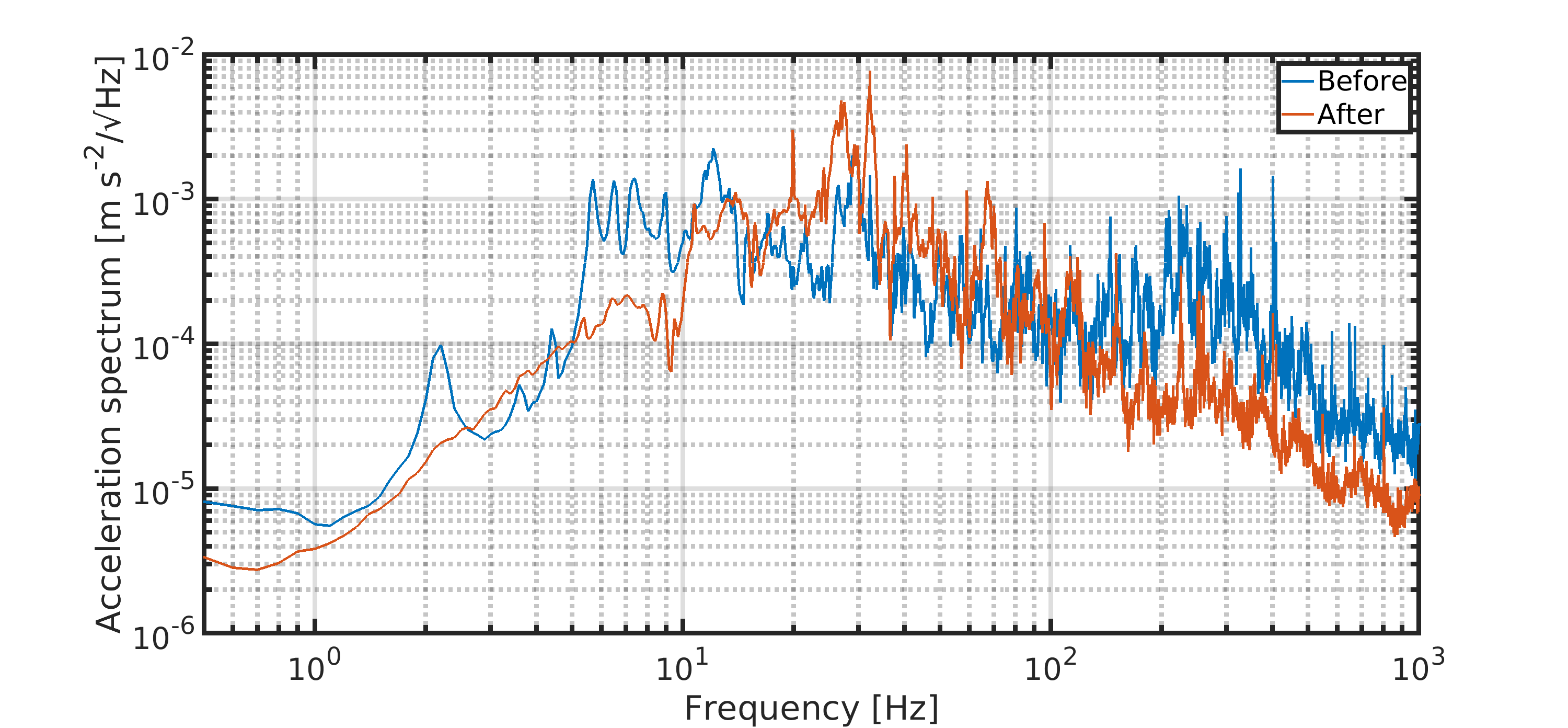}
    \caption{Mitigation effect of structural damping applied to the air ducts of the SAS room. Spectra before mitigation are shown in blue, after mitigation in red. Top-left: acoustic noise inside the SAS room. Top-right and bottom: vibration of the supply and return air duct sections. }
%    \caption{Effect of improving sound insulation of the AHU room and wrapping of the SAS air ducts. Acoustic noise in the SAS room sensibly reduced.  \textcolor{blue}{Promomeria: ventilatore vecchio con belt 1790 cm, fan 280 cm, motor 150 cm, rpm 1440., inverter 41 Hz}}
\label{fig:SAS_ducts}
\end{figure}

%This is the case for example of the air ducts which are made of reinforced aluminum sheets 2 mm think \textcolor{red}{CHECK with DAVIDE.} 
%Diciamo che air ducts non sono sorgente attiva ma fungono da link strutturali di trasmissione del rumore di vibrazione, e possono eventualmente amplificarlo (in corrispondenza di risonanze proprie) o convertirlo in rumore acustico.
% osserviamo risonanze nella figura.
Air ducts are a critical element. They can act as vibration transmission path, but also as vibro-acoustic transducers. Turbulent air flow inside ducts can induce vibration, with possible amplification
in correspondence of ducts mechanical modes.
The ducts of the NEB HVAC are cylindrical tubes with a diameter of 800 mm, reducing to 400 mm after entering the hall, and they are made of 8 mm thick stainless steel foil. Some 90 degree turns are present which in principle might enhance turbulence.
They are rigidly connected to the walls. One good-practice action would have been seismic decoupling (Section \ref{sec:decoupling}) ducts from the building walls but it would have been 
a too invasive operation for the ongoing Virgo commissioning. 
The action undertaken aimed instead to reducing the overall duct vibrations.
The supply and return air ducts in the NEB-SAS, and the first section of the supply air duct in the experimental hall was wrapped with a dense viscous material, 3.5 mm thick, 7 kg/m² surface density \cite{ Airducts_covering, NEB_airducts_insulation,VIR-1314B-21}. %namely TECSOUND SY70, 
% surface weight
%\footnote{The soundproofing specifications of the high-density polymer used for the acoustic isolation test were as follows: TECSOUND 100 (5 mm thick, 10 kg/m²) was applied to the NEB AHU box, while TECSOUND 70 (3.5 mm thick, 7 kg/m²) was applied to the air ducts in the NEB-SAS and NEB-Hall.}
The finished installation is shown in the bottom row of  Figure \ref{fig:soundproofing} and the achieved mitigations are illustrated in Figure \ref{fig:SAS_ducts}. Ducts vibrations reduced considerably and low frequency modes dampened. Notably, ducts dampening led to a significant reduction of acoustic noise inside the SAS room in the band 1 Hz to 1 kHz. This is taken as evidence that the ducts, acting as acoustic transducers, were in fact the dominant sound source inside the SAS room at those frequencies. No significant reduction in seismic or acoustic noise was observed in the hall, indicating that this path either was not a dominant contributor.

%============================================================
\subsection{Airborne noise}
\label{sec:airborne}
% fan replacement
% https://logbook.virgo-gw.eu/virgo/?r=57814

%\textcolor{red}{Un piccolo problema: la cura del rumore trasmesso tra AHU room e SAS possiamo considerarlo un "airborne" noise path??}

%Airborne noise refers to sound pressure waves transmitted through the aerial medium.
%One potential transmission path is the propagation of sound generated within the AHU into the experimental hall.

This section illustrates mitigation measures addressing sound noise propagating in air, either across partitions -via air–structure–air coupling- or through ducts.

The AHU is located in a technical room enclosed by concrete walls; an additional concrete wall separates this room from the SAS, and a lightweight partition wall further separates the SAS from the hall, as shown in Figure~\ref{fig:3Dmap} (right). Despite this multi-layer separation, an aperture in the concrete walls, designed to accommodate the air ducts, was suspected to significantly degrade the overall sound insulation performance.

\begin{figure}[htp!]
    \centering
    \includegraphics[width=0.35\linewidth]{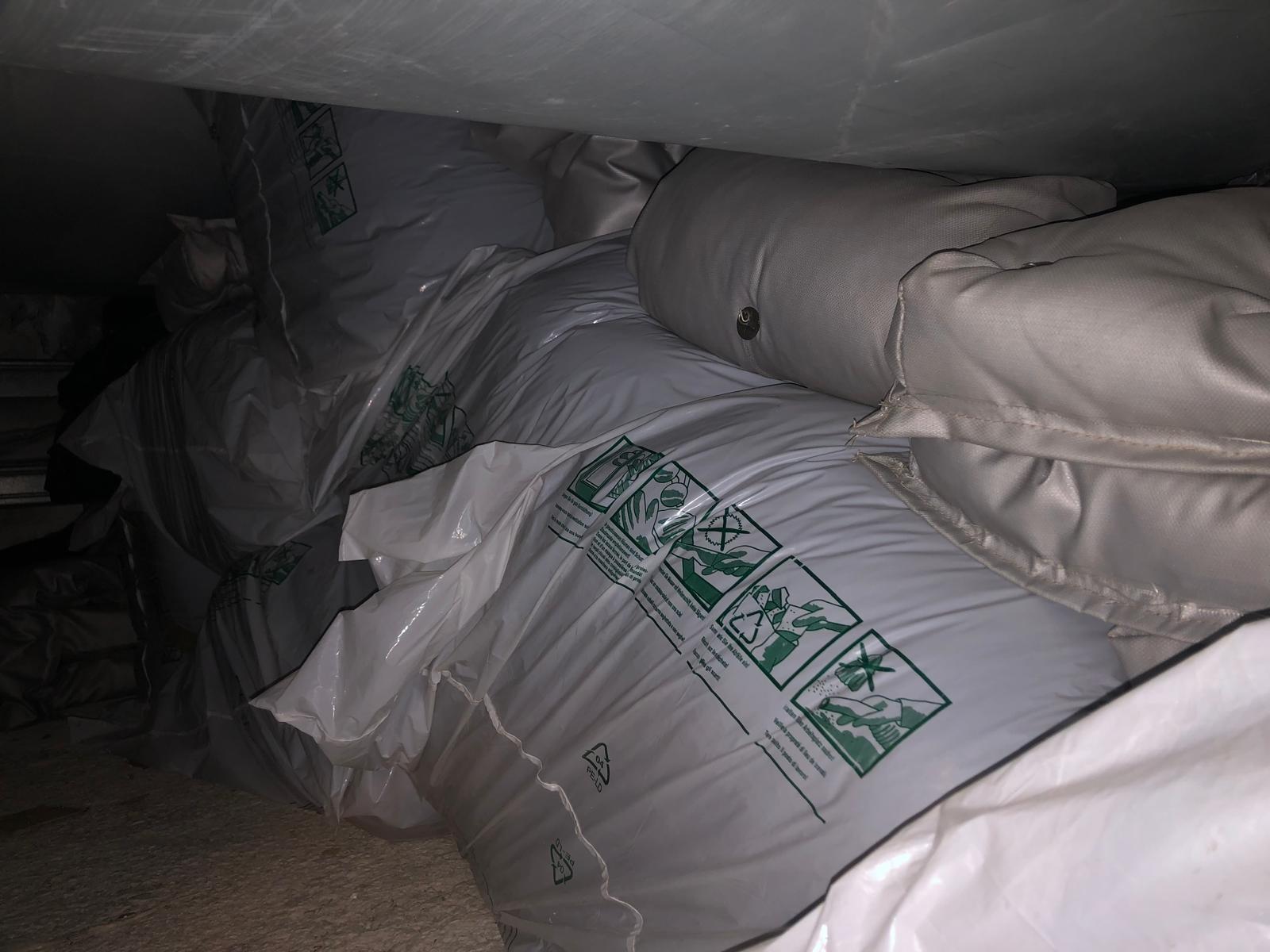}
    \includegraphics[width=0.25\linewidth]{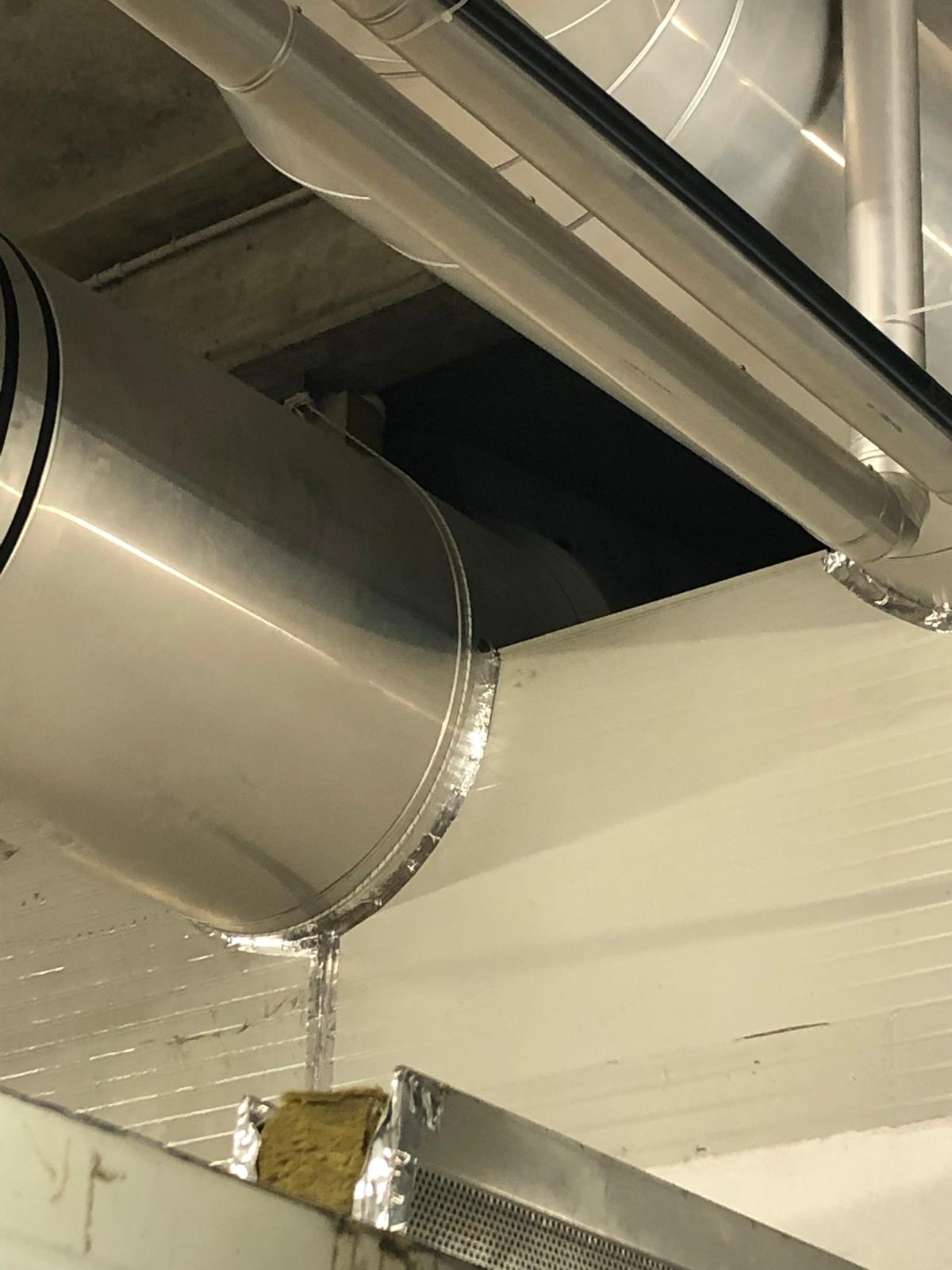}
    \includegraphics[width=0.35\linewidth]{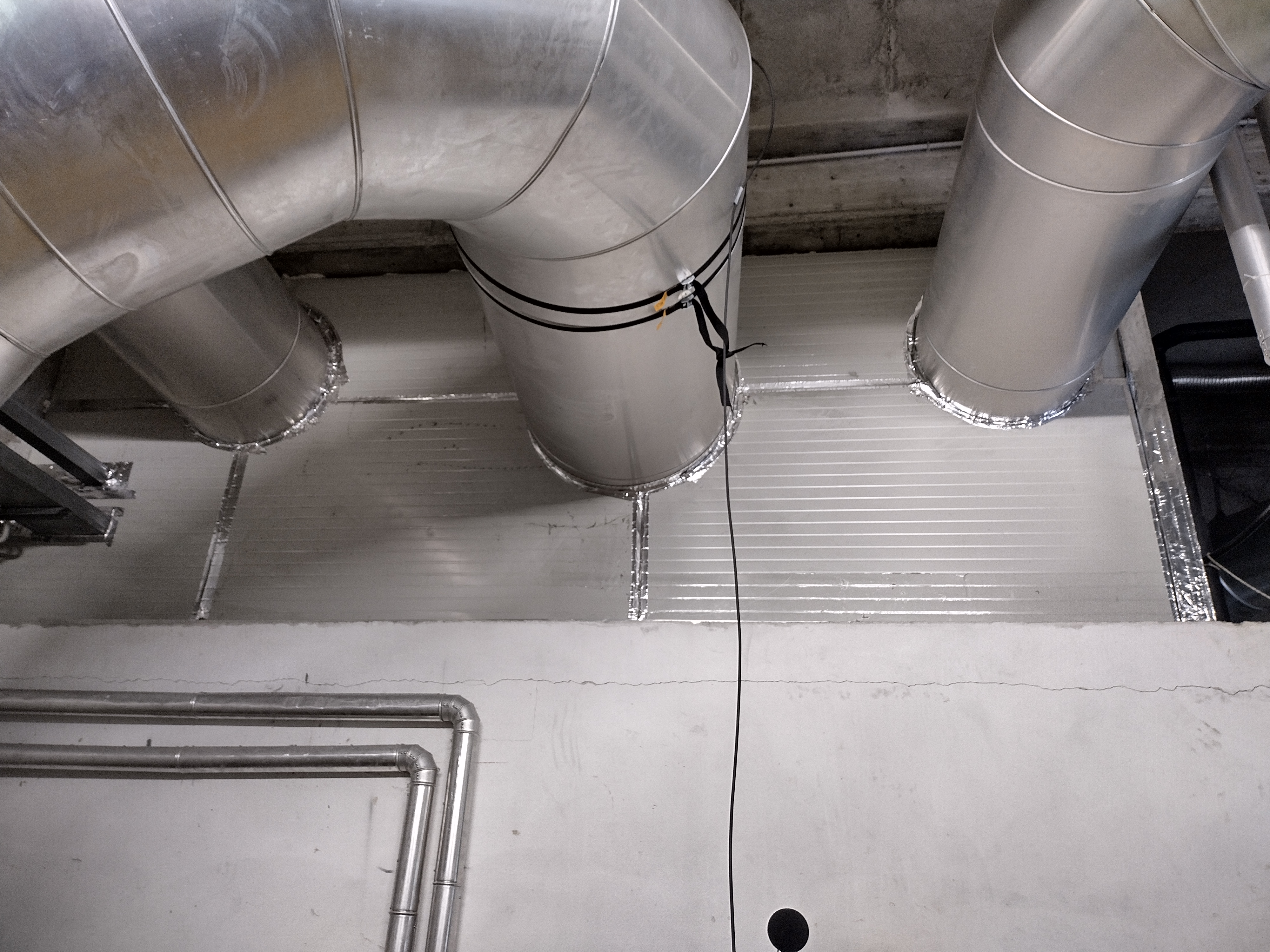}
    \caption{The images illustrate the placement of acoustic treatment materials (fiberglass bales and panels) used for sealing the aperture and improving sound insulation of the AHU room.}
    \label{fig:roomIsolation}
\end{figure}

The existing aperture was filled with fiber-glass bales \cite{NEB_AHUroom_insulation} 
and finished with a lightweight panel,  
as shown in Figure \ref{fig:roomIsolation}.
This intervention was carried out immediately after the duct structural damping work. It did not result in any appreciable reduction of noise levels, indicating that airborne sound transmission through the walls was not a dominant noise path at the time of the intervention.

Another path to consider is airborne noise generated by the fan and reaching the experimental hall by propagating inside the air ducts. The duct disconnection test described in Section~\ref{subsec:Ducts_disconnection} provided clear evidence that this is indeed a relevant transmission path.

It is well established in the literature that centrifugal fans with backward-curved blades are quieter than those with forward-curved blades; however, published studies generally report performance only in the audible frequency range, above approximately 100~Hz \cite{Jiang2023, Zhang2020}. Given that the original AHU fan was of the forward-curved type, it was subsequently replaced with a backward-curved model with similar operational specifications \cite{NewFanComparison}. The corresponding technical parameters are reported in Table~\ref{tab:NEBAHUparameters}.

\begin{figure}[ht!]
    \centering
     \includegraphics [width=.8\textwidth]{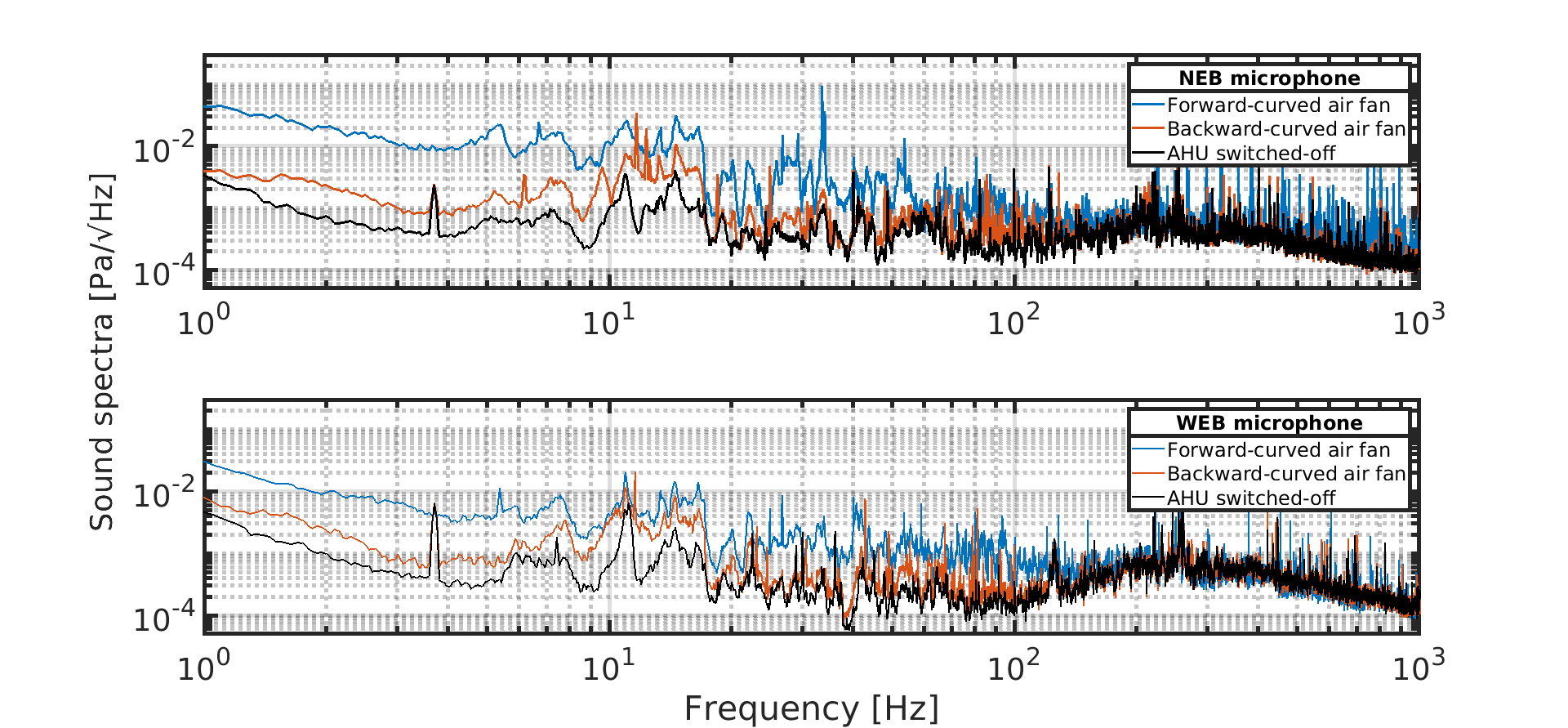}
    \caption{Acoustic spectra of the NEB (top) and WEB (bottom) experimental hall before (blue) and after (red) the replacement of forward-curved air fan with the backward-curved air fan. Both motor inverters were operating at nominal frequencies (50 Hz). Black curves correspond to the HVAC switched OFF.}
    \label{fig:fanreplacement_acoustic}
\end{figure}

Figure~\ref{fig:fanreplacement_acoustic} illustrates the effect of this intervention \cite{NewFan}. Broadband acoustic noise was substantially reduced, starting from a few hertz and up to 50~Hz, with a broadband reduction of order of 20~dB in the 1-10~Hz and 10~dB in the 10–50~Hz frequency bands. The same substitution was implemented for the twin AHU located in the West End Building, yielding similar results.

\begin{figure}[ht!]
    \centering
     \includegraphics [width=.8\textwidth]{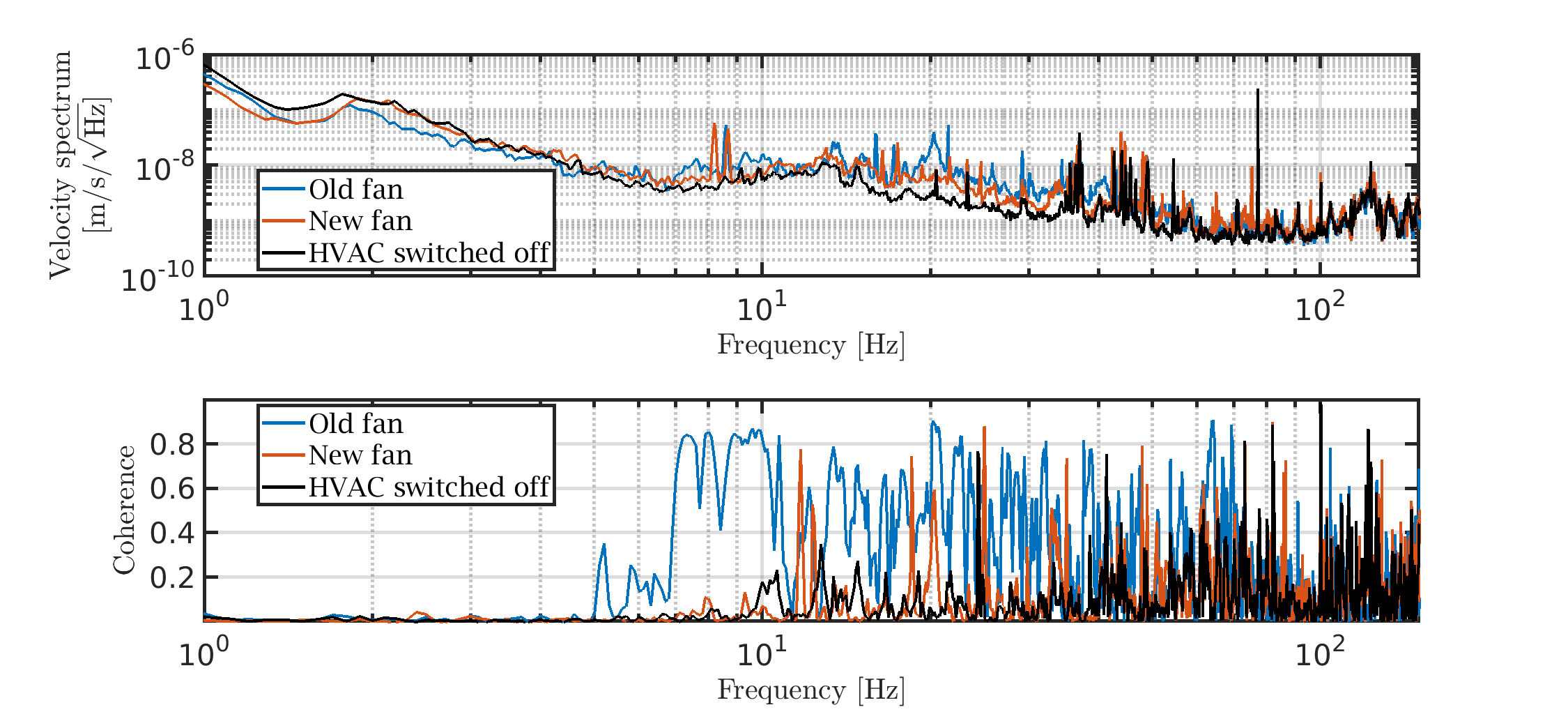}
    \caption{Top. Seismic spectrum of the NEB experimental hall floor vertical seismometer before (blue) and after (red) the replacement of forward-curved air fan with the backward-curved air fan. Both motor inverters were operating at nominal frequency (50 Hz). Below 7 Hz, seismic spectra are dominated by external anthropogenic sources.
    Bottom. Coherence between the seismometer and the microphone in the hall, with the forward-curved blades air fan (blue) and with the new backward-curved blade profile fan (red).
    In both graphs, the black curves correspond to the HVAC being switched off.}
\label{fig:fanreplacement_seismic}
\end{figure}

Concurrently, a reduction in seismic noise was observed,   as shown in Figure~\ref{fig:fanreplacement_seismic}.  The vibration amplitude of the hall floor decreased by a factor of 2 in the frequency range 7–70~Hz. A plausible noise path is that airborne acoustic noise in the hall induces vibrations in the lightweight scaffolding structure surrounding the vacuum chamber and resting on the floor.

\begin{table}[ht!]
    \centering
    \resizebox{0.9\textwidth}{!}{%
\begin{tabular}{ccc} 
\toprule
Parameters &  Value  \\
\midrule
Fan pulley diameter & 250 mm\\
Motor pulley diameter &  118 mm\\
Motor rotation speed & 1455 r.p.m. \\
Belt length  & 1450 mm\\
Number of belts & 2 \\
%Brand & Nicotra ADN 400 K \\
\bottomrule
    \end{tabular}
\hspace{1.5cm}
\begin{tabular}{ccc}
\toprule
Parameters &  Value  \\
\midrule
Fan pulley diameter & 280 mm\\
Motor pulley diameter &  150 mm\\
Motor rotation speed & 1440 r.p.m. \\
Belt length  & 1790 mm\\
Number of belts & 2 \\
%Brand & Nicotra RDH 500R\\
\bottomrule
    \end{tabular} 
    }
    \caption{Design technical parameters of the
    backward-curved air fan (left) and forward-curved air fan (right): diameter value of fan and motor pulleys, motor rotation speed, length and number of the belts, respectively.
    }    
    \label{tab:NEBAHUparameters}
\end{table}

%====================================================================

\subsection{Vibration noise from the water circulation system}
\label{sec:Seismic_water}

% cappello su rumore vibrazione associato 
% a water sys e trasmesso sia in via-strutturale, sia via water
% diciamo subito di rilevanza manutenzione periodica

%%% --- structural-noise
% molle pompe (chillers)
% disconnessione pipes / through holes
% manutenzione pompe
%% AGGIUNGI FOTO molle e through holes.

%\textcolor{red}{
%Questo si sposta dove si parla di water distribution sys: Water chillers and air compressors contribute to the broadband noise with loud \textcolor{red}{(peak amplitude ...)} and short \textcolor{red}{(approximately 1 second long)} seismic bursts occurring every few minutes at each start-up of the motors.}

%The hot and cold water production and circulation system for Virgo's North terminal building is schematized in \textcolor{red}{Figura ... se siamo daccordo a mettere questa figura}.
%Fatta eccezione per molle smorzanti (frequenza cutoff 2~Hz) sotto chillers case, l'impianto idraulico originale non implementava alcun tipo di accorgimento per limitare la propagazione del rumore di vibrazione.
%Abbiamo implementato: molle sotto pompe acqua, giunti di gomma sulle tubature, disconnesso tubature dalle pareti tramite molle, aperto passaggi (through-holes) nelle pareti.
%
Virgo HVAC systems employ independent hot- and cold-water circuits. The cold-water circuit includes a pair of chillers operating in series and two in-line pumps (one serving as a standby unit) that circulate water through metallic pipelines to the water-to-air heat exchanger coil inside the AHU. The hot-water circuit consists of a single boiler and a similar dual-pump configuration.

Chillers and boilers operate in alternating on/off cycles lasting from a few minutes up to several tens of minutes, depending on outdoor temperature. The compressors of the chillers generate short pulses of seismic and magnetic noise (see Section~\ref{sec:Magnetic_noise}) during activation. The pumps produce nearly monochromatic noise at the motor rotation frequency, superimposed on a broadband background.

Noise can propagate either through floors and walls as structural (structure-borne) noise, or as pressure fluctuations within the fluid as water-borne noise. Both propagation paths have been considered and are discussed below.
%Noise propagates through floors and walls - the so-called structural noise - or as fluid pressure noise, the so-called water-born noise. Both aspects have been addressed.

With the exception of low cut-off damped springs under the chiller casings, the original water circulation system lacked specific measures to limit the propagation of structure-borne vibration noise. Structural decoupling and damping of the vibrating elements of the water distribution system were therefore implemented following the approaches described in Sections~\ref{sec:decoupling} and~\ref{sec:damping}.

\begin{figure}[ht!]
    \centering
    \includegraphics [width=.3\textwidth]{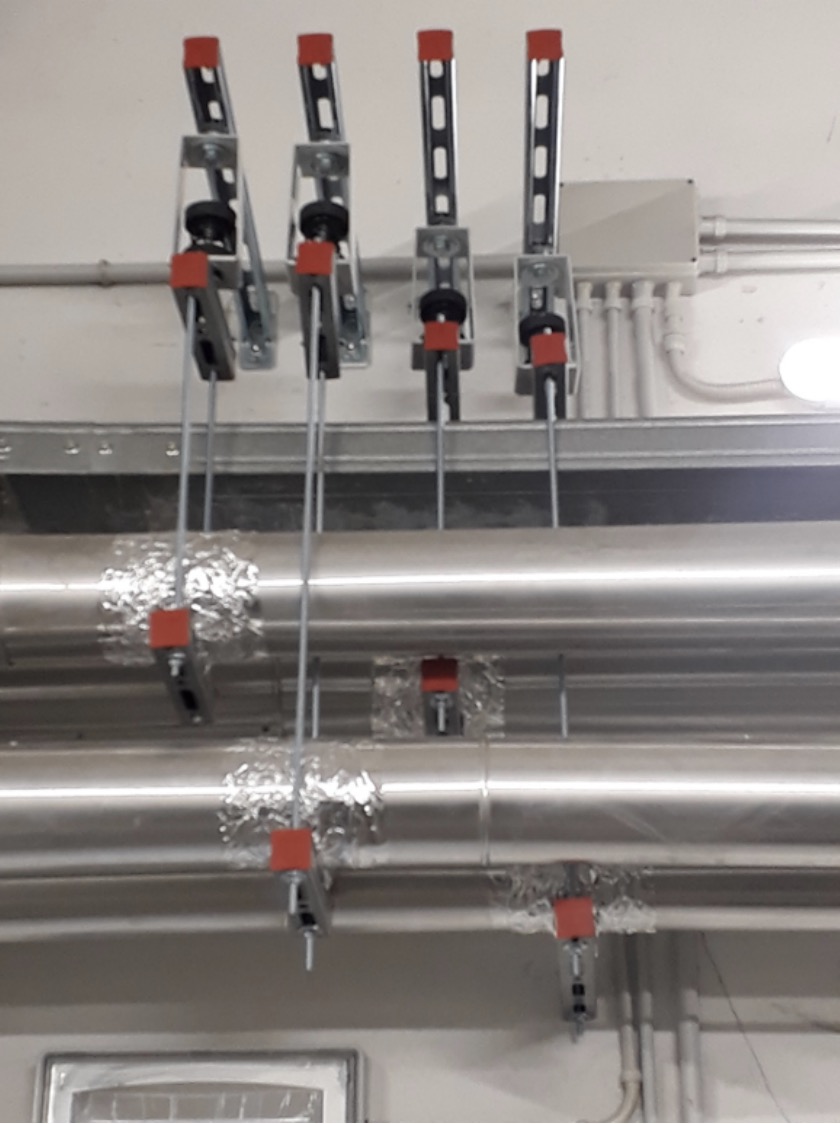}
        \includegraphics [width=.3\textwidth]{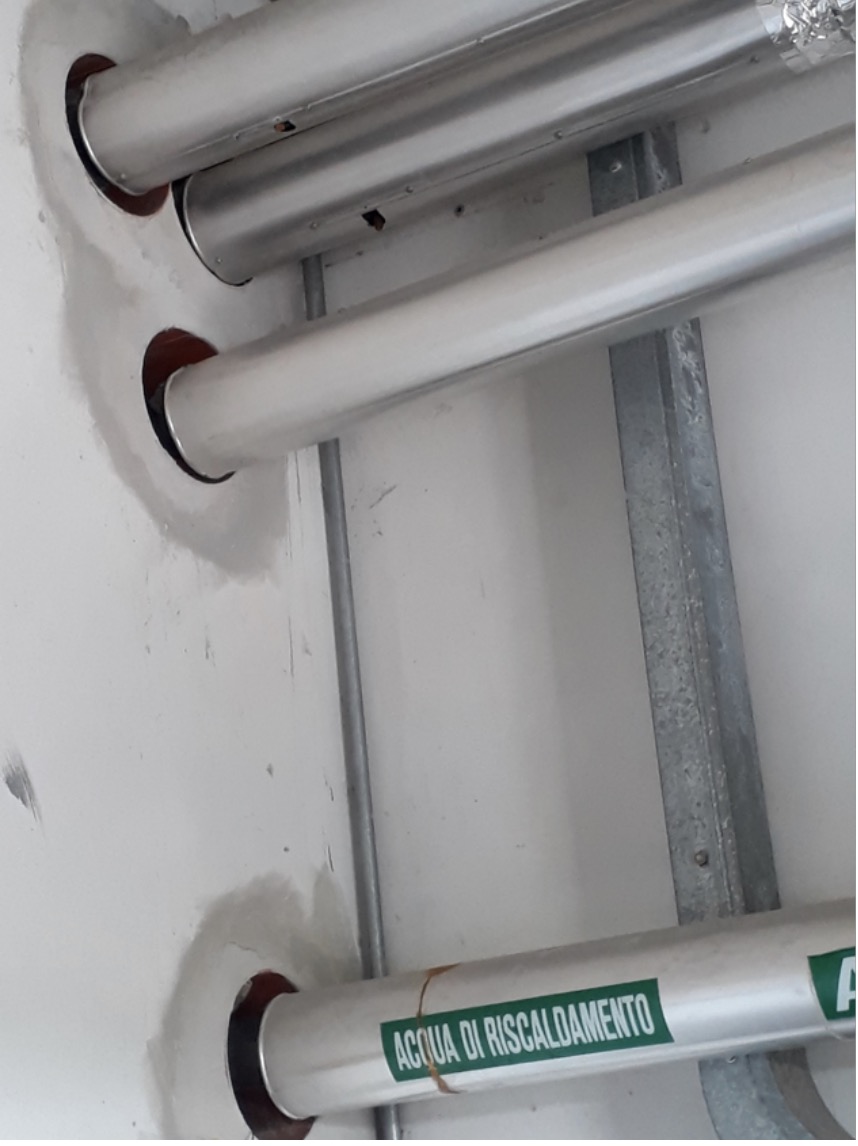}
             \includegraphics [width=.3\textwidth]%{Figure/picture_new_water_pumps.jpg}
             {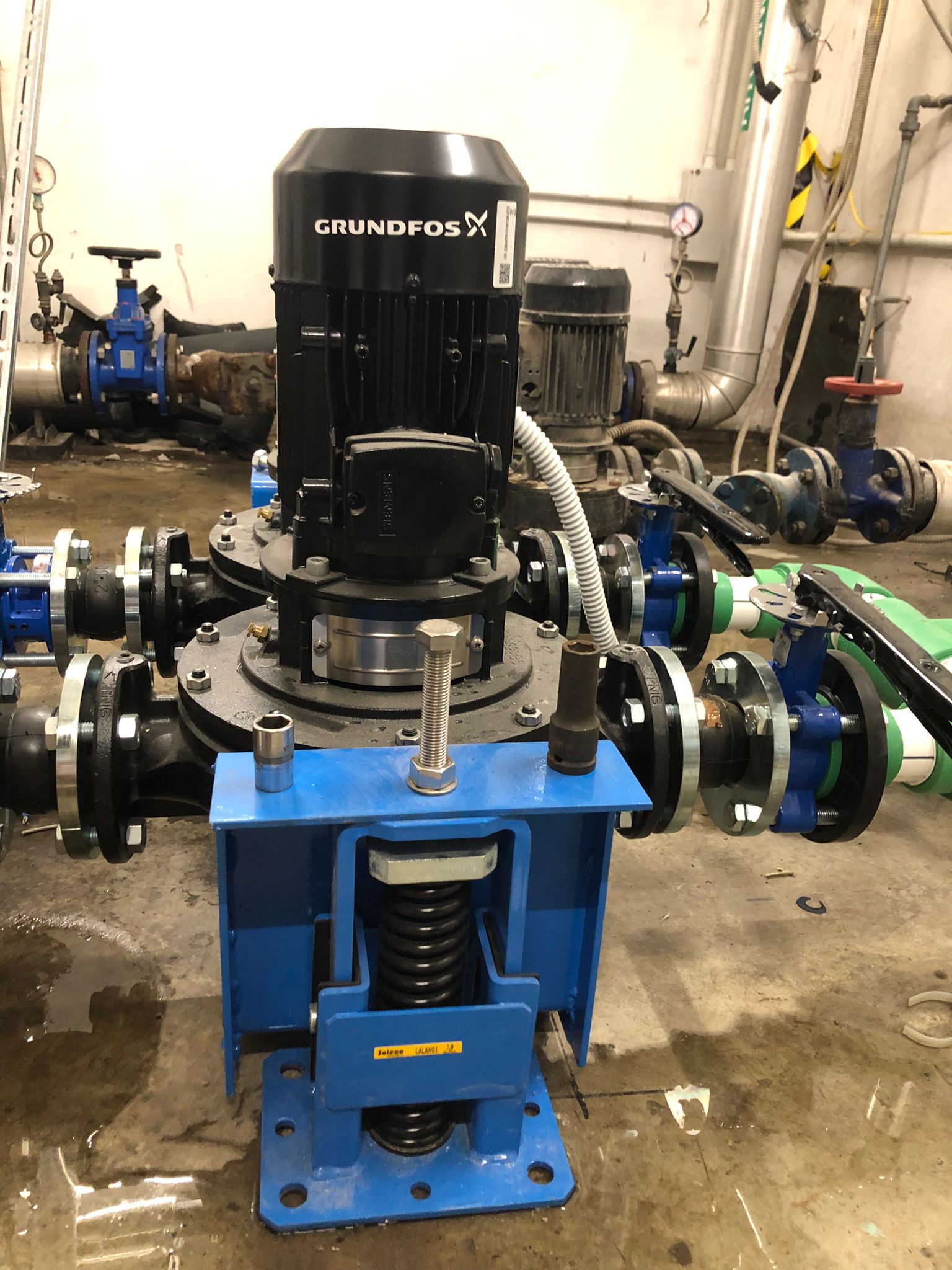}
    \caption{Water pipes structural noise mitigation actions. Left: decoupling from walls with soft damped springs. Middle: through-holes in the walls. Right: decoupling water pumps with springs and rubber joints.}
    \label{fig:foto_pipes}
\end{figure}

\begin{figure}[ht!]
    \centering
    \includegraphics[width=0.35\linewidth] 
{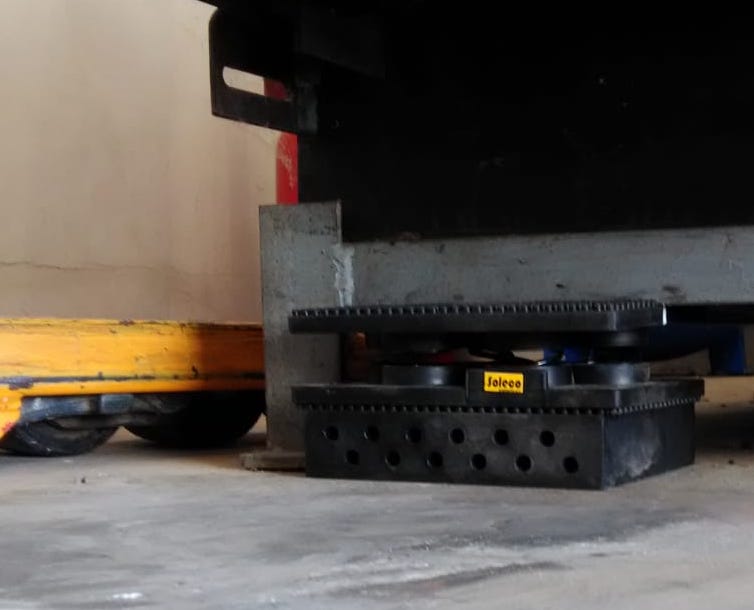}     \includegraphics[width=0.55\linewidth] 
{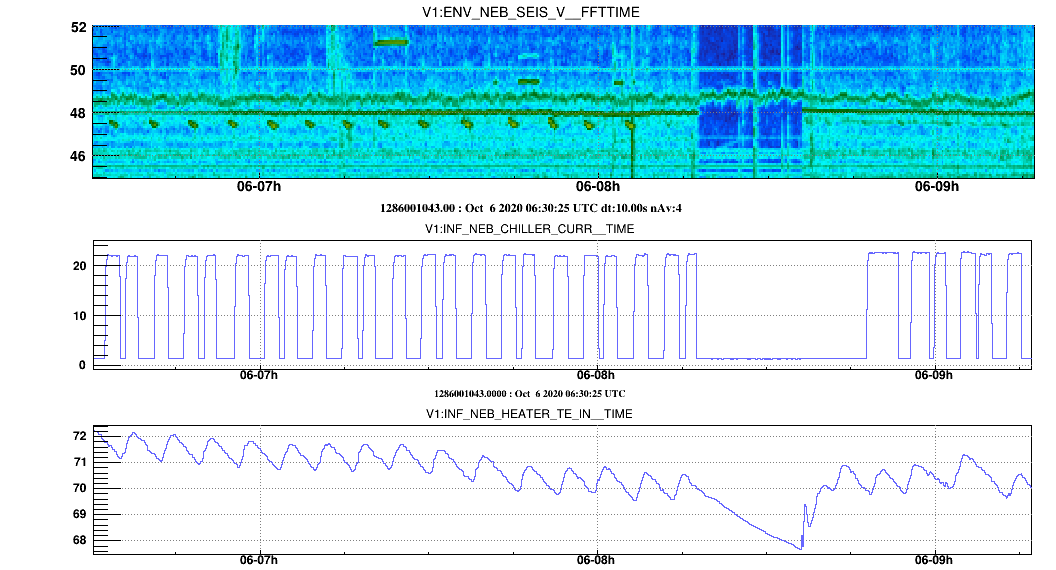} 
    \caption{Left. Seismic isolation installation of boiler casing.  Right: top - spectrogram of NEB hall seismometer; middle - water chiller current monitor;
    bottom - hot water temperature monitor.
    The transient noise at 47 Hz disappears after the pads installation work which occurred between 8:15 and 8:45.
    This plot was produced using the Virgo {\it dataDisplay} software tool \cite{DataDisplay}.
    % elogs 49582, 49755.
    %\textcolor{red}{I dati VIM sembrano non piu' disponibili e comunque non hanno la risoluzione temporale necessaria. Si potrebbe usare questo grafico importandolo in matlab. I Trend invece ci sono.}
    }
    \label{fig:boiler_pads}
\end{figure}

\begin{figure}[ht!]
    \centering
    \includegraphics [width=.8\textwidth]%{Figure/figura_water_sys_mitigations_NEB.png}
    {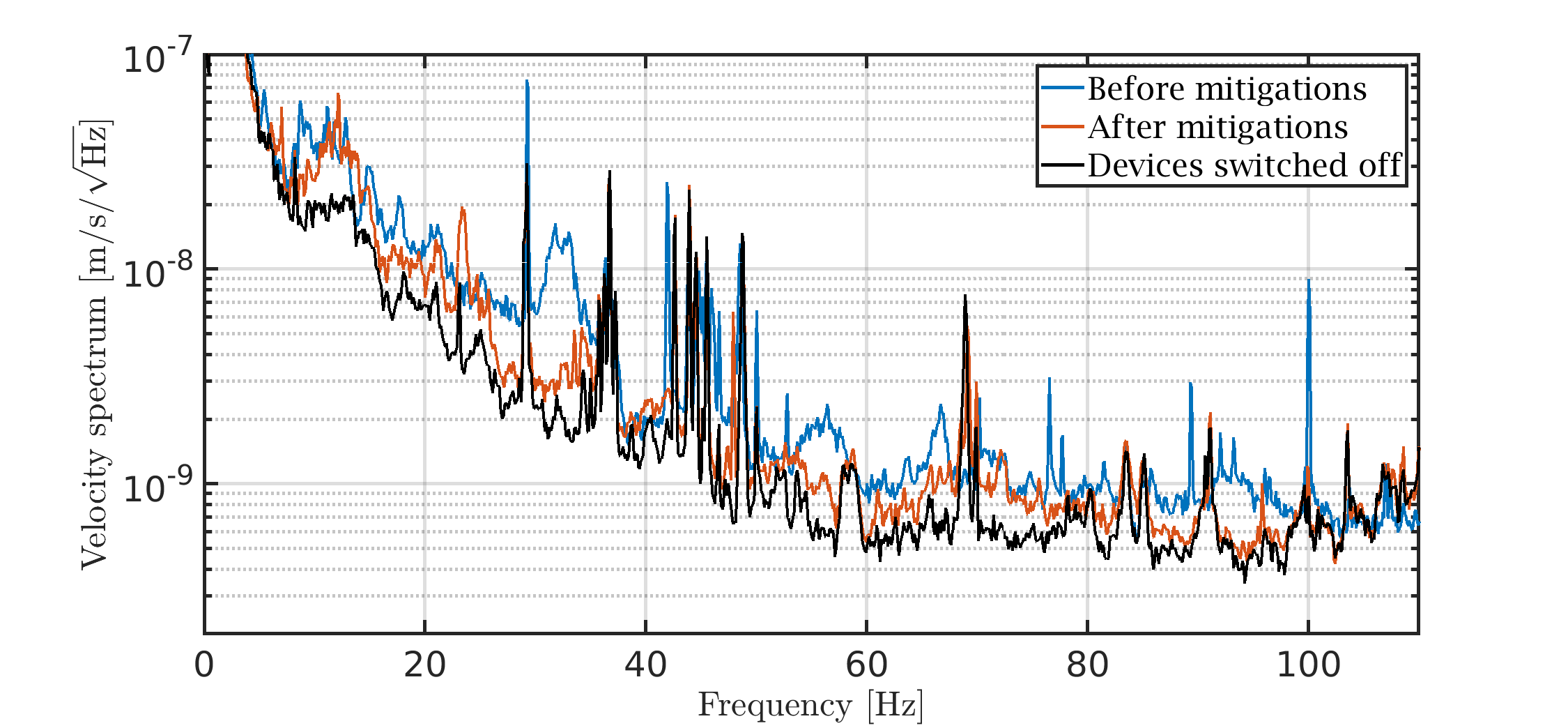}
       \caption{Seismic noise reduction after the mitigation works of the water distribution of the North air handling unit. Note that in order to measure the impact of the sole HVAC water system, the AHU whose seismic contribution is dominant has been kept off during these data recordings.
    Peaks at 47.9~Hz and its first harmonic at 95.8~Hz, which are seen in the red curve but not in the black curve, are associated to the double-pole motors of the water circulation pumps.
    }
    \label{fig:water_system}
\end{figure}

Mitigation actions included mounting boilers and water pumps on damped springs, inserting soft links (rubber joints) between pumps and pipes, decoupling pipes from walls using springs, and enlarging wall penetrations (Figure~\ref{fig:foto_pipes}). Figure~\ref{fig:boiler_pads} illustrates one of these interventions, which eliminated the intermittent 47~Hz noise from the boiler cooling fan. Overall, the implemented measures significantly reduced the seismic noise level of the experimental hall floor, as shown in Figure~\ref{fig:water_system}.
%\textcolor{red}{VERIFICATO CHE ANCHE i CHILLERS erano spenti, per entrambe le date.}

%\textcolor{red}{I grossi picchi che spariscono a 21 Hz, 29 Hz, e tra 80 e 90 Hz SONO dovuto a pompe VUOTO. Da aggiungere info nella caption.} %\textcolor{brown}{RIFARE la figura con i dati salvati, eventualmente scegliere OFF migliore.}
%\textcolor{violet}{Da 26 Ottobre 2022 ci sono NUOVE pompe che girano a 24 Hz. GUARDARE nei dati dell'ultimo switch off (DATI salvati 1 Feb 2024) quale e' l'effetto di queste nuove pompe ed eventualmente decidere se dire qualcosa.}

%textcolor{brown}{Dire della manutenzione. Prendere da primo report.}
\noindent
%Pumps components, like bearings, gaskets, rotors, are subject to weariness which, in addition to increasing the risk of malfunctioning or failure, also impacts on the emitted noise. 
%\textcolor{blue}{Maria, volendo lasciare in questo posto le segeunti frasi, si potrebbe scirvere così: In addition to structural coupling mechanisms, the noise level is also strongly influenced by the mechanical state of the pumps. }Components of pumps, including bearings, gaskets, and rotors, are prone to wear, which, in addition to elevating the risk of malfunction or failure, also contributes to increased noise emissions. Up to a factor three reduction of the pump rotation frequency vibration amplitude, was measured after pumps returned from a maintenance operation \cite{Maintennace_waterpump}.
%\textcolor{red}{Potremmo aggiungere spettrogramma relativo a sostituzione NEB hot pomp 4-poli, con riferimento a elog 57463 del 21 Ottobre 2022. Si vede che picchetto si sposta da 47.5 a 24.4 Hz.}

%DATI della figura
%----
%Before: (49565) 3  Oct. 2020 - 
%SwitchOff_20201003_1285732818_16200.gwf
%-----
%After: (56617) 3 Agosto 2022 - 
%SwitchOff_20220802_1343483459_7200_WEB_AHUwaterpumps.gwf
%SwitchOff_20220803_1343569558_6000_NEB_AHUwaterpumps.gwf
%

%%%% ----- water-borne noise
% - Test slowdown of water pump
% - test chiusura valvola

Vibrations can also propagate as pressure waves within the fluid flowing inside water pipes. Evidence of this mechanism was observed during a test in which the rotational speed of one water pump was reduced while monitoring fluid pressure fluctuations \cite{Pierini}.
% hot pump slowdown - WEB
% 16 June 2021 -elog 52167  MANCANO dati -a meno che Pierini non abbia salvato .mat
% 2 July 2021  - https://logbook.virgo-gw.eu/virgo/?r=52402
% DATI ci sono: 
%SwitchOff_20210702_1309257018_5400_WEB_water_pumps_tests.gwf

\begin{figure}[ht!]
    \centering
   \includegraphics[width=0.8\linewidth] {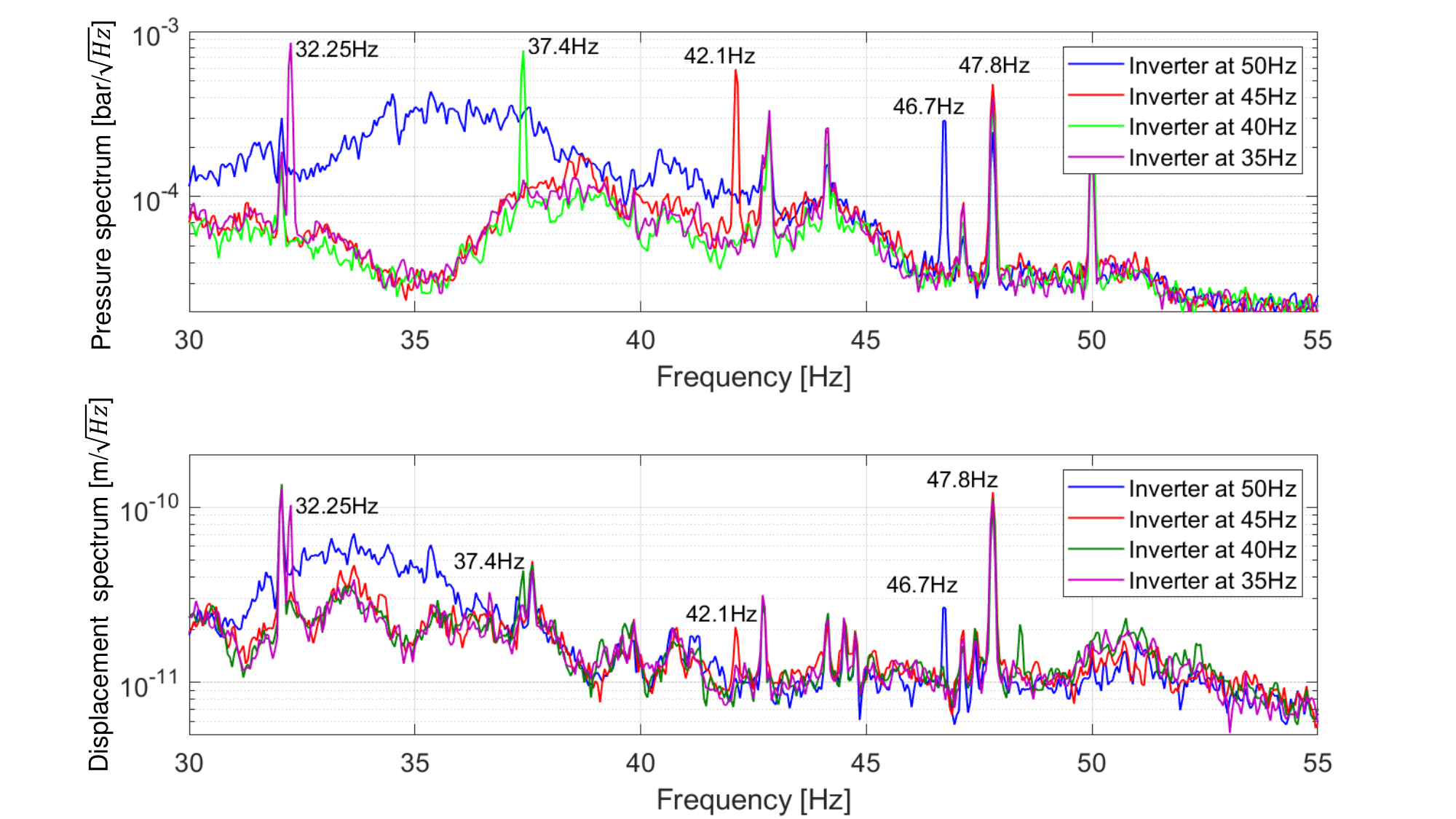}      \caption{Evolution of spectral noise during the WEB hot water pump slowdown test. The colored curves correspond to inverter frequency setting of 50 Hz, 45 Hz, 40 Hz, and 35 Hz.
Top: water pressure inside the pipe. Bottom:
seismic displacement noise of the WEB experimental hall floor.
In both graphs, marked peaks are the noise associated to the pump rotation.}
    \label{fig:waterpump_slowdown}
\end{figure}

The slowdown test was performed on a hot-water pump normally operating at 46.7~Hz, by progressively reducing the inverter frequency from 50~Hz to 35~Hz in 5~Hz steps. The minimum usable inverter frequency was found to be 40~Hz, as further reduction caused the water pressure in the circuit to drop below operational limits, leading to chiller shutdown. When the pump operated at 45~Hz, a clear reduction of seismic noise was observed in the experimental hall in the 30–40~Hz range. This reduction correlated with a decrease in water pressure fluctuations measured by a piezoelectric pressure transducer (Keller, model PA-21SR) installed along the supply pipe. Figure~\ref{fig:waterpump_slowdown} illustrates these measurements.
%\textcolor{red}{RIFARE figura usando dati del slowdown text del 2 luglio 2021. Ma in WEB hall SEISM si vede meno effetto...}
%\textcolor{brown}{Chiediamo a LORENZOP se ha salvato .mat della figura...}

% Vecchie figure
\begin{comment}
    \begin{figure}[ht!]
    \centering
    \includegraphics[width=0.8\linewidth] 
    %{Figure/slowdown_waterpump.PNG}
{Figure/LorenzoP/Web_slowdown_heatpres.png}     \includegraphics[width=0.8\linewidth] 
%{Figure/web_flux.PNG}
{Figure/LorenzoP/Web_slowdown_seis.png}      \caption{Evolution of spectral noise during the WEB hot water pump slowdown test. The colored curves correspond to inverter frequency setting of 50 Hz, 45 Hz, 40 Hz, and 35 Hz.
Top: water pressure inside the pipe. Bottom:
seismic displacement noise of the WEB experimental hall floor.
In both graphs, marked peaks are the noise associated to the pump rotation.}
    \label{fig:waterpump_slowdown}
\end{figure}

\end{comment}

% sensore: Keller, model PA-21SR piezo, "Sealed Gauge" type, 0-4bar range, 4-20mA output, 8-28V input.

A second case study of water-borne noise concerns vibrations of cold-water pipes serving the AHUs located in both end experimental halls, which are occasionally used to provide clean airflow into the vacuum chambers during human access to the interferometer test masses. These AHUs are normally kept off; however, residual water in the pipes—rigidly attached to the building walls—can transmit vibrations generated by the connected chillers.

\begin{figure}[ht!]
    \centering
    \includegraphics[width=0.8\linewidth]{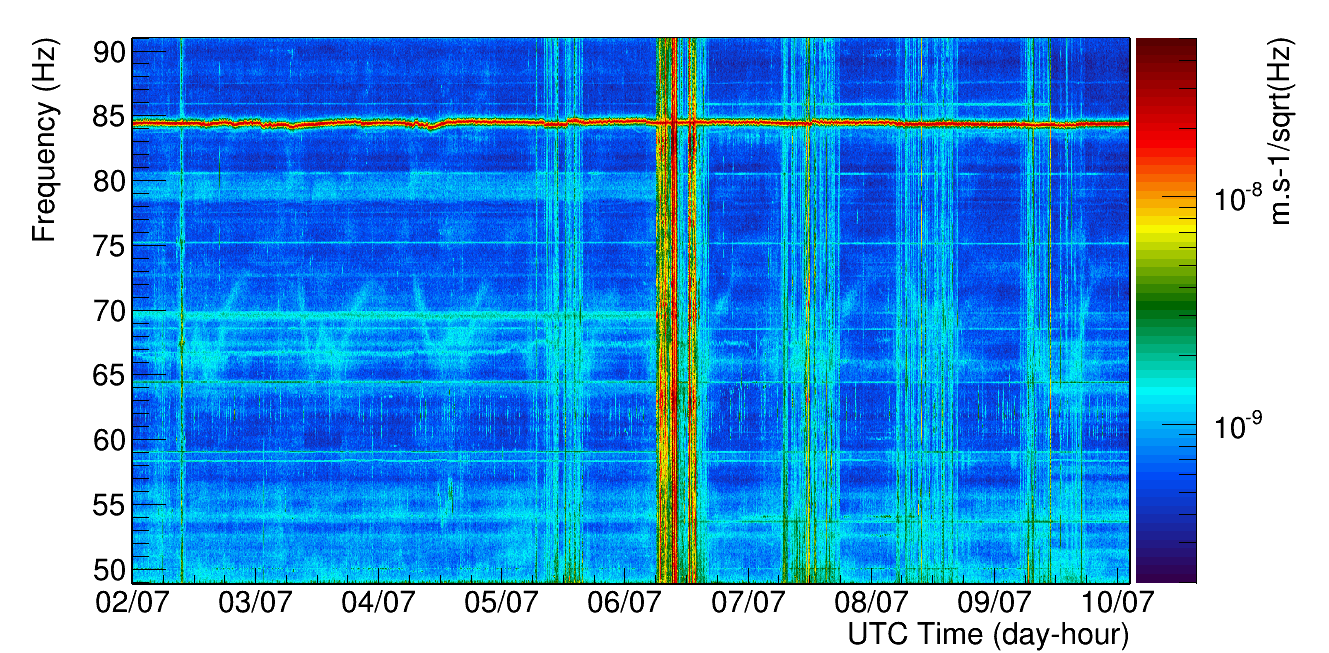}
    \caption{Time-frequency plot of the seismometer on the NEB experimental hall floor, showing a reduced vibration in the frequency band 65-80 Hz after the action of closing the valves of the unused water pipes, performed between 6:00 and 14:00 UTC of July 6$^{th}$, 2021. This plot was produced using the Virgo {\it dataDisplay} software tool \cite{DataDisplay}.}
    \label{fig:pipes_valves}
\end{figure}

Pipe vibrations, and the resulting vibration transmitted to the hall walls and floor, were reduced by closing a valve upstream of the AHU \cite{NEB_AHUCR, WEB_AHUCR}, thereby blocking the transmission of water-borne vibrations within the fluid. As a result, the vibration level of the pipe measured inside the experimental hall was reduced nearly to the plant-off level, with a concurrent mitigation of floor vibration in the 65–80~Hz frequency band, as shown in Figure~\ref{fig:pipes_valves}.
In principle, all water pipes should also have been structurally decoupled from the walls; however, this solution was deemed too invasive. 

Moreover, the operational condition of the pumps themselves was found to significantly affect noise emission: pump components such as bearings, gaskets, and rotors are subject to wear, which increases both failure risk and noise levels. Following maintenance operations, a reduction of up to a factor of three in the vibration amplitude at the pump rotation frequency was measured~\cite{Maintennace_waterpump}.

%\textcolor{red}{FIGURA che mostra la riduzione di vibrazione della pipe. SEMMAI rifare la misura.}\\
%\noindent
%\textcolor{blue}{Maria: il test è stato fatto sia a NEB che a WEB però mi sembra che sul guralp di NEB si veda meglio l'effetto. In entrambi i casi abbiamo meno vibrazione sulle pipe. Quale test mettiamo?}
%%% Rifacciamo la figura usando VIM  spectrograms:
%% NEB, 6 e poi 12 luglio 2021 (elog https://logbook.virgo-gw.eu/virgo/?r=52556)
%%% qui si vede effetto della valvola aperta/chiusa
%%% https://vim-online.virgo-gw.eu/resources_archive/2021-07-07/resources/20210707_spectro_spectro_ENV_NEB_SEIS_V_50_105_604800.jpg
%% WEB,  5 ottobre 2021 installazione bellow e valvola 
%        (elog https://logbook.virgo-gw.eu/virgo/?r=53584)
%        confronto 5, 7 e 18 ottobre (valvola aperta vs chiusa)
%        https://logbook.virgo-gw.eu/virgo/?r=53584
%%% Qui si puo' estrarre effetto del solo bellow.
%%% https://vim-online.virgo-gw.eu/resources_archive/2021-10-05/resources/20211005_spectro_spectro_ENV_WEB_SEIS_V_25_55_86400.jpg

\begin{comment}

\begin{table}[ht!]
    \centering
    \resizebox{.4\textwidth}{!}{%
\begin{tabular}{ccc}
\toprule
Parameters &  Value  \\
\midrule
Fan pulley diameter & 250 mm\\
Engine pulley diameter &  118 mm\\
r.p.m. & 1455 \\
Belt length  & 1450 mm\\
Num. of belts & 2 \\
\bottomrule
    \end{tabular}
    }
    \caption{Technical parameters of NEB AHU system are reported: diameter value of fan and engine pulleys, revolution per minute (rpm), length and number of the belts, respectively.}
    \label{tab:NEBAHUparameters}
\end{table}
\quad

\end{comment}

\section{Noise investigations}
\label{sec:noise_investigations}

This section describes dedicated tests aimed at assessing how the HVAC system couples acoustic, seismic, and magnetic noise into the interferometer. The activities included duct disconnection to discriminate between seismic and acoustic transmission paths, a controlled slowdown of the AHU to evaluate the dependence of acoustic noise on fan speed, and a measurement campaign to characterize the magnetic noise generated by the AHU system. The results guided the definition of targeted mitigation strategies.

\subsection{Ducts disconnection test}
\label{subsec:Ducts_disconnection}

% PRIMA: test disconnessione condotti
% riduzione acustico hall con pannello rigido
% dimostra rilevanza percorso acustico condotti
% picco 15 Hz (fan) e armoniche sparisce da acustico hall
% (ipotesi Lionel: vibro-acustic noise menzionato in qualche REFERENCE)
Low frequency acoustic noise generated by the AHU can follow different paths to the experimental hall: (i) sound generated inside the AHU room can be transmitted through separating walls; (ii) sound can be emitted by vibrating air ducts; 
(iii) sound can also be generated by turbulent air motion inside ducts or at the fan and be transmitted through the ducts; finally, (iv) sound can be generated at the outlets. 
%\textcolor{red}{Manca di elencare qualche percorso importante? (N.B. qui parliamo solo della parte di suono.)}
To assess the relative contribution of these paths, an experiment was conducted in which selected transmission routes were temporarily and selectively interrupted \cite{DuctDisconnection}.

\begin{figure}[htp!]
    \centering
    \includegraphics[width=0.45\textwidth]{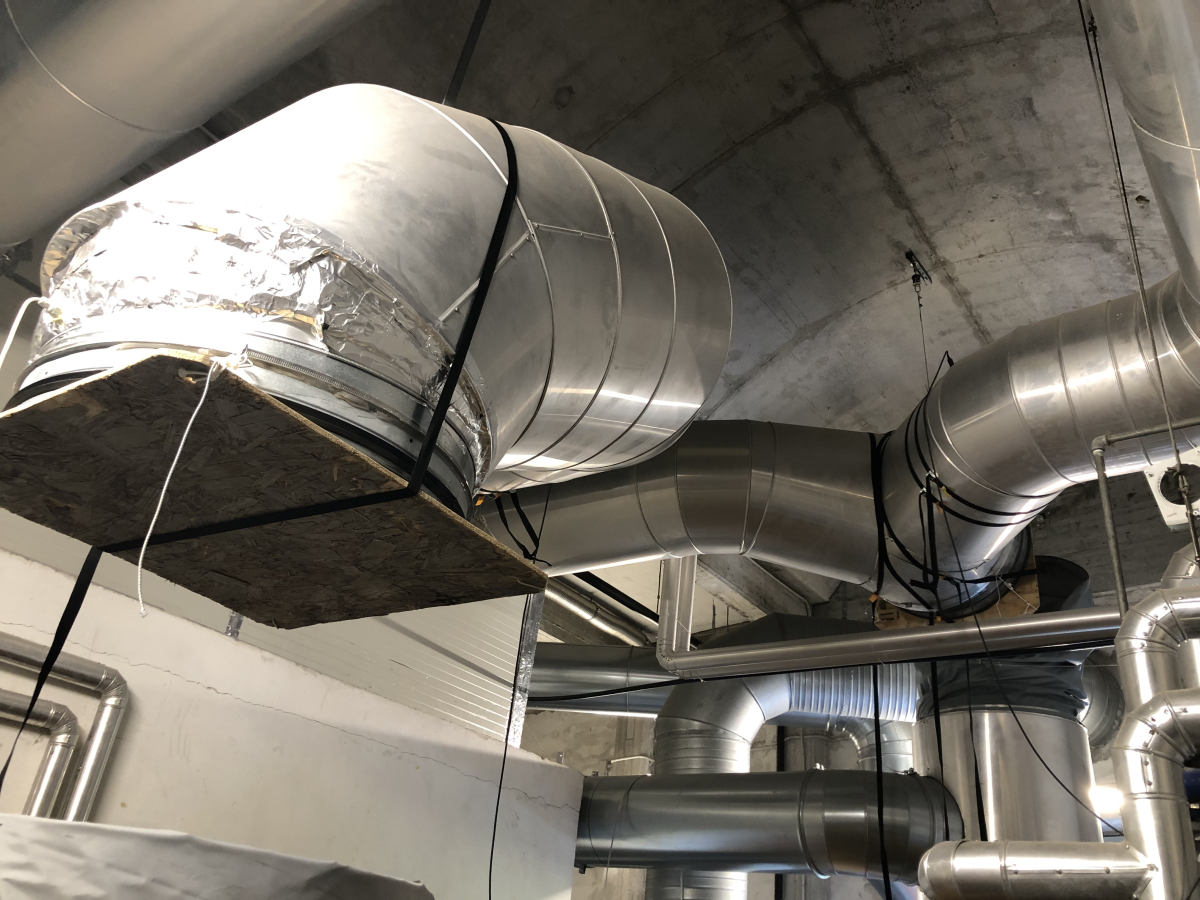}
    \includegraphics[width=0.30\textwidth]{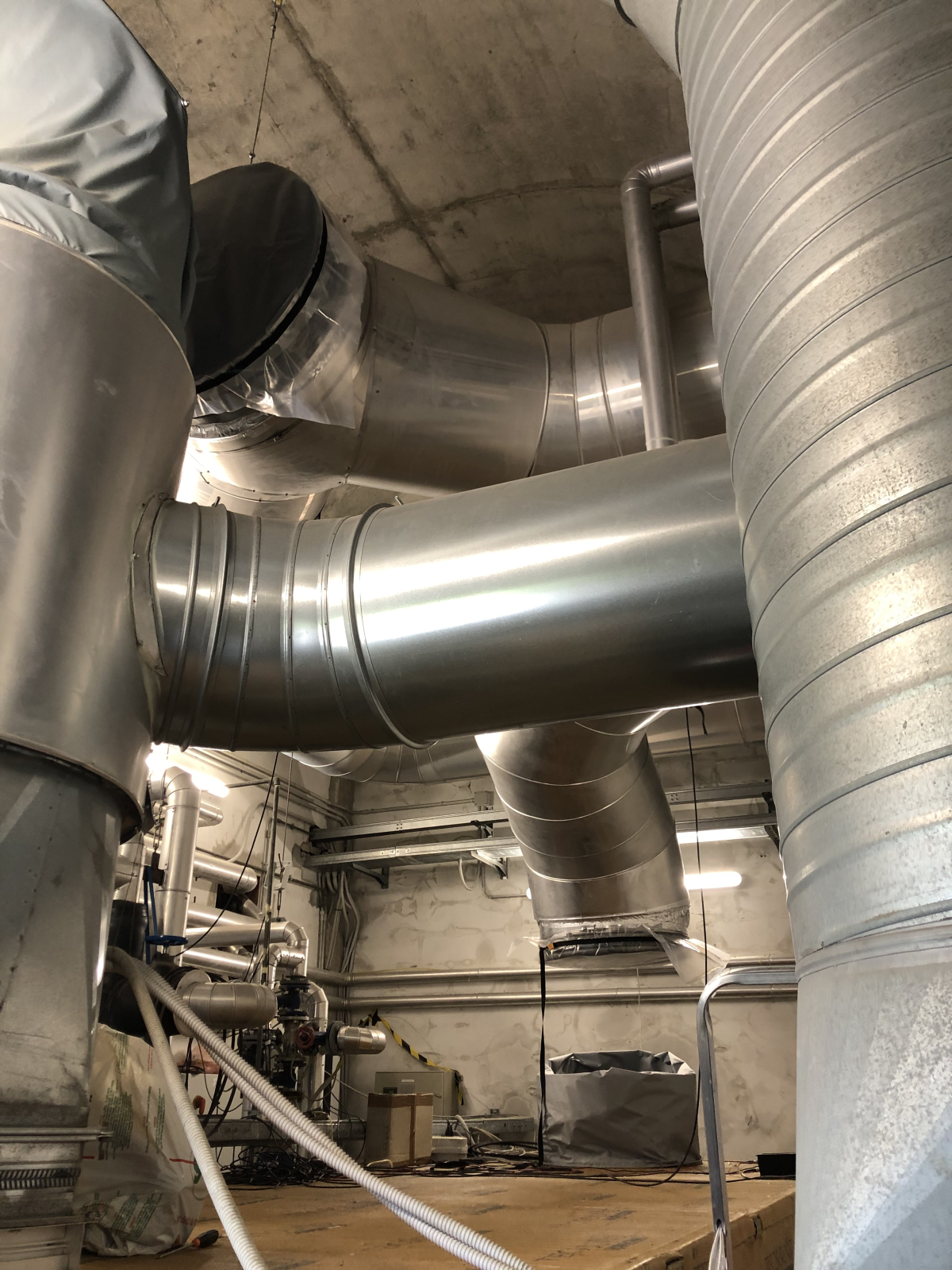}
    \caption{Disconnection of the NEB air ducts: supply and return air ducts were covered by rigid wooden boards (on the left) and plastic foils (on the right).}
    \label{fig:duct_disconnection}
\end{figure}

\begin{figure}[htp!]
    \centering
    \includegraphics [width=0.9\textwidth]{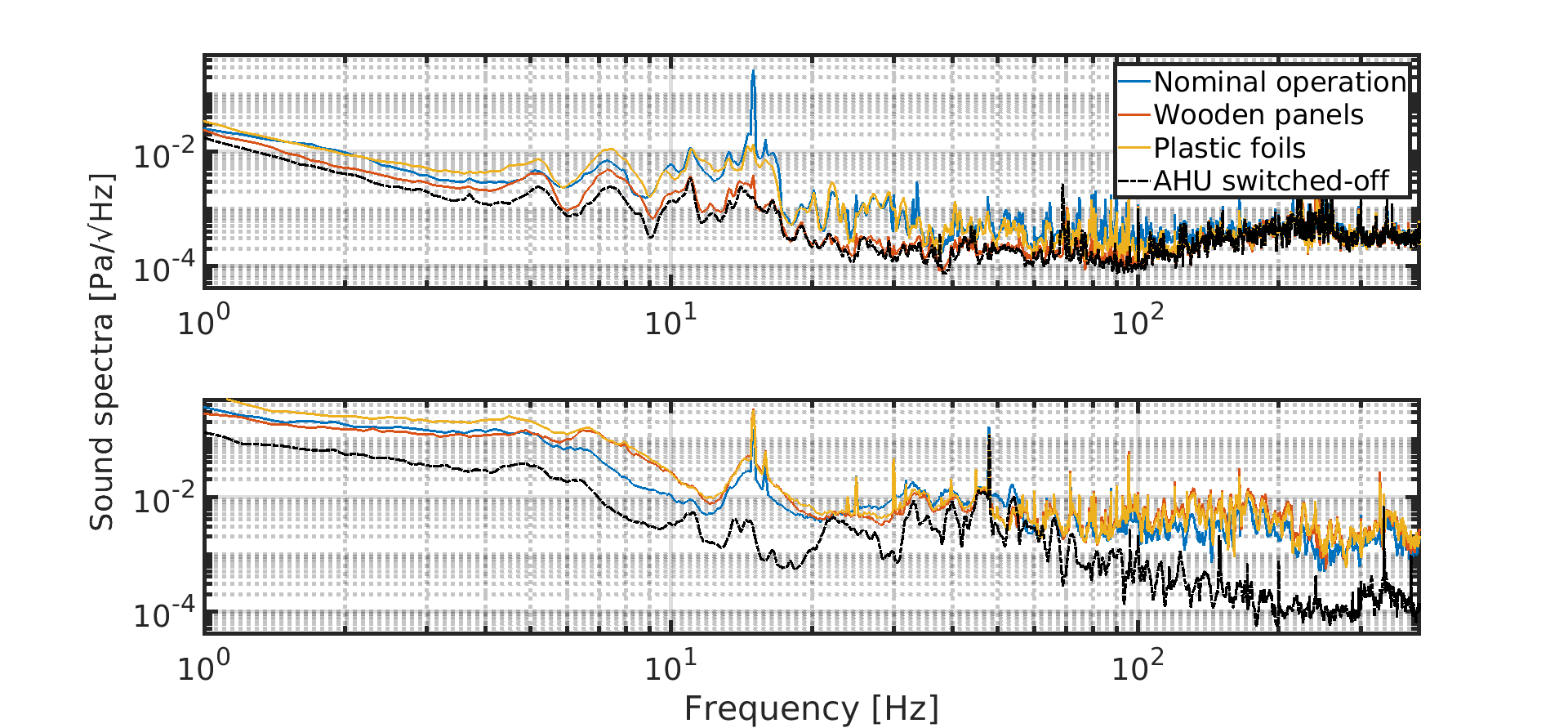}
    \caption{Acoustic spectra measured in the NEB experimental hall (top) and in the NEB AHU room (bottom) during the ducts disconnection experiment. The spectral line at 15 Hz corresponds to the fan rotational frequency during a temporary AHU configuration, slightly different from that reported in the left column of Table \ref{tab:NEBAHUparameters} (motor pulley = 150 mm; belt length = 1407 mm; inverter frequency = 50 Hz). This line disappears when the ducts are disconnected, indicating a vibro-acoustic origin.}
    \label{fig:disconnectionTest_mics}
\end{figure}

Specifically, the supply and return air ducts were disconnected from the AHU casing, and their openings were alternately sealed with a plastic foil and with a rigid wooden plank (see Figure \ref{fig:duct_disconnection}). During the test, the AHU operated under nominal conditions with the inverter set to 50 Hz.
%\textcolor{red}{Sposterei questa footnote nella caption della Tabella 2 sotto, almeno la parte che descrive il calcolo della belt frequency. Gli altri numeri forse non e' essenziale citarli qui?} \textcolor{blue}{I valori sono stati riportati poiché il picco a 15 Hz menzionato nella Fig 21 risale a questi specifici valori di pulley. Il test venne fatto il 30 Nov 2022 con il nuovo ventilatore ma con in corso il cambio di diametro motor pulley. Questo valore e quello delle belt non sono poi quelli definitivi come riportato nella tabella 1. Concordo che le  formule possono esser riportate nella tabella dello slowdown.}
%\footnote{Technical parameters are as follows: fan pulley diameter = 250 mm; motor pulley diameter = 150 mm; belt length = 1407 mm; motor speed = 1455 rpm. With inverter at 50 Hz: fan frequency $\simeq$ 24.91\,Hz; motor frequency $\simeq$\,14.97 Hz; belt frequency $\simeq$\,7.95 Hz.}. 
As shown in Figure \ref{fig:disconnectionTest_mics}, the noise level in the hall remained essentially unchanged with respect to the configuration with connected ducts when the plastic foil was used, whereas a significant reduction was observed with the rigid wooden plank.

This test primarily targeted sound transmission inside the ducts: the rigid plank effectively blocked airborne sound propagation, while the plastic foil did not. These results indicate that airborne noise transmitted through the ducts is the dominant path from the AHU to the hall. Consequently, mitigation efforts focused on reducing fan noise, as discussed in Section \ref{sec:airborne}.

%This test primarily affected the noise path associated with the transmission of sound pressure waves inside the ducts: the rigid wooden plank  effectively blocked this transmission, whereas the plastic foil did not. These results indicate that the dominant noise path from the AHU to the hall is air-borne sound transmitted through the ducts. Based on this evidence, mitigation efforts were prioritized to reduce sound emissions from the fan (described in Section \ref{sec:airborne}).

\noindent

% DATA:
% https://logbook.virgo-gw.eu/virgo/?r=52125
% June 11 2021 - NEB new air duct disconnection test - elog 52125 (includes also tapping on air ducts)
%/data/prod/envmon/HVAC/SwitchOff_data/SwitchOff_20210611_1307433618_6000_NEB_air_duct_disconnection_tests.gwf

%=================================================================
\subsection{Fan slowdown}
\label{subsec:slow_down}

\begin{figure}[htbp!]
    \centering
    \includegraphics [width=.8\textwidth]{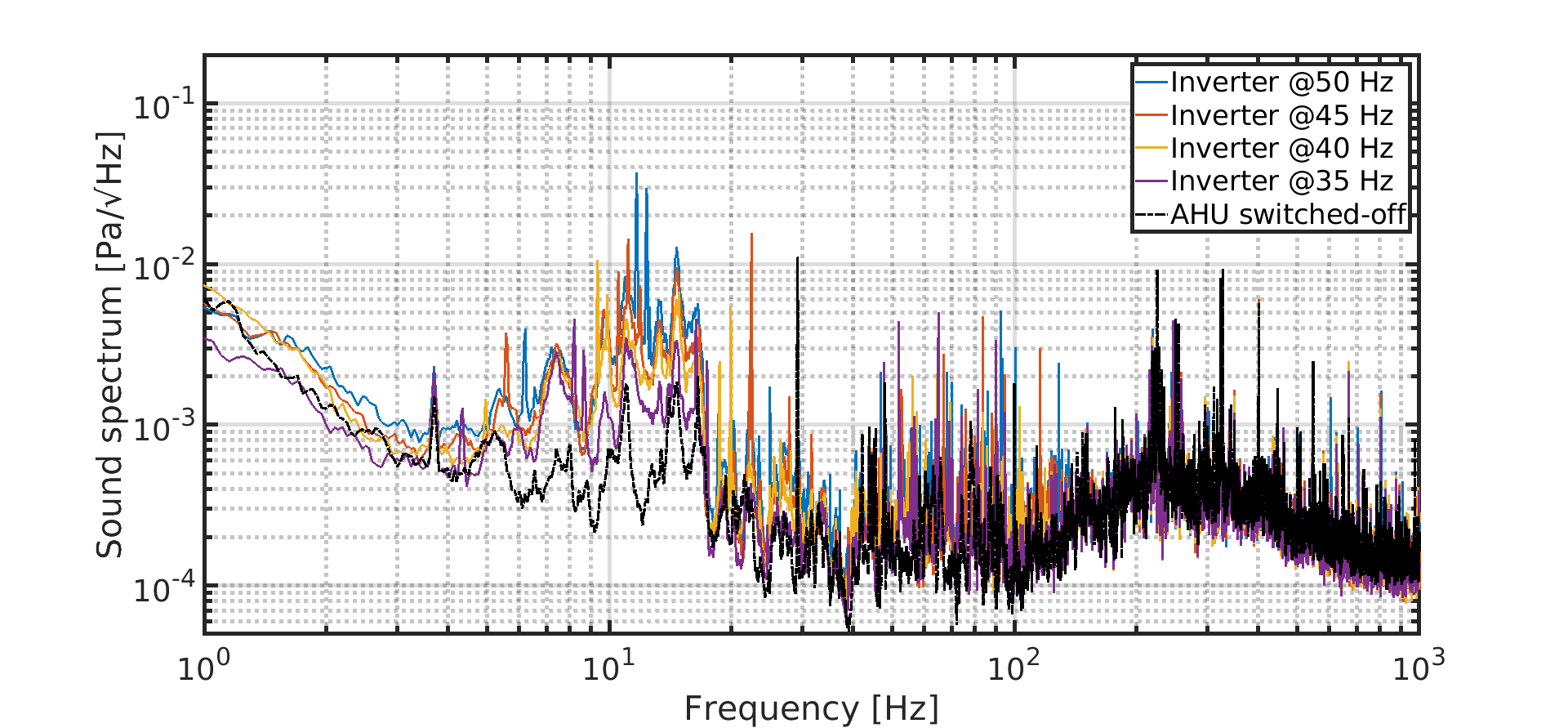}
    \caption{Acoustic spectra recorded by the microphone inside the NEB experimental hall.
The blue, red, yellow, and violet curves correspond to the operation of the backward-curved air fan controlled by the inverter at 50 Hz (fan speed: 24.95 Hz), 45 Hz (10.53 Hz), 40 Hz (19.96 Hz), and 35 Hz (17.47 Hz), respectively. The black curve represents the spectrum with the AHU switched off.}
    \label{fig:fan_slow-down}
\end{figure}

Fan speed is a key parameter in acoustic noise generation. High rotational speeds induce the formation of air vortexes and turbulent flow in the vicinity of the fan and inside the ducts, leading to broadband low-frequency noise \cite{Jiang2023}. A controlled reduction of the AHU fan speed was therefore implemented by adjusting the variable-speed drive \cite{NewFanComparison}. The inverter frequency was decreased from the nominal 50 Hz to 35 Hz in steps of 5 Hz, corresponding to fan rotation frequencies of 11.7 Hz, 10.5 Hz, 9.4 Hz, and 8.2 Hz, respectively (see Table \ref{tab:NEBAHUslowdown}). As shown in Figure \ref{fig:fan_slow-down}, the broadband noise below 100 Hz measured in the experimental hall progressively decreased during the test.

\noindent
Ambient parameters were monitored throughout the experiment: temperature fluctuations remained within nominal limits, while the overpressure was reduced but remained sufficient to meet operational requirements. 

\noindent
A substantial reduction in acoustic noise was achieved by decreasing the fan speed. Further reduction, however, was limited by several factors: (i) compliance with the fan operating-point specifications; (ii) prevention of motor overheating; (iii) preservation of the experimental hall environmental conditions—including temperature and overpressure—as well as cleanliness, the latter being determined by the required air-change rate; and (iv) outdoor temperature conditions during the hot season.

%In addition, a limitation of the present NEB and WEB HVAC system is that 
%it cannot sustain the fan low frequency operation during the hot season. When outdoor temperature is above 30 degree Celsius, the fan speed needs to be increased to ensure stable ambient conditions. 

% elog di riferimento: https://logbook.virgo-gw.eu/virgo/?r=59391
% 
% si riferisce anche una notevole riduzione della vibrazione dei condotti

\begin{comment}
\begin{table}[ht!]
    \centering
    \resizebox{.9\textwidth}{!}{%
\begin{tabular}{|c|c|c|c|c|c|c|}
\toprule
Inverter frequency &  Motor Frequency & Fan frequency & Belt frequency  \\
\midrule
50 Hz (nominal) & 24.95 Hz & 11.70 Hz & 6.18 Hz\\
45 Hz & 22.27 Hz & 10.53 Hz & 5.56 Hz\\
40 Hz  & 19.96 Hz & 9.36 Hz & 4.95 Hz\\
35 Hz & 17.47 Hz & 8.19 Hz & 4.33 Hz\\
\bottomrule
    \end{tabular}
    }
    \caption{Measured parameter values during the slow-down test of the NEB AHU, performed after the installation of the backward-curved air fan.} 
    \label{tab:NEBAHUslowdown}
\end{table}
\end{comment}

\begin{table}[h]
\centering
\resizebox{1.0\textwidth}{!}{
\begin{tabular}{|c| cc| cc| cc|}
\hline
& \multicolumn{2}{c|}{Motor frequency (Hz)}
& \multicolumn{2}{c|}{Fan frequency (Hz)}
& \multicolumn{2}{c|}{Belt frequency (Hz)} \\

Inverter frequency (Hz)
& Theory & Measured
& Theory & Measured
& Theory & Measured \\

\hline
50 (nominal) & 24.25 & 24.95 & 11.45 & 11.70 & 6.20 & 6.18 \\
45        & 21.82 & 22.27 & 11.30 & 10.53 & 5.58 & 5.56 \\
40        & 19.40 & 19.96 & 9.16 &  9.36 & 4.96 & 4.95 \\
35        & 16.97 & 17.47 &  8.00 &  8.19 & 4.34 & 4.33 \\
\hline
\end{tabular}
}
\caption{Measured parameter values during the slowdown test of the NEB AHU, performed after the installation of the backward-curved fan. 
The motor rotational frequency scales with the inverter frequency as 
$f_{\rm motor}=f_{\rm motor}^{(50)}(f_{\rm inv}/50)$, 
where $f_{\rm motor}^{(50)}$ is the motor rotational frequency at the nominal inverter frequency of 50~Hz. 
The fan frequency follows from the pulley transmission ratio,
$f_{\rm fan}=f_{\rm motor}(D_{\rm motor}/D_{\rm fan})$, 
and the belt frequency is given by 
$f_{\rm belt}=(\pi D_{\rm fan} f_{\rm fan})/L_{\rm belt}$, 
where $D_{\rm motor}$ and $D_{\rm fan}$ are the motor and fan pulley diameters and $L_{\rm belt}$ is the belt length, all reported in Table~\ref{tab:NEBAHUparameters}.}
\label{tab:NEBAHUslowdown}
\end{table}

\subsection{Magnetic noise}
\label{sec:Magnetic_noise}
% topics:
% fan and other vibrating objects (by Earth modulation) - FIGURA ok
% sidebands 50Hz - Figura ok
% glitches
% (inverter)
% grounding noise (REF)

%\textcolor{red}{Questo primo paragrafo dovrebbe andare nella section 2. O forse c'e' gia' quindi qui dovremmo eliminarlo.} \textcolor{blue}{Nella sezione due è citato in modo sintetico. Provo a spostarlo}
% Numerous studies have shown the sensitivity of GW interferometers to electromagnetic fields~\cite{Envpaper2020, Nguyen_2021}.
%Ambient magnetic fields, specifically in the low-frequency range ($\lesssim 100$~Hz), 
%\textcolor{violet}{perche' "specialmente sotto i 100 Hz" ? Stiamo pensando all'attenuazione delle camere da vuoto?}
%can interact with magnetic actuators placed along anti-seismic suspensions, or with permanent magnets placed on suspended benches (e.g. Faraday isolators) or, as is the case with Virgo, directly on magnets glued on the test masses.
%There have also been reports of magnetic coupling occurring through cables and connectors \cite{Handbook_Fiori}. %\textcolor{violet}{ricordo una entry di LIGO/Schofield?}

HVAC systems generate both continuous and impulsive magnetic noise. This section presents representative examples of both.

\begin{figure}[ht!]
    \centering
    \includegraphics [width=.8\textwidth]{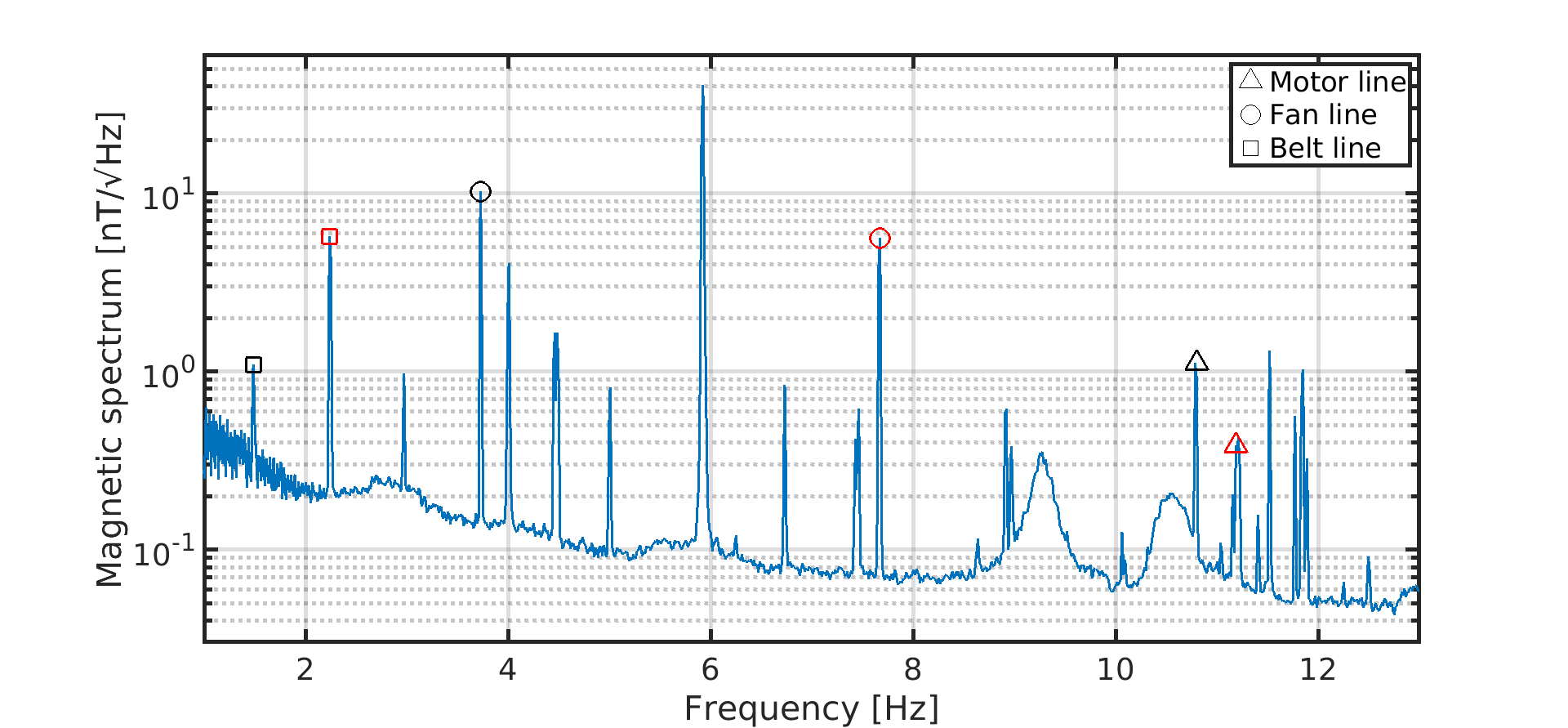}
    \caption{Magnetic spectrum of the probe positioned at a few meters from the AHU. Markers indicate the lines whose frequency matched the rotation frequency of the fan, the motor, and the belts. }
    \label{fig:magnetic_noise}
\end{figure}

\begin{figure}[htp!]
    \centering
    \includegraphics[width=0.25\linewidth]{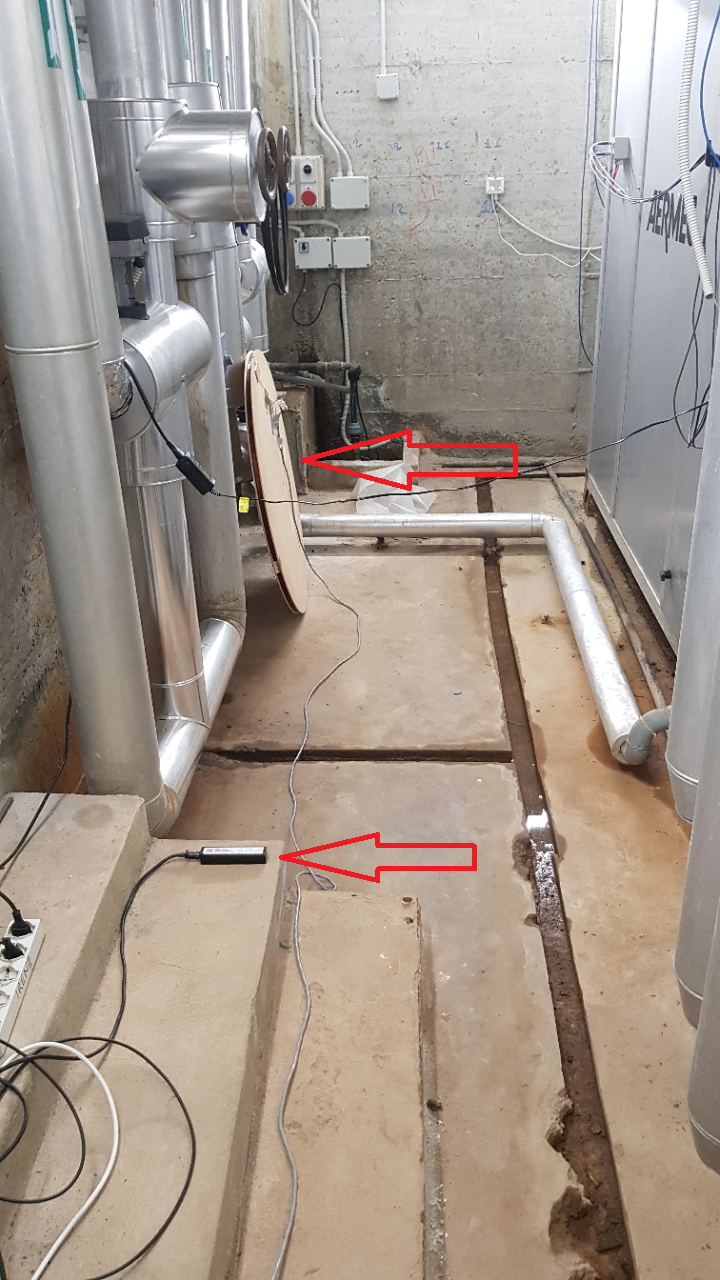}
        \includegraphics[width=0.6\linewidth]{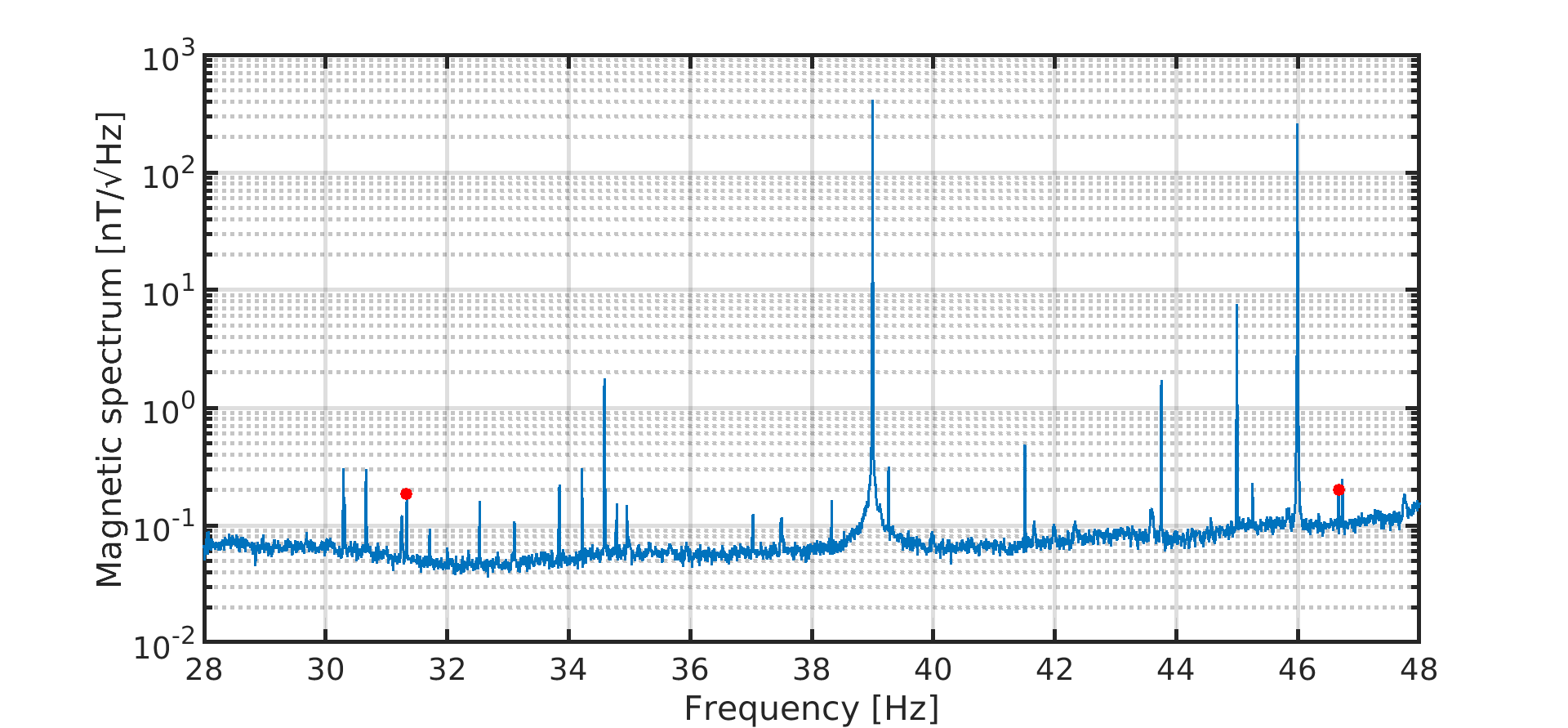}
    \caption{Left. Installation of the portable coil and magnetometer in front of the CEB AHU. %Right. Spectrum of the magnetic probe; the red dots indicate the sidebands at ~$\sim$31.3 Hz and ~$\sim$46.7 Hz.  
    Right. Spectrum of the magnetic probe. The injected line exhibits sidebands at $\pm,7.67$ Hz, corresponding to the rotation frequency of the supply fan. The features visible around $\sim$31.3 Hz and $\sim$46.7 Hz are the resulting sidebands of the injected line, highlighted with red markers.}
    \label{fig:CEBAHU_magninjline}
\end{figure}

Magnetic spectral lines associated with the AHU fan, motor, and belts were identified in ambient magnetic spectra, as shown in Figure~\ref{fig:magnetic_noise}.
These lines originate from large rotating metallic structures (e.g., the AHU fan) that modulate the Earth’s magnetic field. This hypothesis was tested by injecting a 39~Hz magnetic field using a portable 1~m diameter coil \cite{CEBAHU_magneticline}. As shown in Figure~\ref{fig:CEBAHU_magninjline}, the injected line exhibits sidebands at the supply fan rotation frequency, confirming the modulation mechanism.

%\textcolor{red}{Can we say something about how far these magnetic lines propagate.... Mi pare Lorenzo fece una indagine.}

%% DATI Sniffing magnetico /data/prod/envmon/HVAC/magnetico_AHU/CEB_HALL_AHU_20220805_*_sniffing_mag_probe.gwf

%and the same family of spectral lines was observed to rise symmetrically forming so-called sidebands on both sides of the injected frequency \cite{CEBAHU_magneticline}.
%Figure~\ref{fig:CEBAHU_magninjline} shows the result of this test, namely the injected line exhibits sidebands at~$\pm$7.67 Hz, the supply fan’s rotation frequency.

%/data/prod/envmon/HVAC/magnetico_AHU]: ls iniezioni_bobina/CEB_HALL_AHU_20220920_1347713118_7200_coil_injection_test.gwf

\begin{figure}[ht!]
    \centering
    \includegraphics [width=.7\textwidth]{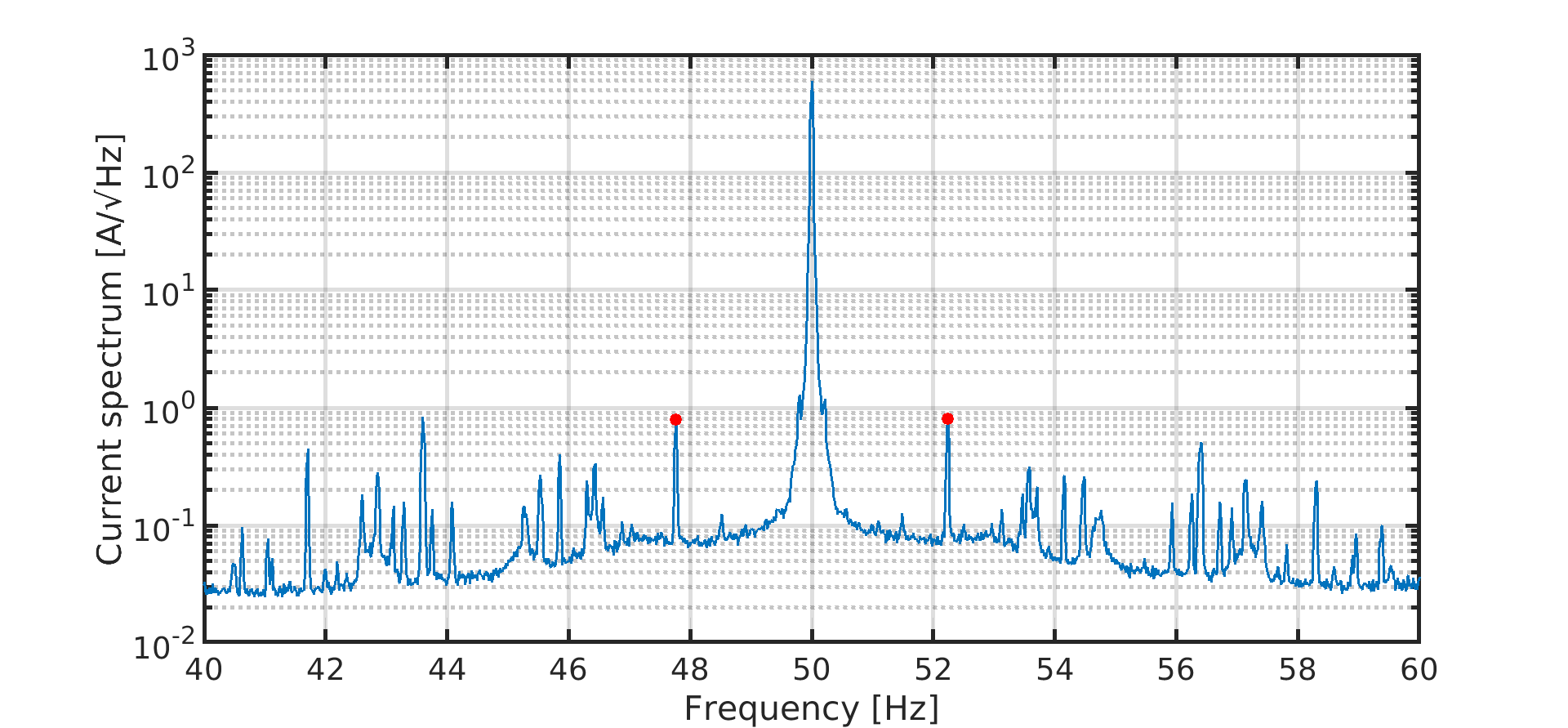}
    \caption{Spectral noise of the electrical mains current monitor. Red dots indicate the $\pm 2.24$~Hz sidebands of the 50~Hz mains line, corresponding to the belt revolution frequency of the CEB hall supply AHU. %\textcolor{red}{Mi pare che in uno switch off avessimo identificato altre linee sidebands associate a AHU... o forse no?}
    }
    \label{fig:CEB_AHU_sidebands}
\end{figure}

Sidebands around the 50~Hz mains frequency observed in the magnetic spectra in Figure~\ref{fig:CEB_AHU_sidebands} are also attributable to the AHU, as they vanish when the unit is switched off. The likely coupling path involves periodic voltage ripples in the mains supply induced by AHU power surges, as discussed in \cite{LIGO_DetcharO2}. The resulting noisy currents propagate through the experimental power distribution network, whose cables radiate the associated magnetic fields. Unlike the local modulation mechanism, this effect is not confined to the vicinity of the AHU and is therefore potentially more detrimental for the interferometer. Indeed, these sidebands were close to limiting Virgo sensitivity during the O3 observing run~\cite{Envpaper2020}.

Another class of magnetic disturbances consists of short-duration ($\sim$0.1~s), high-intensity transient signals, commonly referred to as glitches. These events are typically generated during the start-up of heavy, periodically switching loads, such as chillers and air compressors in the HVAC systems of the terminal buildings. The coupling mechanisms are presumed to be the same as those described above.

\begin{figure}[htp!]
    \centering
    \includegraphics[width=1.0\linewidth]{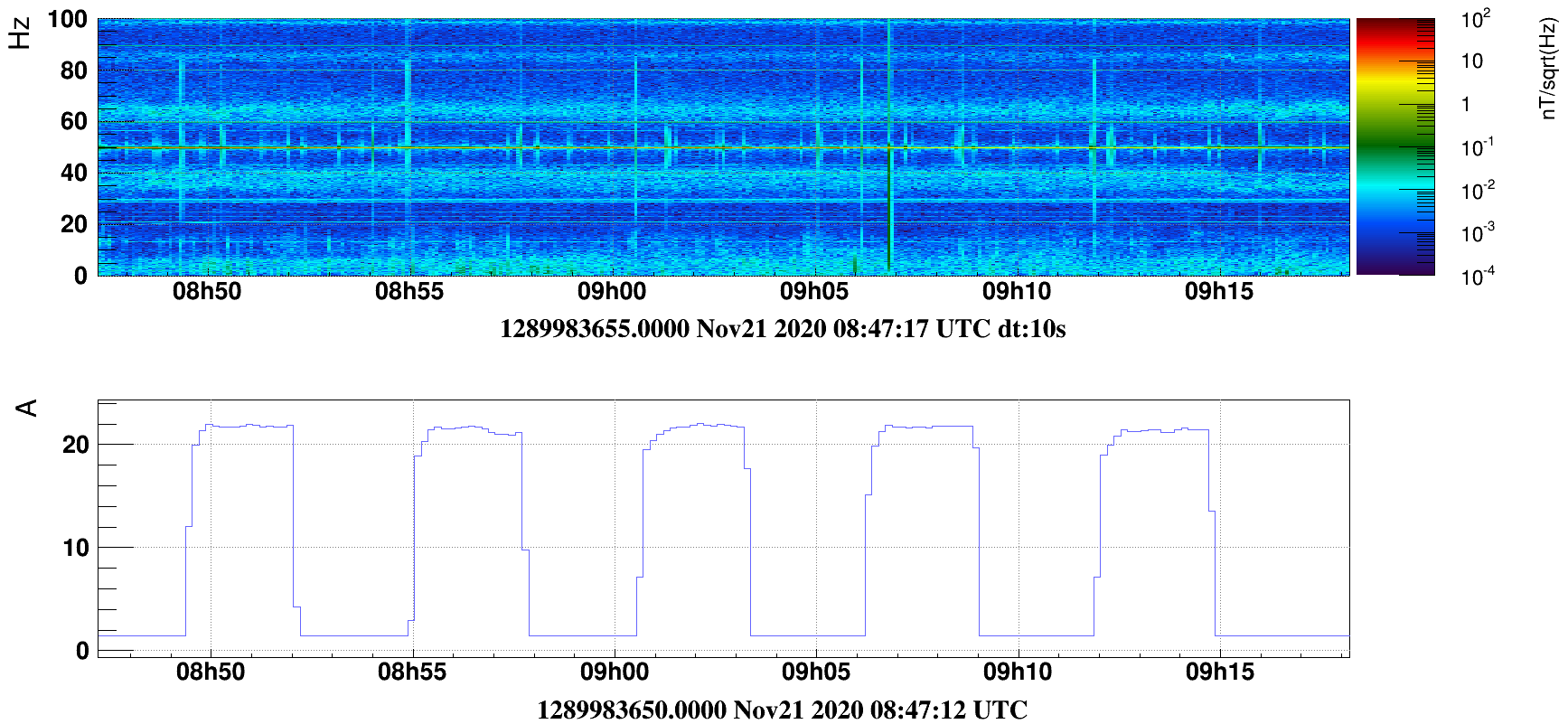}
    \caption{ Top. Time-frequency of the magnetometer inside the NEB hall. Bottom. Current monitor of the NEB water chiller. This plot was produced using the {\it dataDisplay} software tool \cite{DataDisplay}.
    }
    \label{fig:chiller_glitch}
\end{figure}

Although such glitches could in principle mimic transient gravitational-wave signals, they have not been observed in the detector strain sensitivity, but only in magnetometers located within the experimental halls. They often occur with a periodicity correlated with ambient air temperature oscillations or thermal fluid cycles. Magnetic glitches associated with chillers and air compressors typically populate the 20–80~Hz and 40–60~Hz frequency bands, respectively, as shown in Figure~\ref{fig:chiller_glitch}.

Magnetic interference from HVAC systems and similar infrastructure can generally be mitigated by applying established good-practice guidelines, including:
(i) minimizing the loop area between forward and return conductors (e.g., phase and neutral in single-phase systems, or between phases in three-phase systems);
(ii) using twisted-pair cables for single-phase power distribution;
(iii) routing power and signal cables in separate trays;
(iv) electrically isolating heavy or noisy loads (e.g., AHU motors) from experimental power lines by means of dedicated isolation transformers;
(v) avoiding the routing of noisy power cables in proximity to sensitive experimental areas;
(vi) selecting low-EM-emission devices and verifying electromagnetic compatibility under operational conditions (e.g., for UPS systems);
(vii) mitigating start-up inrush currents through the use of soft-starter devices.

% possibili mitigazioni ?
%\textcolor{red}{Descrivere possibili eventuali azioni di mitigazione (cavi intrecciati, canaline schermate... possiamo citare due ARTICOLI indicati da Chincarini ~\cite{BravoRodriguez_2019, Canova_2009})}.
%These kind of actions were judged too invasive and not compatible with the Virgo commissioning, therefore none could be implement.

%\noindent
%These actions are typically invasive and should be considered in the laboratory project phase. 
Over time, several mitigation measures have been implemented at Virgo, leading to a significant reduction of ambient magnetic noise in the CEB experimental area between 10~Hz and a few~kHz. In particular:
(i) the CEB electrical system was renovated during the Advanced Virgo installation, including cable relocation away from the experimental area and the adoption of twisted-pair wiring \cite{eleSYS_upgrade};
(ii) the CEB UPS was replaced with a lower inter-harmonic-noise model;
(iii) during the O3 run, an isolation transformer was installed to separate the neutral conductors of the Virgo Mode Cleaner and Central Buildings, eliminating magnetic noise generated by the Mode Cleaner HVAC system and radiated inside the central experimental hall \cite{Envpaper2020};
(iv) soft-starter devices installed on the water chillers of the Mode Cleaner Building resulted in an approximately threefold reduction of the peak current amplitude \cite{VIR-0181C-21}.

\section{Conclusions}
\label{sec:Conclusions}

Technical solutions were devised and implemented to reduce the low-frequency noise produced by the Virgo HVAC systems, with the goal of mitigating risks for the Advanced Virgo plus, Phase I project. A significant reduction in both acoustic and seismic noise levels has been achieved in the two Virgo terminal halls. This improvement is primarily due to the replacement of the forward-curved fan with a backward-curved unit, which resulted in a significant broadband acoustic noise reduction in the 1–50 Hz frequency band and a factor of two reduction of the hall floor vibration in the 7–70 Hz range.
In parallel, dedicated studies were performed to characterize HVAC noise emissions and to identify the main propagation paths of vibro-acoustic and magnetic disturbances.

In the system under study, the dominant acoustic coupling path toward the halls was identified as the direct transmission of airborne noise from the centrifugal fan through the duct network. Seismic noise was mitigated using standard commercial solutions, including damped springs designed to provide a low-frequency cut-off around 3 Hz. Vibrations transmitted along the ducts were further reduced by applying high-density damping materials. A secondary but still relevant contribution arose from water-borne pressure fluctuations generated by circulation pumps, propagating through the pipe network and exciting pipe vibrations. This effect was mitigated by reducing pumps speed and isolating unused pipe branches. No specific mitigation actions were required for electromagnetic noise from the HVAC system; however, general design guidelines were derived to limit emissions and transmission paths in future installations.

Future upgrades should consider larger-diameter fans operating at lower rotational speed to maintain the required airflow while reducing noise emissions. Improvements in the thermal insulation of conditioned areas would decrease the overall thermal load of the HVAC system, allowing lower airflow rates and air velocities. Proper sealing of the halls is also essential, as it enables the required over-pressure to be maintained with reduced inlet airflow.

Noise generated by HVAC systems remains a major concern for facilities hosting next-generation gravitational-wave detectors, which aim to minimize infrastructure-induced disturbances. Underground facilities will impose additional constraints but are expected to face similar low-noise challenges. The results presented here are intended to provide practical guidance and design recommendations for future low-noise infrastructures.

% Ringraziamenti
\ack{
The authors would like to thank Davide Soldani (EGO) for his technical support during the installations and for insightful discussions on noise mitigation strategies, Andrea Paoli (EGO) for his assistance in providing technical information, and Rosario De Rosa (Universit\`{a} di Napoli “Federico II” and INFN Sezione di Napoli) for stimulating scientific discussions throughout the experimental work and data analysis.
%for stimulating scientific discussions on the investigations and the results obtained.
}

%\funding{Sample text inserted for demonstration.}
% This section is a list of funder names and grant numbers

%\roles{Sample text inserted for demonstration.}
% List author names and the contributions made to the article, using terms from the NISO Contributor Roles Taxonomy (CRediT) https://credit.niso.org

%\data{Sample text inserted for demonstration.}
% For more information on IOP Publishing's research data policy see: https://publishingsupport.iopscience.iop.org/questions/research-data/

%\suppdata{Sample text inserted for demonstration.}

%\section*{References}
\bibliographystyle{iopart-num}
\bibliography{Bibliography}

\end{document}